\documentclass[useAMS,usenatbib]{mn2e}

\usepackage{graphicx}
\usepackage{times}

\begin{document}
 
\title[The entropy core in simulated galaxy clusters.] {The entropy core in galaxy clusters: numerical and physical effects in cosmological grid simulations.}
\author[F. Vazza]{F. Vazza$^{1}$%
\thanks{%
 E-mail: vazza@ira.inaf.it} \\
$^{3}$ INAF/Istituto di Radioastronomia, via Gobetti 101, I-40129
Bologna, Italy}
\date{Accepted ???. Received ???; in original form ???}
\maketitle

\begin{abstract}
A flat distribution of low gas entropy in the core region
of galaxy clusters is a feature commonly found in Eulerian cosmological
simulations, at variance with most standard simulations of Smoothed Particles Hydrodynamics fashion.
From the literature it is still unclear whether  
this difference is entirely due to numerical artifacts (e.g. spurious transfer
from gravitational energy to thermal energy), to physical mechanisms (e.g.
enhanced mixing in Eulerian codes) or to a mixture of both.
This issue is related to many still open lines
of research in the characterization of the dynamical evolution of the baryons
in galaxy clusters: the origin of the cool core/non-cool core bi-modality,
the diffusion of metals within galaxy clusters, the interplay between
Active Galactic Nuclei and the Intra Cluster Medium, etc.

In this work, we aim at constraining at which extent the entropy core is affected by numerical effects, and which are the physical reasons for its production in cosmological
runs.
To this end, we run a set of 30
high resolution re-simulations of a $\sim 3 \cdot 10^{14}M_{\odot}/h$ cluster of galaxies
with a quiet dynamical history, using modified versions of the cosmological Adaptive-Mesh-Refinement code ENZO and investigating many possible (physical and numerical) details
involved in the production of entropy in simulated galaxy clusters.

We report that the occurrence of a flat entropy core 
in the innermost region of massive cluster is mainly due to 
hydro-dynamical processes resolved by the numerical code (e.g. shocks and mixing motions) and that additional spurious effects of numerical origin (e.g. artificial heating due to softening effects) affect the size and level of the entropy core only in a minor way.

Using Lagrangian tracers we show that the entropy profile of non-radiative simulations is produced by a mechanism of {\it ``sorting in entropy''} which takes place with regularity during the cluster evolution. The evolution of tracers
illustrates that the flat entropy core is caused by {\it physical}
mixing of subsonic motions (mostly driven by accreted sub-clumps) within the shallow inner cluster potential.

Several re-simulations were also produced for the same cluster object 
with the addition of radiative cooling, uniform pre-heating at high redshift ($z=10$)
and late ($z<1$) thermal energy feedback from AGN activity in the cluster, in order 
to assess the effects of such mechanisms on the final entropy profile of the cluster.
We report on the infeasibility of balancing the catastrophic cooling (and recovering a flat entropy profile) by means of the investigated trials for AGN activity alone,  
while for a sub-set of pre-heating models, or AGN feedback plus pre-heating models,
a flat entropy distribution similar to non-radiative runs can be obtained
with a viable energy requirement.
Complementary analysis are presented also for a major merger cluster, obtaining
similar results and achieving a generally good consistency with X-ray  data for the entropy distribuion in real galaxy clusters.

\end{abstract}

\label{firstpage}

 
 \section{Introduction}
\label{sec:intro}

The presence of a flat entropy core in the center of non-radiative galaxy clusters simulated 
with Eulerian grid codes, and its complete absence in the core of galaxy clusters simulated with
Lagrangian approaches (such as Smoothed-Particle-Hydrodynamics codes, SPH), has been object of
an interesting debate in the last few years and among a number
of different groups (e.g. Frenk et al.1999; Voit et al.2005; Dolag 
et al.2005; Lin et al.2006; Wadsley et al.2008; Tasker et al.2008; Mitchell
et al.2008; Springel 2010; Abel 2010).

The issue of the real inner entropy profile of non-radiative 
galaxy clusters may be loosely connected to the case of real
clusters, which are interested by cooling and energy feedback by
several astrophysical sources; nonetheless, this topic represents 
one of the main diagnostic to compare cosmological simulations of
galaxy clusters performed with different numerical approaches. 

\bigskip

The evolution of gas in the simulated Universe, when only shock
heating is present as {\it explicit} source of energy dissipation, represents
a rather simple scenario to study the thermodynamics of cosmic baryons starting
from simple initial conditions.
Understanding the systematics that affect the generation of entropy in the
different codes, even in this rather idealized setup, would represent a 
useful step forward in all research topics dealing with the numerical
treatment of non-reversible processes in astrophysical plasmas at all
scales. 
In non radiative simulations of galaxy clusters, the task of understanding the correct
entropy distribution in the innermost cluster regions
is made complicated by a few circumstances: 
a) the radius of the entropy core produced in grid codes
is rather small $r_{core} \sim 0.1 R_{vir}$ (where $R_{vir}$ is the cluster virial
radius), and even the most resolved cosmological simulations can concentrate only
a moderate number of resolution elements, of the order of $N \sim 10^{2}-10^{4}$, inside this region: therefore resolution and sampling problems may always be present; b) the baryon accreted by a cluster are interested by several dynamical processes across their evolution (shock heating, violent relaxation, gas mixing and 
sloshing of the Dark Matter peak) which distribute entropy in clusters in different ways;  disentangling the various effects within the same galaxy cluster is usually not a trivial task; c) due to the typical radial entropy distribution
in galaxy clusters, {\it physical} mixing driven by matter sub-clumps in-falling from the outside regions, and {\it numerical} mixing or spurious heating from N-body noise would have the same net effect, leading  to an increase of gas entropy within the densest regions in clusters; unluckily, the different numerical methods are prone to numerical
mixing in a ways difficult to quantify. 
Therefore very similar entropy configurations may be degenerate respect to various interplays between physical
and numerical effects along the whole cluster evolution, and
specific numerical tests aiming at the close comparison between re-simulations
of the same objects with different numerical methods are highly desirable in this respect.

From the literature, the first clear indication of a fundamental difference between the results of SPH codes and grid ones in galaxy cluster simulations
was presented in the Santa Barbara Comparison project (Frenk et al.1999). In this work,
evidence was obtained that the innermost entropy  radial distribution in cluster simulated in grid
methods such as ENZO contain a nearly isoentropic core inside $\sim 0.1 R_{vir}$, at variance
with SPH codes; later works basically confirmed this trend also at higher resolutions
(e.g. Voit et al.2005; Wadsley et al.2008). Several reasons were   
suggested to interpret this discrepancy: over-mixing in grid codes (e.g. Wadsley
et al.2008), spurious N-body heating from DM particles in the cluster core 
(e.g. Lin et al.2006; Springel 2010), Galilean invariance in the gravity solver of
grid codes (Tasker et al.2008; Robertson et al.2010), lack of physical mixing in SPH codes (e.g. Dolag 
et al.2005; Agertz et al.2007; Abel 2010), pre-shocking in SPH (e.g. O'Shea et al.2005)  etc.
To date, the most detailed analysis of the generation of entropy has been 
presented by Mitchell et al.(2009), by studying idealized cluster mergers with a non-cosmological setup with the SPH code GADGET2 (Springel et al.2005) and the grid code FLASH (Fryxell et al.2000).
The authors provided striking evidences that the mechanism
at work in setting the different entropy level between SPH and grid codes takes place 
at the time of the closest encounter between the colliding structures, and it is related to the suppression of
mixing in SPH because of artificial viscosity, which highly suppresses hydro instabilities and mixing  motions respect to grid codes. 
It would be now interesting to extend the results of this seminal paper to fully
cosmological simulations, and to the case of clusters in which the mass is assembled
with realistic and different dynamical history (major mergers or regular smooth accretions).

\bigskip

The present paper is first devoted to constrain at which extent all the numerical
effects cited above may affect also the specific distribution of gas entropy in realistic
and high resolution cosmological simulations. Secondly, this paper
is devoted to focus on the physical mechanisms for the generation and the spreading
of gas entropy within clusters, in relation to gravitational mechanisms (e.g. shock
heating and mixing motions) and to non-gravitational ones (e.g. radiative
cooling and energy feedbacks from astrophysical sources such as AGNs). 
To this end we produced a set of 30 re-simulations
of the same galaxy cluster, with a final mass of $M \approx 3.1 \cdot 10^{14}M_{\odot}/h$ and a very quiet dynamical history for most of its evolution
(complementary results for a major merger cluster are shown in the Appendix).

The first part of this work (Sec.\ref{sec:numerics})  explores many of the possible
numerical mechanisms which may lead to the formation of an inner entropy core in {\it cosmological}
cluster simulations, by using customized re-simulations with the adaptive mesh refinement 
code ENZO (Norman et al.2007). 
In detail, Sec.\ref{subsec:amr} discusses the role of the mesh refinement
strategy adopted in the simulation; Sec.\ref{subsec:dual} estimates the role
played by cold unresolved flows and N-body gravitational heating, making changes
to the ``dual energy formalism'' method adopted in ENZO; Sec.\ref{subsec:smooth}
investigates the influence of the softening length in the computation of 
the gravitational force in the PM method; Sec.\ref{subsec:gas_res} 
compares re-simulations adopting different maximum resolution to compute hydro-dynamics
of baryons.

In Sec.\ref{subsec:tracers} the {\it physical} generation of entropy and the volume spreading of it during cluster evolution is investigated by means of Lagrangian
tracers, injected and evolved in non radiative runs.

Radiative cooling can completely alter the above picture, in systems characterized 
by a cooling time shorter compared to the cluster lifetime (e.g. Katz \& White 1993; Fabian 1994). Indeed the hot gas phase in the core region of these systems is removed by the radiative losses, and the inward motion of the cooling gas would
produce the theoretical ``cooling flow'' scenario.
However, drastic cooling flows are not observed in real clusters, and additional {\it non-gravitational} heating mechanisms are need to restore (or keep) the cooling gas on an higher adiabat (e.g. Evrard \& Henry 1991; Kaiser 1991; Lloyd-Davies, Ponman \& Cannon 2000).
Two of the most promising scenarios in this respect are the pre-heating scenario (e.g. White 1991), in which gas is heated before it collapses within structures, or AGN feedback (e.g. Churazov et al.2000), in which the inner entropy is
raised by the energy released, through different channels, out of the regions
surrounding the growing super massive black hole. 
Many works investigated the above mechanisms with cosmological simulations, 
by implementing non-gravitational heating mechanisms in
SPH or in grid methods (e.g. Borgani et al.2002; Borgani et al.2005; Heinz et al.2006; Sijacki \& Springel 2006; Younger \& Bryan 2007; Burns et al.2008; Mc Carthy et al.2009; Teyssier et al.2010).  
However comparing the outcomes of these studies is made complex by the great number of assumption and
parameters often involved in the modelization of feedback sources. Also considering
the underlying fundamental differences of the hydro methods discussed above, 
the characterization of the inner entropy profile of 
radiative cluster with non-gravitational heating mechanisms at play is a still very open topic of research for simulations.
For instance, qualitatively similar models applied to GADGET (Borgani et al.2002) or ENZO simulations (Younger \& Bryan 2007) has lead to different conclusion about the efficiency of pre-heating at the scale of galaxy clusters or galaxy groups. The ability of cluster merger in quenching (or slowing down) the cooling catastrophe in radiative simulations is also a debated issue (Burns et al.2008; Poole et al.2008). 

In the second part of the paper (Sec.\ref{sec:phys}) we focus on the additional {\it physical} mechanisms
which are able to affect the gas entropy distribution in real galaxy cluster, by studying in detail
the effects of a) radiative cooling (Sec.\ref{subsec:cool}); b) non-gravitational heating by a "uniform" heating mechanism
in the early Universe (Sec.\ref{subsubsec:ph}); c) non-gravitational heating by a central AGN with outflows, within
an already formed cluster (Sec.\ref{subsubsec:agn}-\ref{subsubsec:ph_agn}). 
Our goal is not that of constraining the most likely extra heating mechanism at work in real galaxy cluster, but rather to show what is the net effect of {\it plausible} heating mechanism on the entropy distribution of a cluster with an 
ongoing cooling flow, using a budget for the energy release under control.

Section \ref{sec:conclusion} finally summarizes our discussions of the results and our conclusions, while in the Appendix complementary tests studying the numerical and physical generation of entropy in major merger cluster are reported for completeness.

\section{Numerical Code and Setup}
\label{sec:simulations}

The computations presented in this work were performed using the 
ENZO code, developed by the Laboratory for Computational
 Astrophysics at the University of California in San Diego 
(http://lca.ucsd.edu).

ENZO is an adaptive mesh refinement (AMR) cosmological hybrid 
code highly optimized for high performance computing
(Norman et al.2007 and references therein). 
It uses a Particle-Mesh method to follow
the dynamics of the collision-less Dark Matter (DM) component (Hockney \& Eastwood 1981),
and a Eulerian solver based on the  Piecewise Parabolic 
Method (PPM, Woodward \& Colella, 1984). 

\bigskip
The adopted
simulational setup is the same as in Vazza, Gheller \& Brunetti (2010).
In summary, cosmological initial conditions were produced with 
nested grid/DM particle distributions of increasing in order to 
achieve high DM mass resolution 
in the region of cluster formation; also an implemented mesh
refinement scheme was applied to trigger mesh refinement
based on gas/DM over-density and/or velocity jumps
across cells.
In this scheme, a normalized 1--D velocity jump across 3 adiacent cells in the scan direction (at a given refinement level) is recursively computed as
as $\delta \equiv |\Delta v/v_{min}|$, where $|v_{min}|$ is the minimum velocity, in absolute value, among the 3 cells. The scheme is made manifestely non Galileian invariant by the presence of $v_{min}$; this problem is unavoidable in
this kind of simulations, because in principle every forming shock wave moves on a different reference frame, and a run-time procedure to account for this 
would represent a too large computational effort.
In order to have this effect under control we performed many convergence tests
with idealized and cosmological simulations with ENZO, finding showing that a very
 good numerical convergence is achieved in cluster simulations by fixing $\delta = 3$ (more detailed discussions can be found in Vazza et al.2009; Vazza, Gheller \& Brunetti 2010).

All the re-simulations presented here focus on the evolution
of the same galaxy cluster, which was the most massive one 
of the sample already presented in Vazza, Gheller \& Brunetti (2010).

In order to role of various numerical effects (such as resolution,
softening in the gravitational force, etc.) various re-simulations
starting with the same initial conditions were produced.
Table \ref{tab:tab1} lists the details of all runs performed for this project. 
The re-simulations employing additional physics (such as pre-heating and AGN
feedback) where produced with original implementations made starting from the public 1.5
version of ENZO.

\bigskip

\begin{table}
\label{tab:tab1}
\caption{Main characteristics of the performed runs. Column 1: run identification code;
column 2: cell resolution  at the maximum refinement level; column 3: maximum softening length; column 4: 
mesh refinement strategy:"D"= gas/DM over-density refinement; "V"=velocity jumps refinement; column 5: additional numerical parameter of the simulations.$\eta_{1}$, $\eta_{2}$ and $M_{thr}$ are parameters
involved in the dual energy formalism switch (Sec.\ref{subsec:dual}); ``PH'' means that uniform pre-heating has
been adopted, assuming an entropy increase of $S_{0}$ at $z=10$; ``J'' means that $\epsilon_{jet}$ extra energy  
has been injected at $z \leq 1$ within the cluster, assuming AGN feedback.}
\centering \tabcolsep 5pt 
\begin{tabular}{c|c|c|c|c}
 ID & Max Res. [kpc/h] &  $\epsilon_{soft}$. [kpc/h] & AMR & note \\  \hline
	 	
    R0 & 25 & 50 & DV & non-radiative\\ \hline
	
    R1 & 25 & 25 & DV & non-radiative\\ 
    R2 & 25 & 50 & D & non-radiative\\
    R3 & 25 & 25 & D & non-radiative\\
    R4 & 12.5 & 25 & DV & non-radiative \\
    R5 & 12.5 & 25 & D & non-radiative \\
    R6 & 12.5 & 12.5 & DV  &  non-radiative\\
    R7 &  12.5 & 12.5 & DV & non-radiative \\ 
   R18 &  25 & 100 & D & non-radiative \\ 
   R19 &  25 & 100 & DV & non-radiative \\ 
   R20 &  25 & 12.5 & D & non-radiative \\ 
   R21 &  50 & 50 & D & non-radiative \\ 
   R22 &  50 & 50 & DV & non-radiative \\ \hline
		
    R8 & 25 & 25 & DV & $\eta_{1}=10^{-2}$   \\
    R9 & 25 & 25 & DV & $\eta_{1}=10^{-4}$   \\
    R10 & 25 & 25 & DV & $\eta_{2}=1$   \\
    R11 & 25 & 25 & DV & $\eta_{2}=10^{-2}$ \\
    R12 & 25 & 25 & DV & $M_{thr}=1.1$ \\
    R13 & 25 & 25 & D  & $M_{thr}=1.5$ \\ \hline

    R15 & 25 & 25 & DV & cooling \\
    R16 & 25 & 25 & D & cooling \\
    PH1 & 25 & 25 & DV & cool.+PH($10keV cm^{2}$) \\
    PH2 & 25 & 25 & DV & cool.+PH($100keV cm^{2}$) \\
    PH3 & 25 & 25 & D & cool.+PH($100keV cm^{2}$) \\
    PH4 & 25 & 25 & DV &  cool.+PH($200keV cm^{2}$) \\
    B1 & 25 & 25 & DV  &  cool.+J($10^{58}ergs$) \\
    B2 & 25 & 25 & DV  &  cool.+J($10^{59}ergs$) \\
    B3 & 25 & 25 & D  &  cool.+J($10^{59}ergs$) \\
    B4 & 12.5 & 12.5 & DV  &  cool.+J($10^{58}ergs$) \\
    B5 & 25 & 25 & DV  &  cool.+PH+J($2\cdot 10^{57}ergs$) \\
\end{tabular}
\end{table}

\begin{figure}
\includegraphics[width=0.45\textwidth]{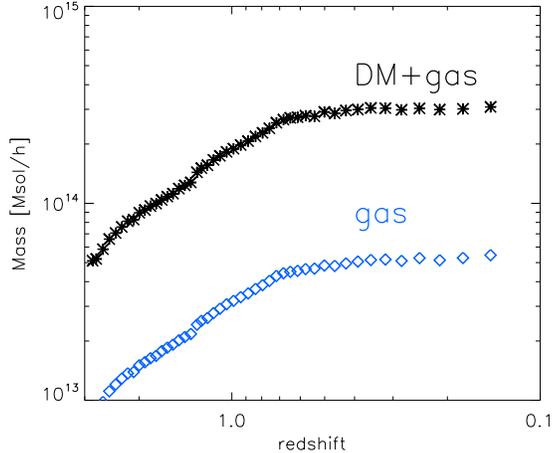}
\caption{Redshift evolution of the Dark matter plus gas mass ({\it black}), and of total gas mass ({\it blue}) inside the virial radius of the galaxy cluster studied in this work.}
\label{fig:mass_evol}
\end{figure}

\begin{figure}
\includegraphics[width=0.485\textwidth]{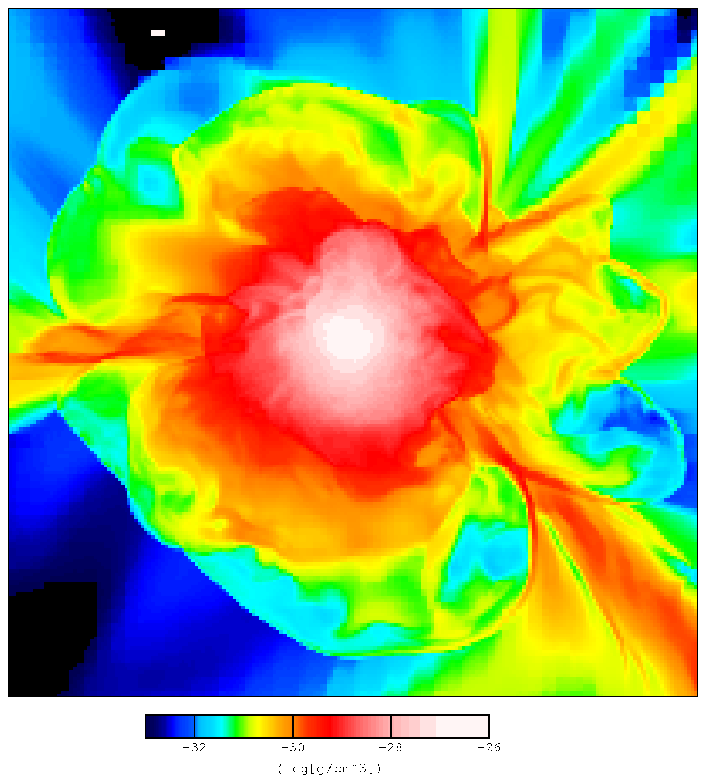}
\includegraphics[width=0.485\textwidth]{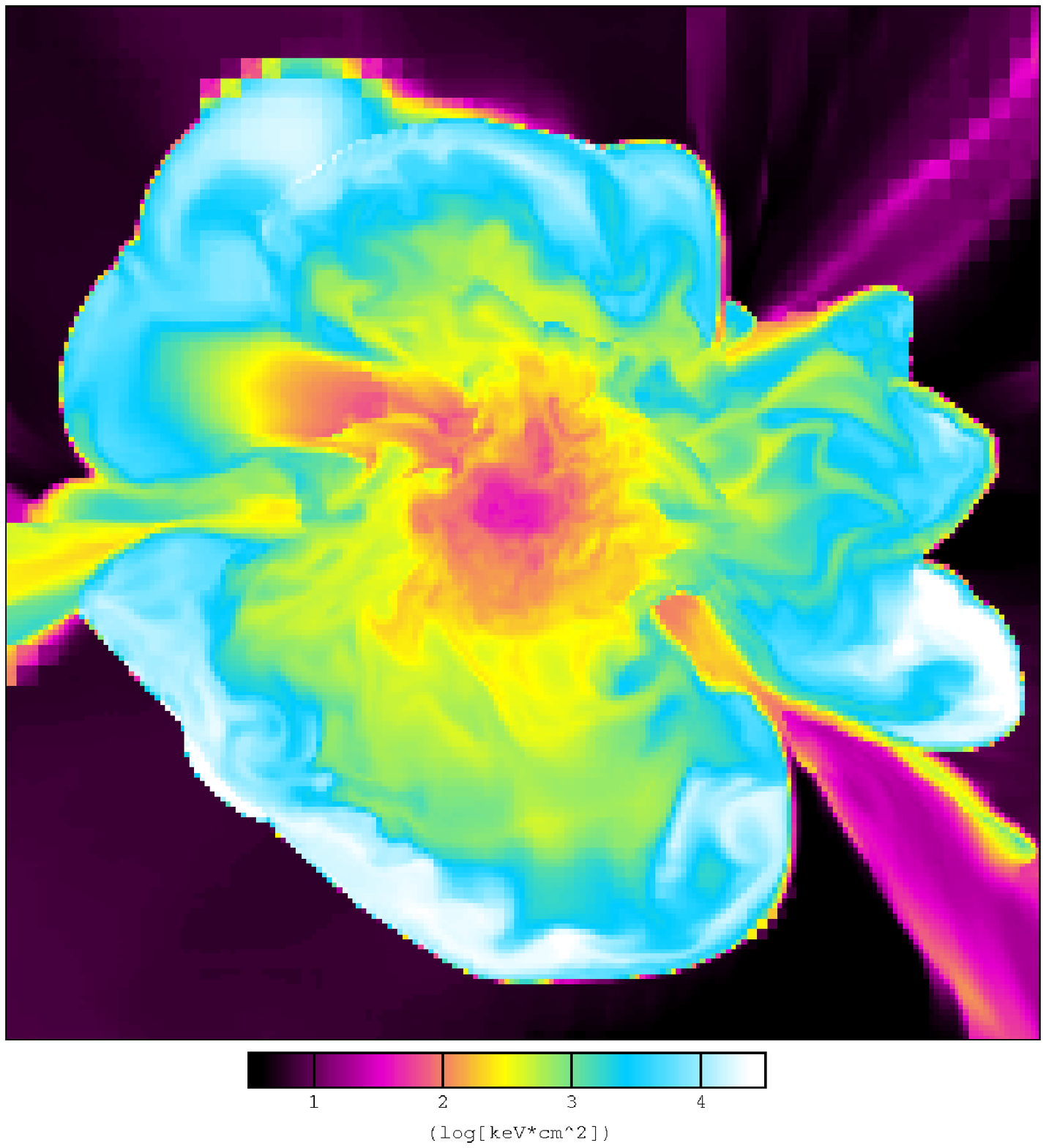}
\caption{Maps of gas density ({\it top}) and gas entropy ({\it bottom}) for a slice
crossing the center of the cluster at $z=0$ (run R0). The side of the 
image is $\approx 5.5Mpc/h$ and the width of the slice is one cell $=25kpc/h$.}
\label{fig:map_r0}
\end{figure}

\begin{figure*}
\includegraphics[width=0.45\textwidth]{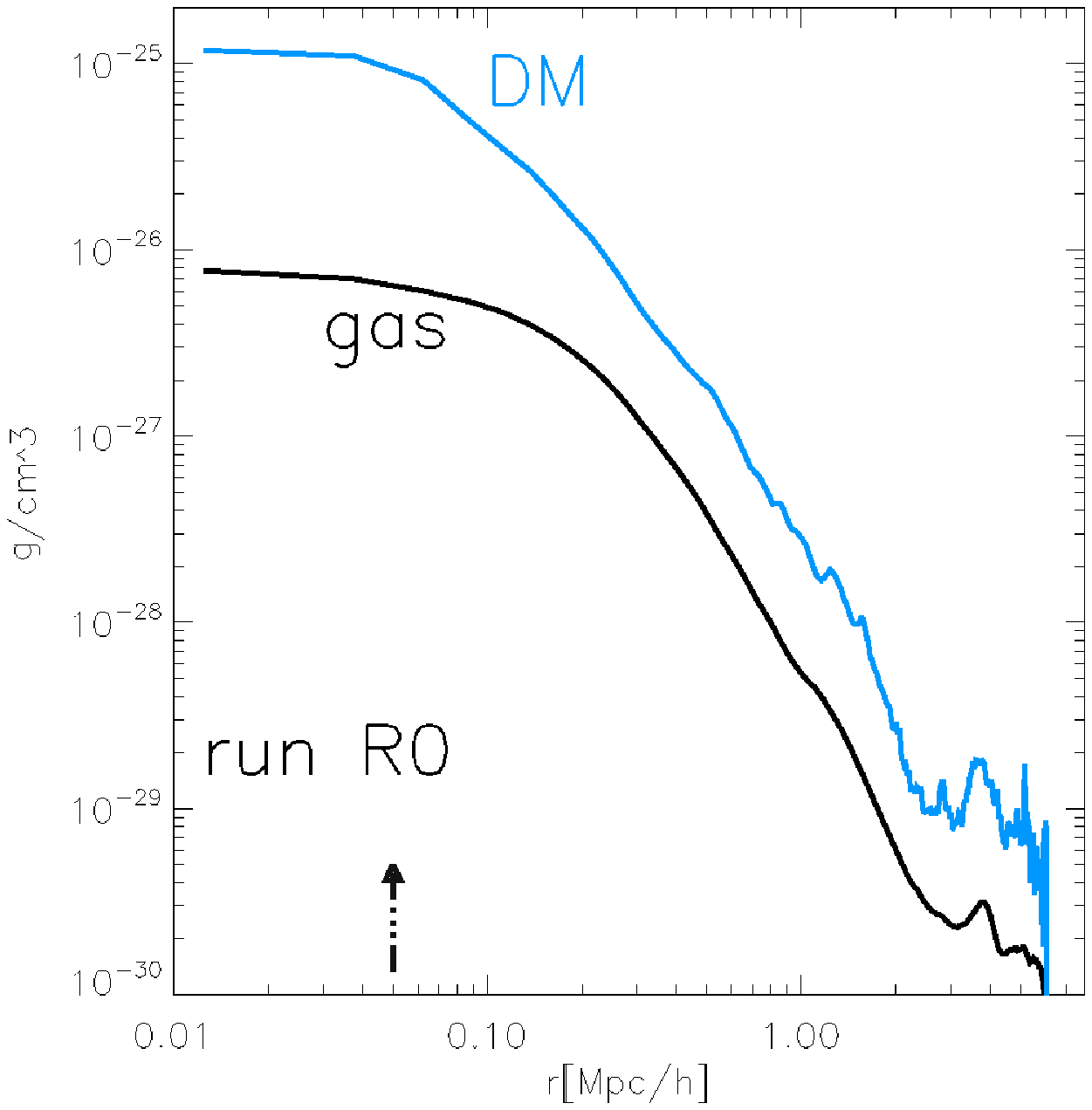}
\includegraphics[width=0.45\textwidth]{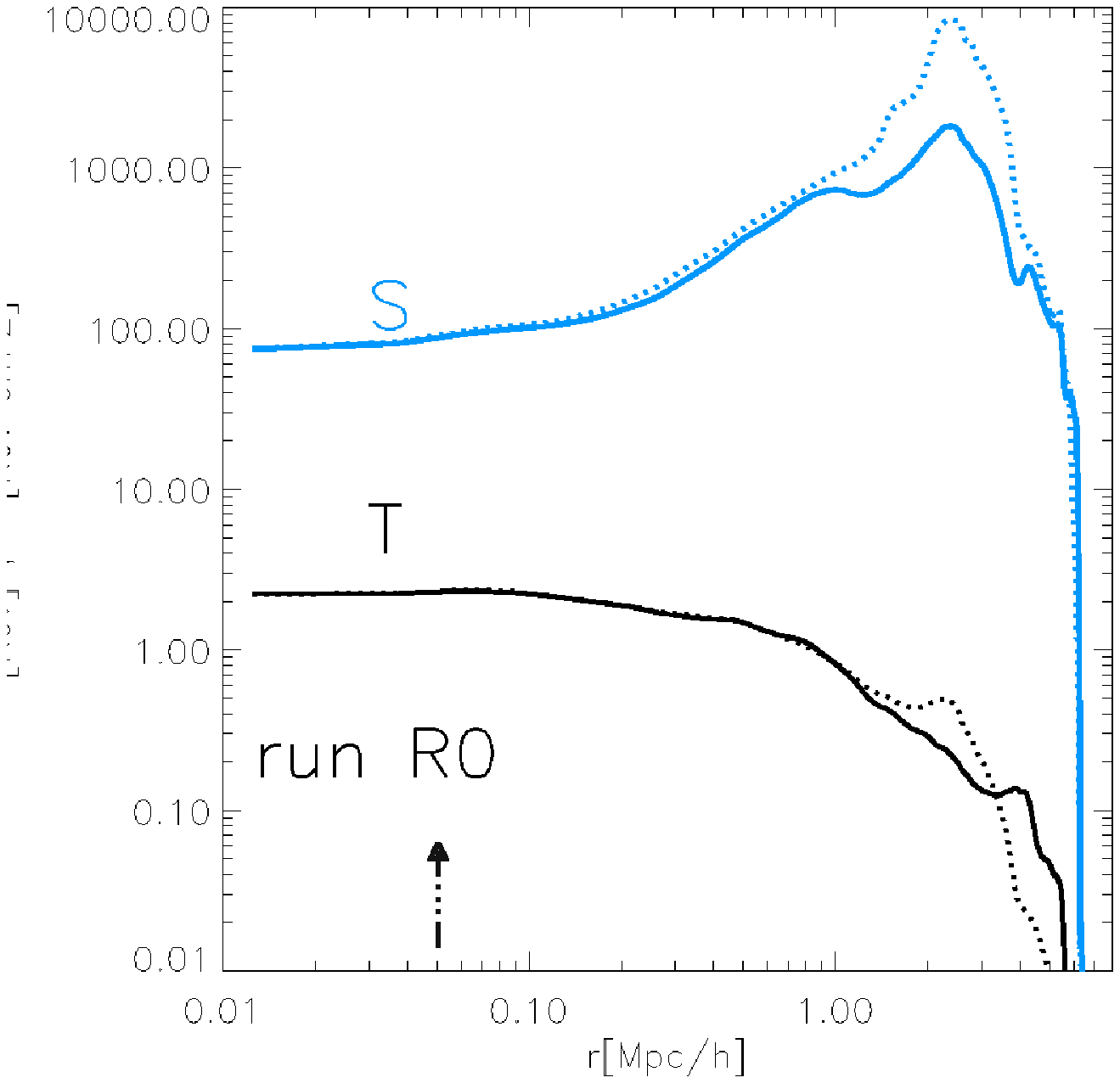}
\caption{Radial profiles for the run R0, showing gas density and DM
density profiles (left panel) and gas temperature and gas entropy (right
panel). In the right panel, the solid lines refer to mass-weighted profiles, while the dashed lines refer to volume-weighted profiles. The vertical arrows show the softening
length adopted in this run.}
\label{fig:prof_r0}
\end{figure*}

\section{Numerical effects on the entropy production.}
\label{sec:numerics}

\subsection{The fiducial run}
\label{subsec:fiducial}
The fiducial cluster run of this project is R0, which has the same
numerical setup and mesh refinement strategy already adopted in 
other works by our group (Vazza et al.2009; Vazza, Gheller
\& Brunetti 2010; Vazza et al. 2010).
The assumed cosmology is the ``Concordance'' $\Lambda$CDM model, with
parameters: $\Omega_0 = 1.0$, $\Omega_{BM} = 0.0441$, $\Omega_{DM} =
0.2139$, $\Omega_{\Lambda} = 0.742$, Hubble parameter $h = 0.72$ and
a normalization for the primordial density power spectrum $\sigma_{8} = 0.8$.
The mesh refinement on gas/DM over-density is triggered since the beginning of
the simulation inside a cubic with the side of $\approx 5 R_{vir} \approx 12Mpc/h$ \footnote{In the following, we will refer to this region as to the 'AMR region'.}, $z=30$, while the additional mesh refinement triggered
by velocity jumps is activated from $z=2$.

The lowest resolution level ($l=0$) inside the AMR region is $220kpc/h$, while 
the maximum refinement level ($l=3$) is set to $25kpc/h$. 
The DM matter particles have a mass resolution of $6.7 \cdot 10^{8} M_{\odot}/h$,
and the comoving gravitational softening is $\epsilon = 50kpc/h$. 

This fiducial cluster run does not include radiative cooling or any heating 
mechanism other 
than gravitational collapse 
or shock heating. 
 
Fig.\ref{fig:mass_evol} shows the redshift evolution 
of the total mass and of the gas mass inside the virial cluster region, 
measured with a spherical over-density method.
The trend of the cluster mass growth and the visual inspection of movies
of its evolution confirms that 
this cluster does not
experience any violent merger event for $z<1.5$, and roughly $\sim 70$ per cent
of its mass has been already assembled at $z \sim 0.6$.
At $z=0$ the cluster has a total mass of $\approx 3.1 \cdot 10^{14}M_{\odot}/h$ , an average temperature of $2.2keV$ and a virial radius of $R_{v}=1.89 Mpc/h$.

In Fig.\ref{fig:map_r0} we show
the maps of gas density and gas 
entropy {\footnote{All throughout this paper, we will refer to $S \equiv \frac{P}{\rho^{\gamma}}$ as to the ``gas entropy'', as usually done in cosmological numerical simulations, where $P$ is the gas pressure, $\rho$ is the gas density within a cell and $\gamma$ is the adiabatic index.}} through the center of the cluster.
The cluster is quite regular in shape, with sharp circular shock
structures and a well defined inner entropy floor of 
size $r_{core} \sim 100kpc/h$ surrounded by the steep increase of the ICM entropy, up to a
 maximum at about $\sim 2Mpc/h$ from the cluster center. 
This cluster may be considered as a ``prototype'' of 
relaxed clusters in the local Universe with intermediate mass, as 
produced by non-radiative cosmological simulations of Eulerian fashion. For 
complementary results on larger mass clusters simulated with the same approach, we
refer the reader to recent high-resolution re-simulations presented in Vazza et al.(2010), and references therein.

Figure \ref{fig:prof_r0} shows the 
radial profiles of gas density, DM density, temperature and entropy 
for run R0 at $z=0$. In the right panel of this Figure we compare gas mass weighted profiles and volume weighted profiles for the cluster run: the two estimates
provide consistent results within $<2Mpc/h$ from the cluster
center. Unless specified, in what follows we will make use of gas density
weighted averages.

We investigated the ``robustness'' of this cluster representation by
adopting several changes in this simple setup, which will be discussed in the 
next Sections.

\begin{figure}
\includegraphics[width=0.48\textwidth]{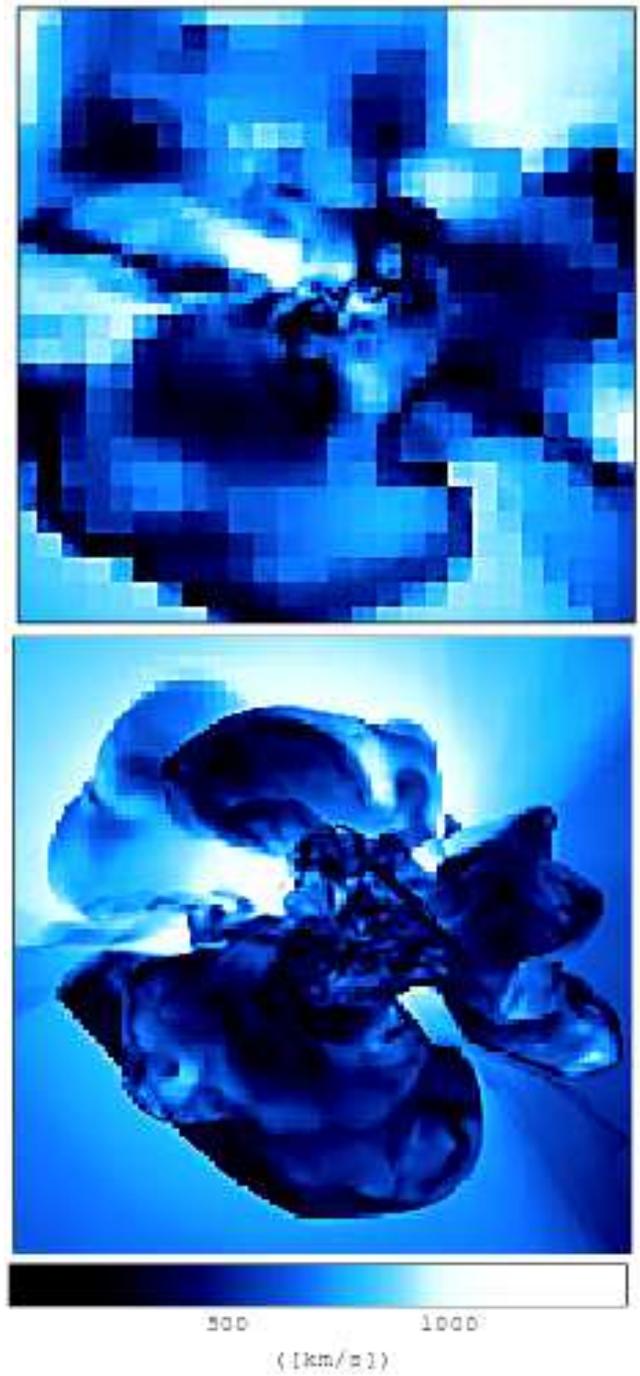}
\caption{Slices showing the absolute value of the 3--D velocity field for run R0 (Top panel) and run R2 (Bottom panel) at $z=0$. The size of the image is as in Fig.\ref{fig:map_r0}.}
\label{fig:map_vel}
\end{figure}

\begin{figure}
\includegraphics[width=0.4\textwidth]{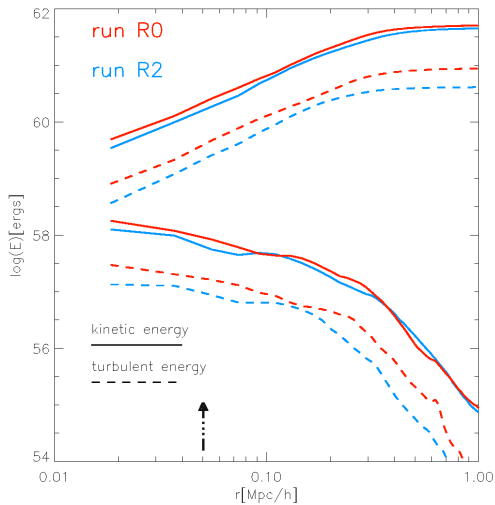}
\caption{Radial profiles of kinetic energy for run R0 (red solid) and for run R2 (blue solid), and of ``turbulent'' kinetic energy for the same runs (dashed
lines). The 4 top lines refer to the total kinetic/turbulent energy inside a given radius, while the 4 bottom lines refer to the values within shells of width $25kpc/h$ at the same radii. The vertical arrow show the softening length adopted in these runs.}
\label{fig:prof_vel}
\end{figure}

\subsection{The role of the mesh refinement strategy.}
\label{subsec:amr}

Shock heating during the gravitational collapse in the forming cluster has a leading role in the production of
the baseline entropy distribution within clusters, since no other
{\it physical} mechanisms of irreversible heating are present in simple 
non-radiative cosmological simulations (e.g. Voit et al.2005). 
Previous works in the literature (Dolag et al.2005; Iapichino \& Niemeyer 2008; Vazza et al.2009; Vazza, Gheller \& Brunetti 2010) have suggested that 
an increase of the level of chaotic motions within simulated clusters is expected as soon as 
the effects of a coarse resolution or artificial viscosity are limited by ad-hoc techniques, and as a result
typically higher entropy level is found in the core region of cosmic structures.
Also, the different ability in modeling shocks and mixing in the central 
phase of cluster mergers has proved to be responsible for the bulk of the difference
between SPH or grid simulations of galaxy clusters (Mitchell et al.2009).
Obtaining a good spatial resolution of chaotic and mixing motion in the ICM is therefore 
crucial and the mesh refinement strategy outlined in 
Sec.\ref{subsec:fiducial} is designed for this purpose (see also Iapichino \& Niemeyer 2008).

We compare the cluster entropy profile in the fiducial run (R0) 
against a re-simulation using the standard mesh refinement based on gas/DM
over-density alone (run R2).
Figure \ref{fig:map_vel} shows the maps of the absolute value of the velocity
field in the two runs, for a slice crossing the center of the cluster at $z=0$.
Similarly to what reported in Vazza et al.(2009), the extra refinement
strategy allows us to preserve a very accurate description of the low density outer accretion region, and on the mixing motions following the crossing of satellites even
if the involved over-density is not enough to trigger other refinements.
The radial distributions of the kinetic energy and ``turbulent'' energy
(we fiducially consider as ``turbulent'' the components of the 3--D velocity field which are characterized 
by a coherence scale smaller than $<200kpc/h$, see Vazza et al.2009 Sec.4.2 for further
details) are shown in Fig.\ref{fig:prof_vel}.
As expected, R0 shows a slightly higher level of kinetic energy all across the cluster volume, and an kinetic energy in small-scale chaotic motions by a factor $\sim 2-3$ 
compared to the standard run R2. 

\begin{figure}
\includegraphics[width=0.4\textwidth]{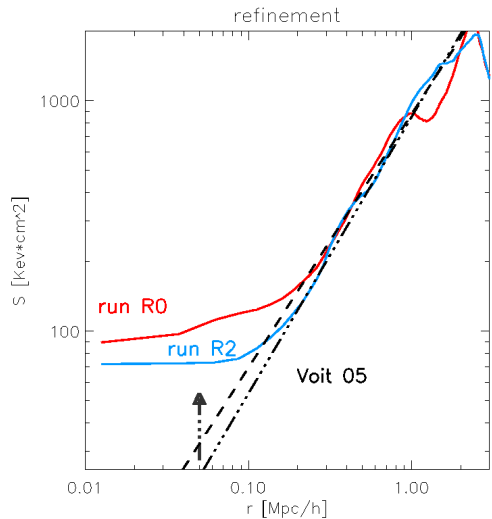}
\caption{Gas entropy profile for run R0 (red) and run R2 (blue). The additional
black lines show the fits profiles presented in Voit et al.(2005), for
$S \propto r^{1.1}$ (dashed) and $S \propto r^{1.2}$ (dot-dashed). The vertical arrow show the softening length adopted in these runs.}
\label{fig:prof_ref}
\end{figure}

The entropy profiles of the two simulations are shown in Fig.\ref{fig:prof_ref}.
The additional black lines show the fit
profile presented in Voit et al.(2005), where a
systematic study of entropy profile in non-radiative cluster simulations
(of SPH and grid fashion) were presented. 
The run with the additional refinement strategy produces a significantly larger
entropy within the core region, by a $\sim 20-30$ per cent (the gap is  $30-50 keV cm^{2}$).
It is interesting to notice that the gas entropy and the gas kinetic/turbulent energy of
run R0 is larger compared to run R2 within approximately the same radius, $r<200kpc/h$
(the softening for the gravitational force here is $50kpc/h$ at the maximum refinement level).

Already from this, one would speculate that enhanced turbulent mixing motions are responsible for the level of gas entropy inside clusters; we will focus on this issue
with more detail in Sec.\ref{sec:phys}.

\begin{figure*}
\includegraphics[width=0.95\textwidth]{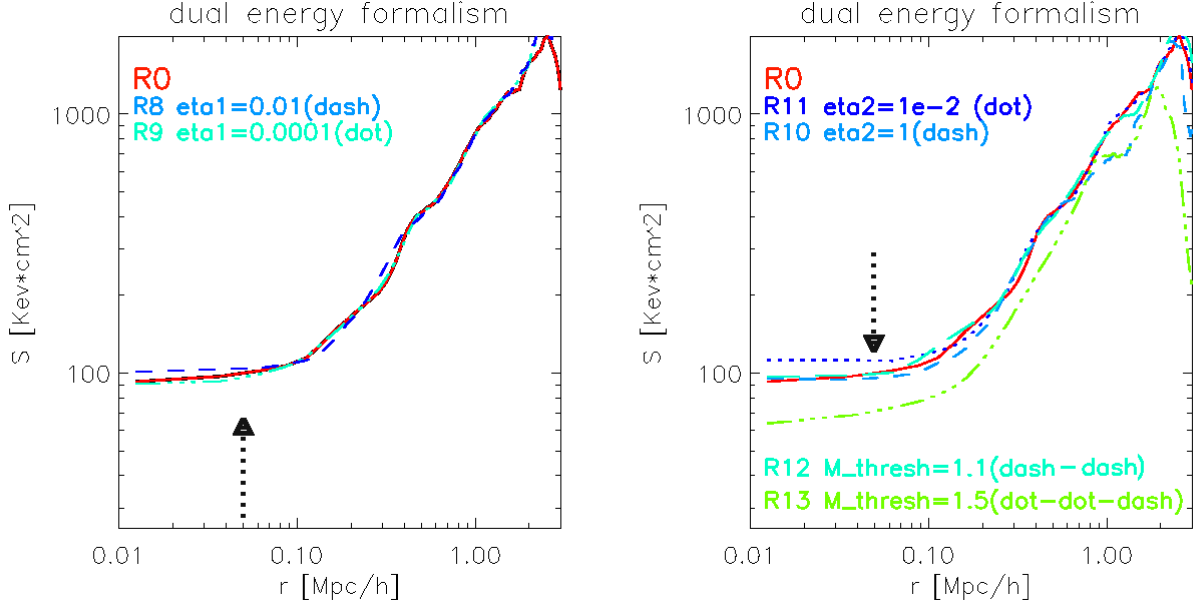}
\caption{Comparison of the entropy profiles for different
choices of the $\eta_{1}$ (R8,R9, left panel) and for different
choices of $\eta_{2}$ (R10,R11, right panel). Additional green profiles are shown
for ENZO re-simulations which adopted an additional switch condition on the Mach number to trigger the dual energy formalism (R12,R13). The vertical arrow in both panel shows the softening length adopted in these runs.}
\label{fig:prof_eta}
\end{figure*}

\subsection{The role of gravitational N-body heating and cold unresolved flows.}
\label{subsec:dual}

It has been suggested in the recent past that the  
entropy core in grid simulation may be due to 
an incorrect handling of the gas thermal energy/entropy, in the case of poorly resolved cold flows dominated by kinetic energy (e.g. supersonic
flows in the rarefied Universe) and due to noise-induced heating in
virialised structures (Lin et al.2006; McCarthy et al.2007; Springel 2010).  This second mechanism may arise because the small-scale
velocity fluctuations induced by the N-body gravitational field can
be readily dissipated  by the mesh-based hydrodynamics,  causing a spurious
heating of the gas.
In Springel (2010) the initial conditions for the
Santa Barbara Comparison (Frenk et al.1999) were evolved with the AREPO code  
using an energy-entropy switch scheme in order to decide if the 
gas entropy had to be
updated as the difference between the total energy and the kinetic
energy (as customarily done in grid codes) or by directly using the entropy 
equation instead. This was done in order to ensure that
in the case of transonic or subsonic motions ($M<M_{thresh}$, with $M_{thresh}=1.1$) the evolution of gas entropy was not dominated by any spurious dissipation, since a {\it conservative} entropy equation was followed.
The level of core entropy in the simulated cluster was found to be 
considerably reduced after the use of this switch, and the conclusion
was that the entropy in the core region of clusters simulated with standard methods
in grid simulations may be significantly affected by this numerical effect.

This issue is very interesting and deserves investigation; to this end we
re-simulated run R0 with different choices for the computation of gas thermal energy.
The strategy customairly  adopted in ENZO is to adopt a ``dual energy formalism'' (Ryu et al.1993; Bryan et al.1995) designed to 
compute the evolution of gas thermal energy whenever the internal energy, $e$, cannot be
calculated correctly as the difference between total, $E$, and kinetic energy (due to the fact that $E>>e$ in highly supersonic flows, and round-off errors may dominate).
We briefly recall here its design:
a first switch condition is followed to update the gas pressure in the case
of very cold flows

\begin{itemize}
\item $p=\rho(\gamma-1)\cdot(E-v^{2}/2) $   if   $(E-v^{2})/E>\eta_{1} $
\item $p=\rho(\gamma-1)\cdot e $     if  $(E-v^{2})/E<\eta_{1}$
\end{itemize}

where $p$ is the gas pressure, $v$ is the modulus of the velocity field and $\gamma=5/3$. 
The second switch is adopted to update the internal gas energy without
advecting numerical errors from each cell's local neighborhood:

\begin{itemize}
\item $e=(E-v^{2}/2)$   if   $\rho(E-v^{2})/max[\rho E]_{neigh}>\eta_{2}$ 
\item $e=p/\rho (\gamma-1)$    if   $\rho(E-v^{2})/max[\rho E]_{neigh}<\eta_{2} ,$

\end{itemize}

where $max[\rho E]_{neigh}$ is the maximum total energy in the (1--D) neighborhood of the cell. 
This approach ensures that $e$ is not contaminated by errors advected by
the total energy formulation (Bryan et al.1995).
However we note that in the above switch conditions the presence of $v$ makes them non Galilelian
invariant, because bulk velocities affect the exact value of $E$ and of
$max[\rho E]$. To keep this problem under control, one need to resort to convergence 
studies by varying the numerical parameters involved in the switch; the customary
values set in ENZO cosmological simulations are $\eta_{1}=10^{-3}$ 
and $\eta_{2}=0.1$.

As a first  step to investigate the role played by the dual energy formalism in the
production of the entropy level in clusters, we performed several tests by re-simulating the fiducial run adopting different choices of the threshold values
involved in the above "switches": $\eta_{1}=10^{-2}$ (R8), $\eta_{1}=10^{-4}$ (R9), or $\eta_{2}=1$ (R10),  $\eta_{2}=10^{-2}$ (R11).

The motivation for this kind of tests is that if the low entropy gas sitting the in the cluster center at $z=0$ is 
reminiscent of the entropy production prior to the cluster 
virialization, then any artifact present in the ``cold'' ($T<10^{5}K$) 
Universe at early redshifts (i.e. due to an incorrect handling of
the gas internal energy, or to features related to non Galileian invariance) will be highlighted by different choices
of the entropy switch. It should be noted that, in absence of a re-heating
UV background due to stars/AGN (e.g. Haardt \& Madau 1996), the baryon temperature outside structures can be as low as $T \sim 1-10 K$ in simulation of this kind.

By comparing the results of these runs with the fiducial one, we report that the net effect of the above changes is quite small, as shown in 
Fig.\ref{fig:prof_eta}: the maximum difference is  found for run R11 ($\eta_{2}=10^{-2}$) and results only in a $\sim 10$ per cent larger entropy inside the cluster core. Cold
unresolved flows are unlikely to be the reason of the entropy core
in grid simulations.

\bigskip

As a second step to investigate further  the possible role played by weak shocks
and extra-heating related to N-body gravitational noise, we 
supplemented the dual energy formalism in ENZO with an additional Galilelian-invariant switch based on the Mach number, with a procedure similar to Springel (2010).

As shown above the standard switch condition of the dual energy formalism
may be affected by the presence of large bulk velocities, which enter
in the total energy $E$, and thus the value of the parameter $\eta_2$ cannot 
be readily realted to the real Mach number of the flow.
For this reason we  implemented an on-the-fly shock in ENZO simulations, which
allows us to apply a Galileian invariant switch condition  (based on the Mach
number) to decide whether or not the total energy equation must be used to update
the gas dynamics.
This method relies on the temperature jumps method of Ryu et al.(2003) and
adopts the following procedure:
a) candidate shocked cells are identified according to the $\nabla \cdot \vec{v}<0$ criterion;
b) local gradients of gas entropy, $\nabla S$ and gas temperature $\nabla T$ are evaluated, and a shock
condition is matched whenever $\nabla T \cdot \nabla S>0$; c) the
shock Mach number is estimated by inverting the maximum temperature jump 
across the shocked cell:

\begin{equation}
T_{2}/T_{1}=\frac{(5M^{2}-1)\cdot(M^{2}+3)}{16\cdot M^{2}}.
\end{equation}

This procedure is followed on the fly during the simulation, and in this method the 
total energy equation is used to compute the gas internal energy only when
$\rho(E-v^{2})/max[\rho E]_{neigh}<\eta_{2}$  {\it and} $M<M_{thresh}$ simultaneously.
This procedure forces the code to update the gas thermal energy only in an adiabatic way, for flows with $M<M_{thr}$;
however we note that this formulation is not totally equivalent to the energy-entropy formalism discussed
in Springel (2010), because the internal energy equation does not explicitly conserve
gas entropy (however it conserves gas energy).
The internal energy equation only depends
on the second order of the gravitational potential, $\Phi$, through the 
coupling with the gas velocity field, $\sim \nabla v \sim \nabla^{2} \Phi$,
and therefore in this way the bulk of the (possible) N-body noise heating
should be greatly reduced.

We performed two re-simulations of R0 adopting $M_{thr}=1.1$ (R12) and  $M_{thr}=1.5$ (R13) (long dashed lines in the right panel of Fig.\ref{fig:prof_eta}).

We verified the difference in the total number of cells advanced
with the total energy equation in the standard switch ($\eta_2$ condition) or with 
the new switch ($M_{thr}$ condition) by applying the two methods to the same snapshot of the R0 run, at $z=2$ and $z=1$. We found with the first approach the number of cells
advanced with the total energy equation is $\sim 70$ ($\sim 63$) per cent of the total 
in the AMR regions at at $z=2$ ($z=1$); with the $M_{thr}=1.1$ condition this
ratio is decreased to $\sim 50$ ($\sim 45$) per cent of cells at $z=2$ ($z=1$), and
with $M_{thr}=1.5$ the ratio is $\sim 40$ ($\sim 30$) per cent of cells at $z=2$ ($z=1$).

When the switch conditions on the Mach number are applied since the beginning of the
cluster simulation, the run with $M_{thr}=1.5$ (run R13) shows a significantly
reduced (by a factor $\sim 50$ per cent) entropy production at all radii from the cluster center, but yet producing a very flat entropy distribution for $r<0.1 R_{vir}$.
However run R13 shows a nearly constant decrease of entropy at {\it all} radii from the cluster center compared to all the other runs. Considering that the distribution
of thermalized energy at shocks in our galaxy clusters simulations is a very steep function of $M$, with a well defined peak around $M \sim 2$ (e.g. Vazza, Brunetti \& Gheller 2009; Vazza et al.2010), the filtering procedure of run R13 is expected to filter out a $\sim 50$ per cent of the energy input from cosmological shocks. Therefore this 
explains why the large value of $M_{thr}=1.5$ removes a significant part of the 
{\it genuine} production of entropy at cosmological shocks, and suggests that the 
entropy level found in the other run is mainly produced by the action of {\it physical}
shocks dissipation.

When the Mach number for the switch is set to the same value used by Springel (2010), $M_{thr}=1.1$ (run R12), the final entropy profile of our cluster is found to be nearly identical to the fiducial run, with a  very similar flat entropy core. This is
somewhat at variance with the findings reported in Springel (2010), where spurious
entropy production was masked out by applying the $M_{thr}=1.1$ condition.
A possible explanation for this may be related to the different on-the-fly strategies adopted to compute in run-time the Mach numbers between cells;
further tests (also using other grid-based codes) may help to understand
at which extent the exact formulation of the Mach number switch can be crucial
to monitor the production of artificial entropy in cosmological cluster simulations.

To summarize what provided by our tests with ENZO AMR, the choice of the flow Mach number
to decide whether or not the total energy must be used to evolve the 
gasdynamics of cells, is found to affect the {\it normalization} of the cluster entropy profile
at $z=0$, but not the {\it shape} of the distribution (at least for the $M_{thr} \leq 1.5$
cases examined). 
Therefore the flat entropy core seems a pretty stable feature produced by ENZO AMR
runs, and does not seem strongly affected by the details of the formalism adopted for the dual energy
equation. At the present stage,  we consider unlikely that the numerical issues discussed above can be fully responsibile for the well known difference in the entropy profiles measured according to SPH or to grid methods.

\begin{figure}
\includegraphics[width=0.45\textwidth]{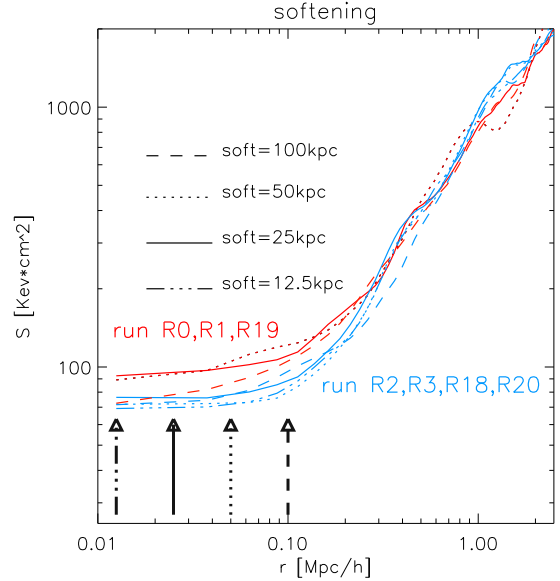}
\caption{Comparison of the profiles obtained for different
choices of the gravitational softening adopted in the N-body calculation
(solid lines: $\epsilon_{soft}=50kpc/h$; dotted lines: $\epsilon_{soft}=25kpc/h$) for
run adopting the standard refinement (R2,R3) or the velocity based one (R0,R1). The vertical arrows show the softening length of the various runs.}
\label{fig:prof_soft}
\end{figure}

\subsection{The role of the smoothing in the gravitational force.}
\label{subsec:smooth}

ENZO uses a particle-mesh N-body method (PM) to follow
the dynamics of collision-less systems (e.g. Hockney \& Eastwood 1981).  
DM particles are distributed onto a regular grid
using the cloud-in-cell (CIC) interpolation technique, forming a spatially
discretized DM density field.  The DM density is then sampled onto the
grid and the baryon density (calculated in the hydro method of the code) is added, 
and the gravitational potential is calculated  on the periodic root
grid using Fast Fourier Transform algorithms to solve the elliptic Poisson's equation.
Since the acceleration is the gradient
of the potential, the values of two potentials in close cells are required to calculate
it, and thus the effective force resolution is about twice of the cell size. To calculate more accurate potentials on sub grids in the case of adaptive
mesh refinement, the DM distribution is resampled onto the finer grids
using the same CIC scheme as for the root grid.
The new boundary values are obtained with the interpolation from 
the gravitational potential on the parent grid, and a multirelaxation
technique is used to obtain the gravitational potential at every point
within the sub grids (e.g. O'Shea et al.2005).

The softening length, $\epsilon_{soft}$, used to compute the gravitational force and to update the DM particle motions is bound 
to be a multiple of the cell size. In  ENZO $\epsilon_{soft}$ can be as small as the finest refinement
level in the volume, or an integer multiple of this.

Since the size of the entropy core is usually not much larger than the typical values of $\epsilon_{soft} $
found in the innermost cluster region of most of AMR simulations, it is possible that the entropy
core is an artifact due to the fact that for scales smaller than $\epsilon_{soft}$,
the gravitational force does not longer obey the $\propto 1/r^{2}$ scaling, but the $\propto 1/(r+\epsilon_{soft})^{2}$
scaling instead {\footnote{We note that however in recent ENZO AMR re-simulation of 
more massive galaxy clusters the scale of the entropy core is found to be $\sim 10$ larger than the gravitational softening (Vazza et al.2010).}}.

We investigated this issue by producing a several runs with identical initial but varying the minimum allowed $\epsilon_{soft}$ to compute gravity forces. 
For the the velocity-based refinement, we compared the fiducial run R0 
with run R1 ($\epsilon_{soft}=25kpc/h$) and with run R19 ($\epsilon_{soft}=100kpc/h$);
for the standard refinement scheme, we compared run R2 with run R3 ($\epsilon_{soft}=25kpc/h$), with run R18 ($\epsilon_{soft}=100kpc/h$) and with R20 ($\epsilon_{soft}=12.5kpc/h$).
Figure \ref{fig:prof_soft} shows that for a softening length of
$\epsilon_{soft} \leq 50kpc/h$  the flat entropy profile is a well converged feature in this cluster run, within a scatter of a $10$ per cent at most, which may be due 
to slightly different timings in the different runs.
For all softening smaller than $\epsilon_{soft}=50kpc/h$, we also
confirm that the two mesh refinement scheme always produce well separated entropy
floor inside $r<200kpc/h$.
For the sake of completeness, the same analysis has
been repeated in the case of a major merger cluster (see Appendix), leading
to consistent results.

Mitchell et al.(2009) performed similar tests using FLASH, and also reported that the resolution used to compute gravity forces does not play an important role in setting the entropy profile inside the cluster, at least for maximum $\epsilon_{soft}<40kpc$. 

In general, we conclude that the flat entropy core in AMR simulations is not due to 
spatial undersampling of the gravity forces, and that the flat entropy core is observed
even when $\epsilon_{soft} \sim 0.05-0.1 r_{core}$ (as in run R20).
We thus conclude that the entropy core in grid simulation is not dependent
on the adopted gravitational softening, for minimum softening length of $50kpc/h$
or smaller.

\begin{figure}
\includegraphics[width=0.45\textwidth]{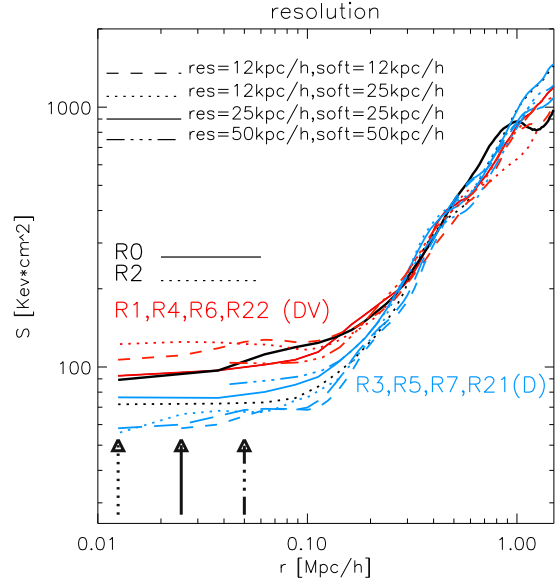}
\caption{Overview of the effects of gas resolution and gravitational
softening in the re-simulations. Red lines: runs adopting the velocity-based
refinement (R1,R4,R6). Blue lines: runs adopting the overdensity-based
refinement (R3,R5,R7). The additional black lines are for the fiducial
run (R0, solid line) and for run R2 (dotted line). The vertical arrows show
the softening length of the various runs.}
\label{fig:prof_res}
\end{figure}

\subsection{The role of gas resolution.}
\label{subsec:gas_res}
The maximum cell resolution adopted to compute the hydrodynamical equations
may play a role in setting the entropy content of a galaxy clusters, 
modifying the dynamics and propagation of shocks waves and modifying the
evolution of chaotic motions in the ICM.
The distribution of shocks
energy and Mach number in grid codes is a quite regular function of the underlying
grid resolution (e.g. Ryu et al.2003; Vazza, Brunetti \& Gheller 2009).
The average Mach number responsible for
most of the thermal energy
dissipation is a well converged quantity, for grid
resolutions better than $\sim 500kpc$, even if the exact distribution
of the high Mach number ($M>10$) tails is dependent on the underlying
grid resolution.
 
However in the tests usually found in the literature for grid simulations the above dependence is coupled with the dependence on $\epsilon_{soft}$ (which may also indirectly affect the production of shocks by changing the morphology of in-falling gas/DM sub-clumps). We aim here to disentangle the two effects, by keeping the maximum softening length fixed to $\epsilon_{soft}=25kpc/h$, bud varying the maximum gas resolution in the computation of fluid-dynamical effects.

In runs R1 and R3 we allow ENZO to refine
up to one level more, $l=4$, keeping the same setup and refinement strategy 
of runs R4 and R5, respectively.
If we compare the solid lines in Fig.\ref{fig:prof_res} (run R1,R3) and the
dotted lines (run R4,R5) we see that the trend with gas resolution is opposite 
in the two mesh refinement strategies: while in the velocity-based refinement
the entropy inside $r \sim 100kpc/h$ is {\it increased} by a $\sim 20-30$ per cent (R4), in the case of the overdensity-based refinement the inner entropy is 
{\it decreased} by a similar amount. The gap in entropy inside $r_{core}$ is of the
order of $\Delta S \sim 30-60 keV cm^{2}$.
We also re-simulated runs R4 and R5, allowing the code to
increase also the softening length up to $l=4$ (run R6, R7, long dashed lines): the reported trends are the same (see red and blue dashed lines in Fig.\ref{fig:prof_res}).
The re-simulations shows that the difference between the two refinement
schemes is always significant: a very flat entropy core inside $r<100kpc/h$
is produced in both cases, but a significant gap is found when comparing the two
strategies, with the implemented refinement scheme producing the larger value.
The trend is confirmed also by the comparison of two runs where the minimum softening
and cell resolution were fixed to $l=2$ ($50kpc/h$), run R22 and run R21 (dot-dashed lines in the same panel).

 The reason for the {\it opposite} trend in the velocity-based scheme can be understood by the increased role of mixing motions, which cause
a slightly more efficient inward transport of higher entropy material from
the cluster outskirt (see also Sec.\ref{sec:phys}). The efficiency of this mechanism is expected to increase as resolution is increased, since the velocity-based strategy is explicitly
designed to reduce the artificial dampening of small scale chaotic motions, which are
partly responsible of diffusive mixing in the ICM.
In the overdensity-based approach {\it clumps} are refined
more and more, and they can deliver low entropy gas in the innermost region
in a more efficient way; an increase of resolution also minimizes the effect of numerical
mixing and let the cold gas phase to survive longer (e.g. Wadsley et al.2008; ZuHone, Markevitch \& Johnson 2009). 
On the other hand in the velocity-based strategy the clumps are also more refined, but they are also more efficiently destroyed, before reaching the cluster center, by ram-pressure stripping at the outer regions and excite chaotic motions which are more long-living since they are not damped by the code; the net effect is an increase of entropy
inside $r_{core}$ in the velocity-based strategy.
This stresses the need of having an accurate description of shocks and turbulent motions of the ICM, since the inner entropy distribution does not only depend on the maximum resolution within the core, but also on the resolution at the 
outer regions, where the bulk of the cluster entropy is produced.
Convergence tests reported in Mitchell et al.(2009) and in ZuHone, Markevitch
and Johnson (2009) suggest that full convergence for the entropy profile in the standard refinement strategy is reached for a maximum resolution of $\sim 10kpc/h$ or smaller. A similar conclusion is likely also the velocity-based strategy, but we could not run so far a re-simulations reaching $l=5$ for computational limitations. However the trend of the red lines in Fig.\ref{fig:prof_res} suggest that convergence is near.

We conclude that even if the different mesh refinement strategies and the adopted maximum refinement levels can produce modification in the level of the inner entropy budget
(up to a factor of 2 at the peak resolution investigated), the flat entropy core is 
present in all cluster resimulations (see also the additional resolution tests in the Appendix). This again calls for a mechanism of {\it physical} nature, which drives an efficient
spreading of entropy enriched gas in the innermost regions of evolved
clusters; this will be investigated in detail in the next Section.

\begin{figure}
\includegraphics[width=0.45\textwidth]{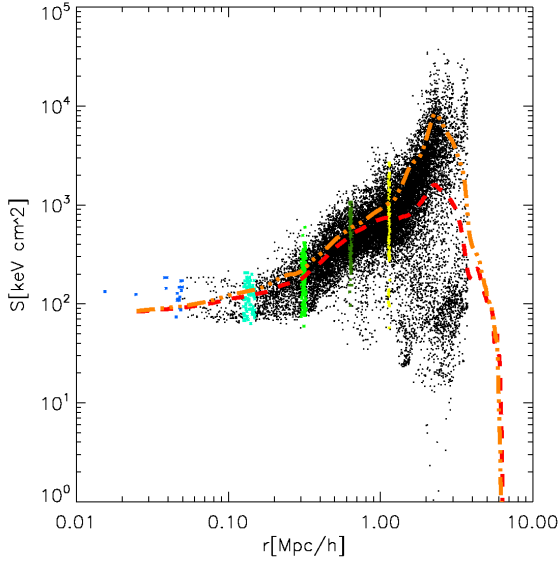}
\caption{Radial entropy distribution for all $N=10^{5}$ tracers evolved in run R0.
The colored sub sample shows the selections used for tests in Sec.\ref{subsec:tracers}.
The additional lines show the average entropy profile of the cluster (red/dashed:gas density weighted profile;
orange/dot-dashed: volume-weighted profile).} 
\label{fig:trac1}
\end{figure}

\begin{figure}
\includegraphics[width=0.45\textwidth,height=0.38\textwidth]{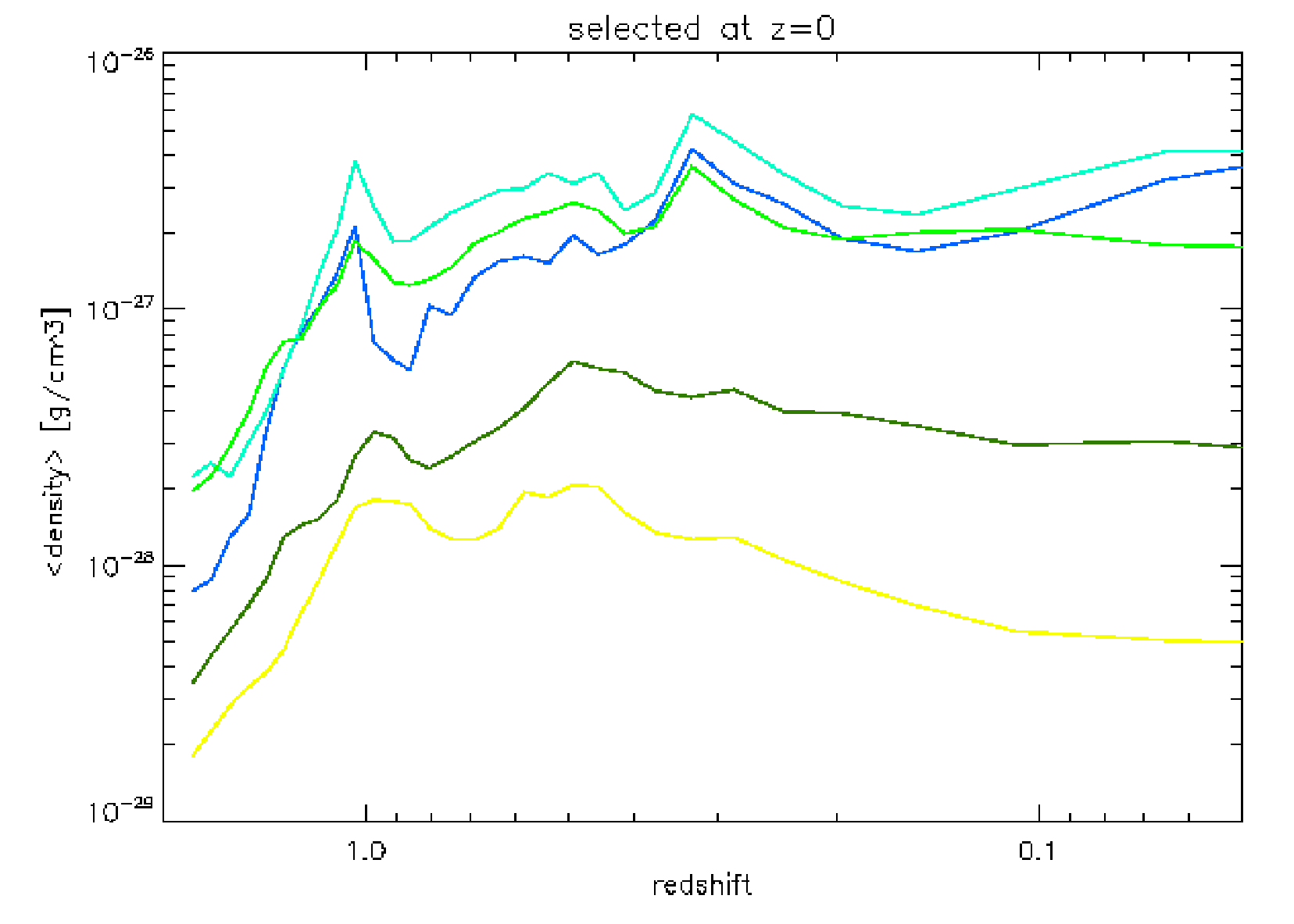}
\includegraphics[width=0.45\textwidth,height=0.38\textwidth]{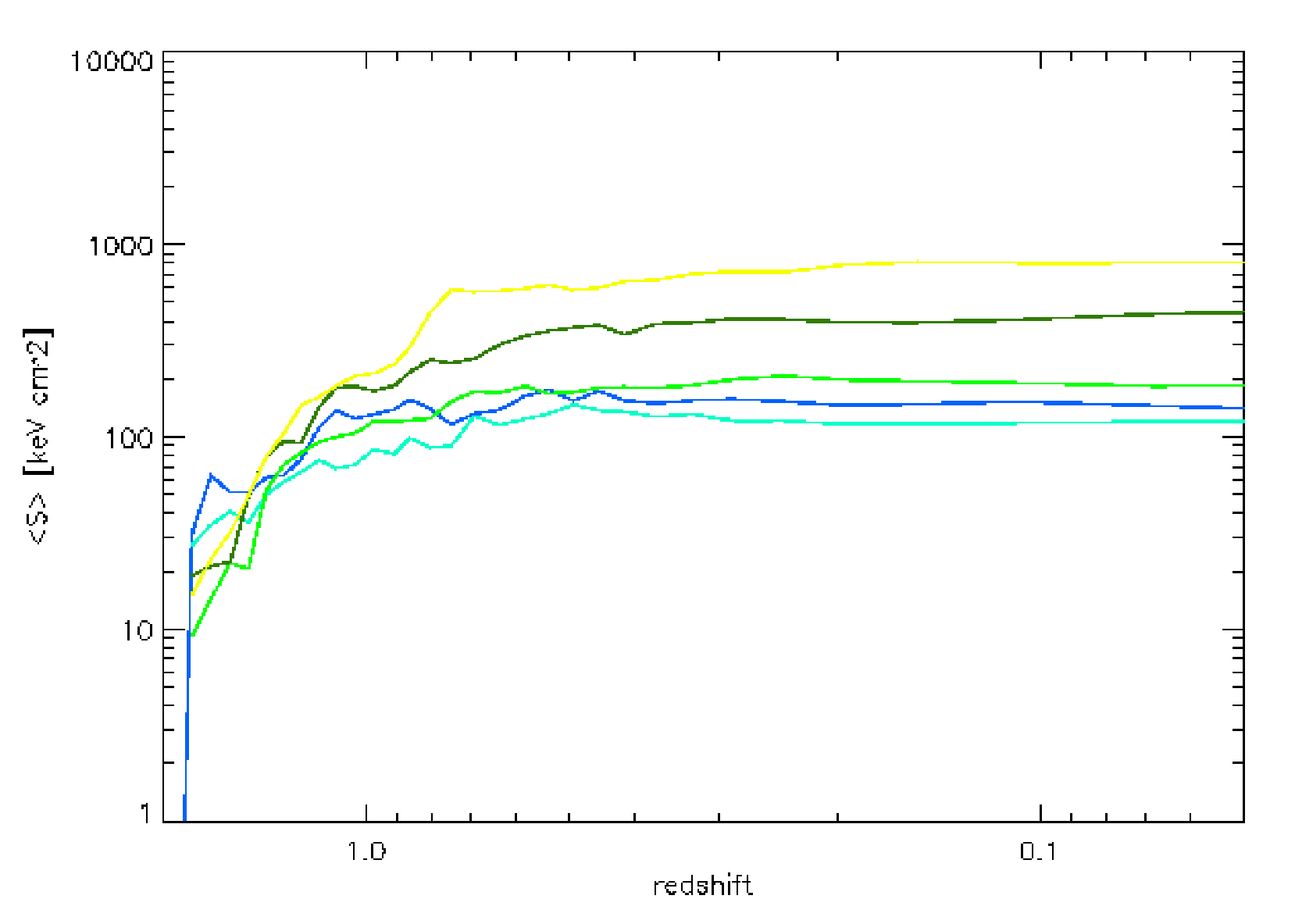}
\includegraphics[width=0.45\textwidth,height=0.38\textwidth]{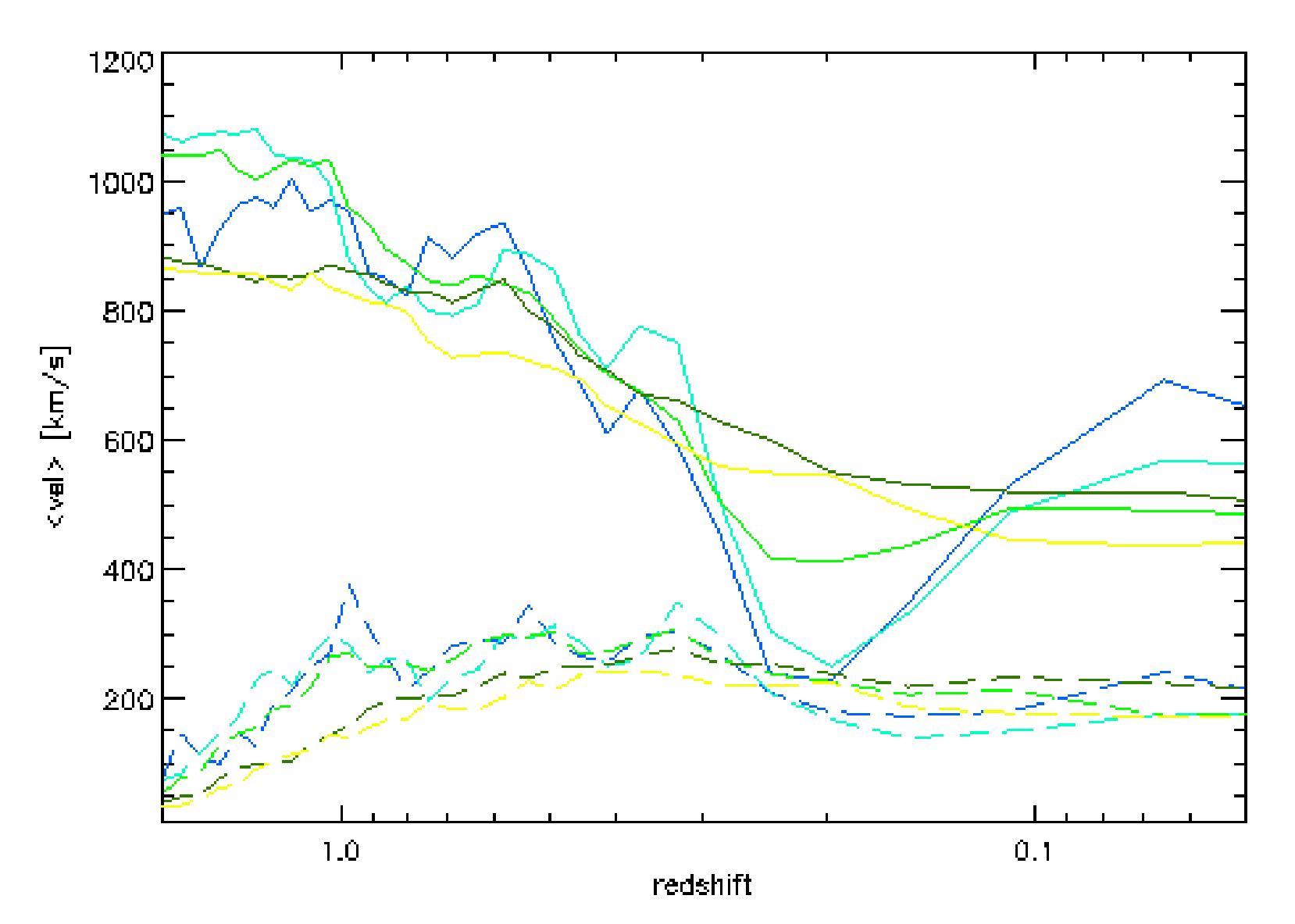}
\caption{Time evolution of gas density (top panel), gas entropy (central panel)
and mean gas velocity (or mean gas turbulent velocity, dashed lines) for the 5 groups
of tracers selected according to their position in the cluster atmosphere at $z=0$, as in Fig.\ref{fig:trac1} (see Sec.\ref{sec:phys} for details).}
\label{fig:trac2}
\end{figure}

\begin{figure}
\includegraphics[width=0.45\textwidth,height=0.38\textwidth]{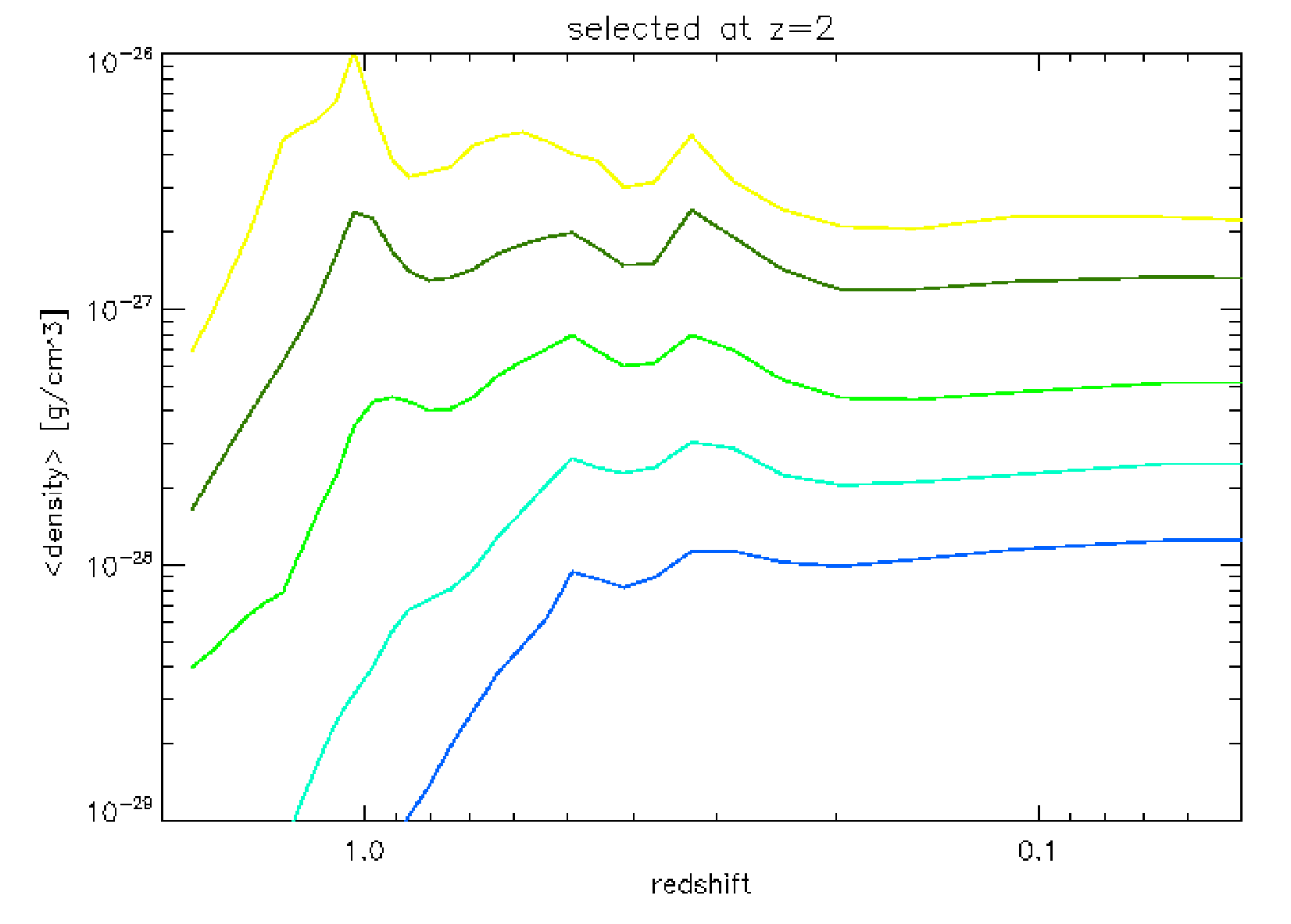}
\includegraphics[width=0.45\textwidth,height=0.38\textwidth]{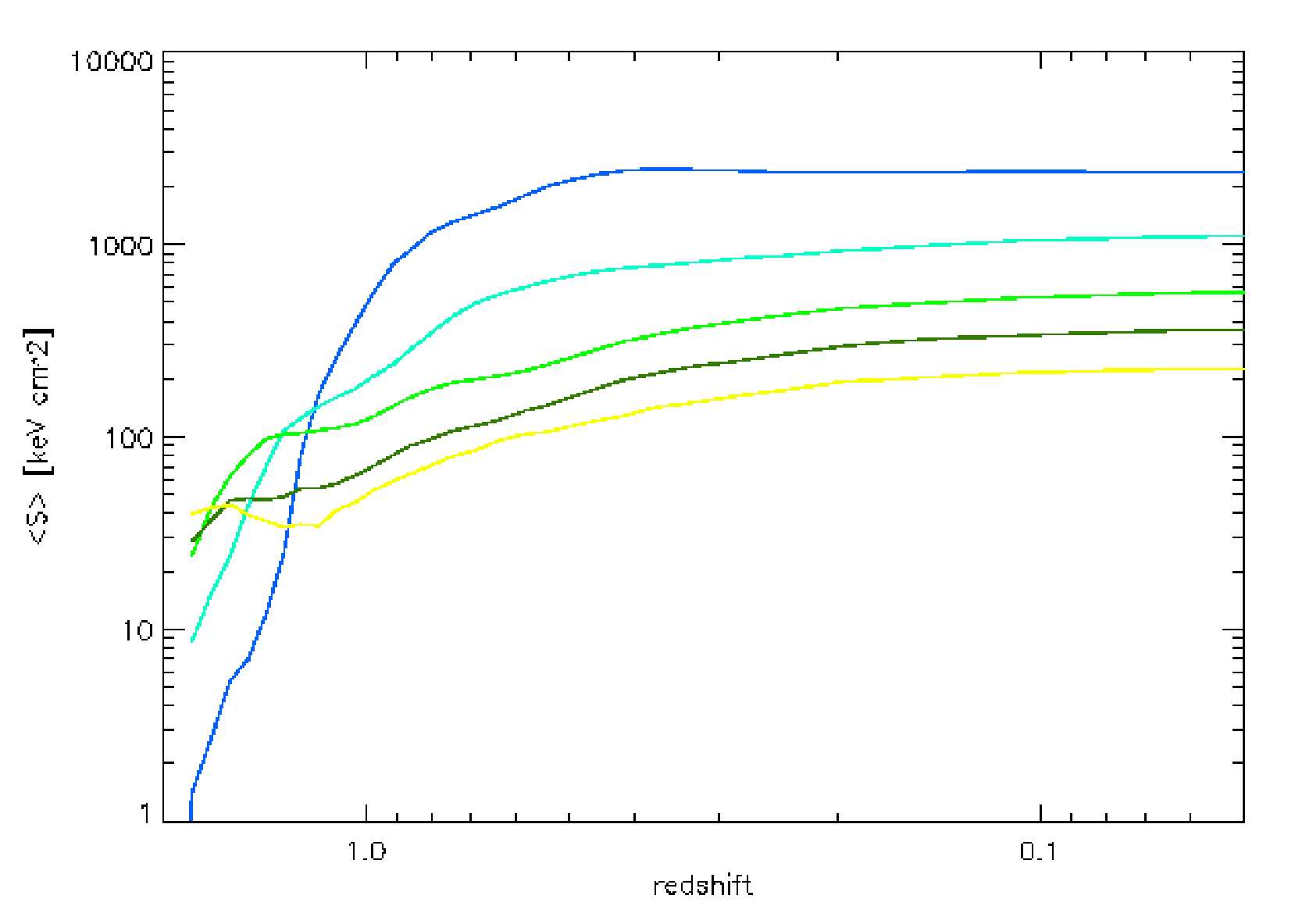}
\includegraphics[width=0.45\textwidth,height=0.38\textwidth]{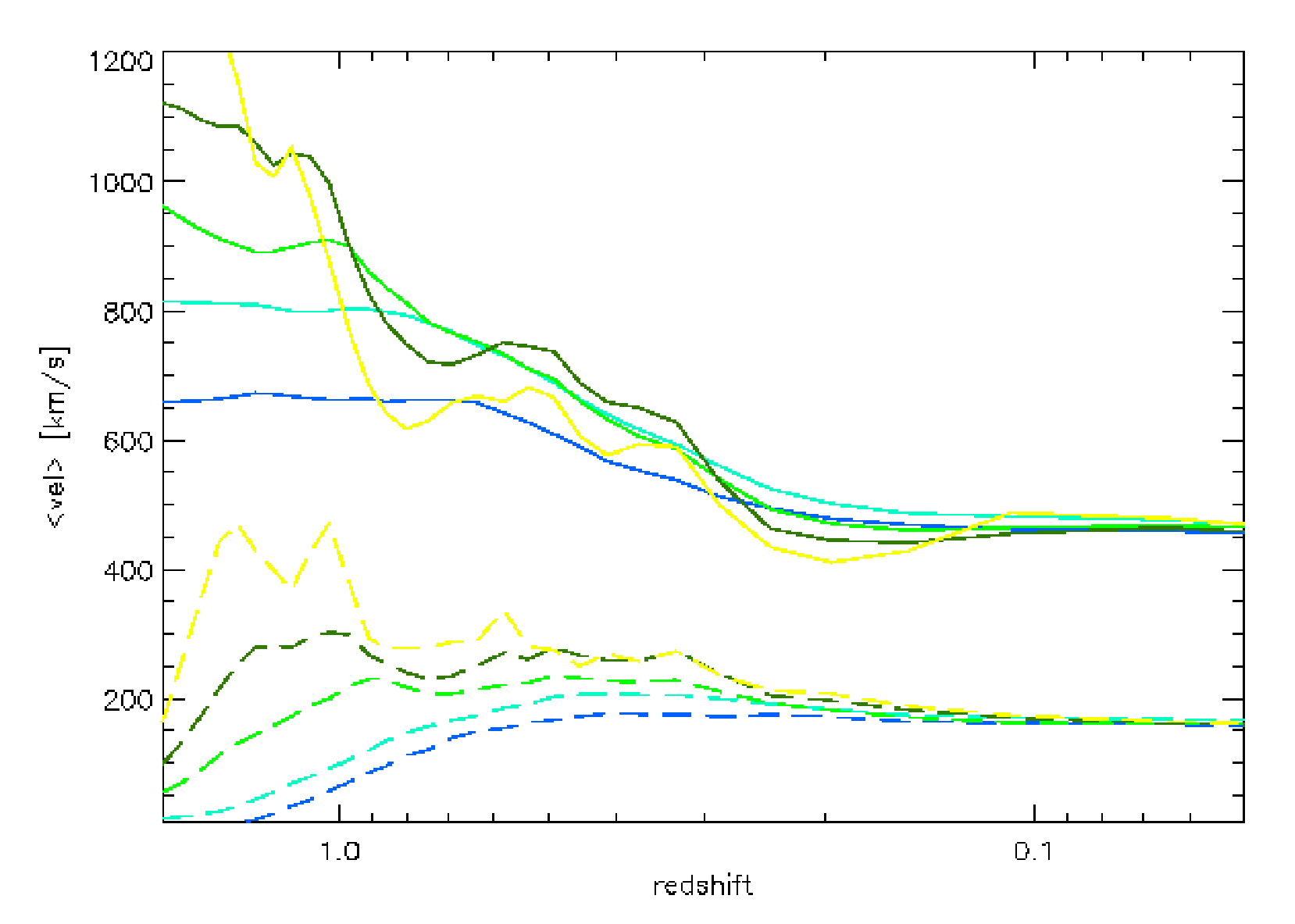}
\caption{Time evolution of gas density (top panel), gas entropy (central panel)
and mean gas velocity (or mean gas turbulent velocity, dashed lines) for the 5 groups
of tracers selected at $z=2$ according to their initial gas density (see Sec.\ref{sec:phys} for details).} 
\label{fig:trac3}
\end{figure}

\section{Physical effects on the entropy production.}

\label{sec:phys}

The numerical tests reported in the previous Sections have shown that
the presence of a regular low gas entropy distribution in galaxy
clusters simulated with grid-based techniques is a not a numerical
artifact, rather but a very stable feature against a number of important
changes in the possible setup of a cosmological simulation at high resolution.

At this point it is interesting to answer to the following questions:
a) what is the main {\it physical} mechanism which sets the inner gas entropy distribution 
in a forming galaxy cluster, in  non-radiative runs? b) Is the
inner gas entropy distribution affected by a more
sophisticated physical modeling of cluster dynamics (e.g. employing radiative cooling)? 
c) What is the effect of other
non-gravitational extra-heating mechanisms (e.g. AGN feedback)
on the the gas entropy profile of clusters?

In the following Section we explore how entropy is build over time in 
the same galaxy cluster analyzed above, by means of a Lagrangian approach based on tracers particles.
In the other Sections,
we analyze how is the entropy floor modified when a more realistic
modeling of cluster physics is considered (e.g. assuming radiative cooling) 
and if extra-heating processes (e.g. uniform pre-heating or AGN
feedback) are capable to reproduce an entropy distribution similar to 
that of the fiducial run.

\bigskip

\begin{figure}
\includegraphics[width=0.45\textwidth]{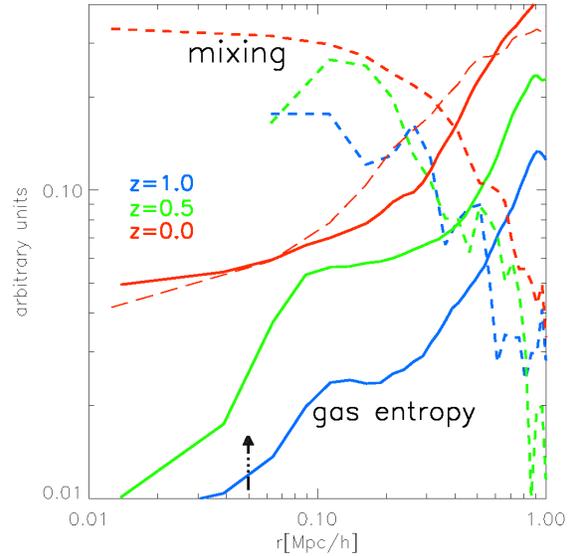}
\caption{Radial profile of the mixing parameters for the tracers in
run R0 (dashed lines) and radial profile of gas entropy (solid lines) , at $z=1$ (blue), $z=0.5$ (green) and $z=0$ (red). 
The long-dashed line shows the radial profile of $0.5-{\it M}$ at $z=0$.
The gas entropy has been dived by $2000$ for a better visualization. The vertical arrow show the softening length adopted in this run.}
\label{fig:mix_prof}
\end{figure}

\begin{figure}
\includegraphics[width=0.45\textwidth]{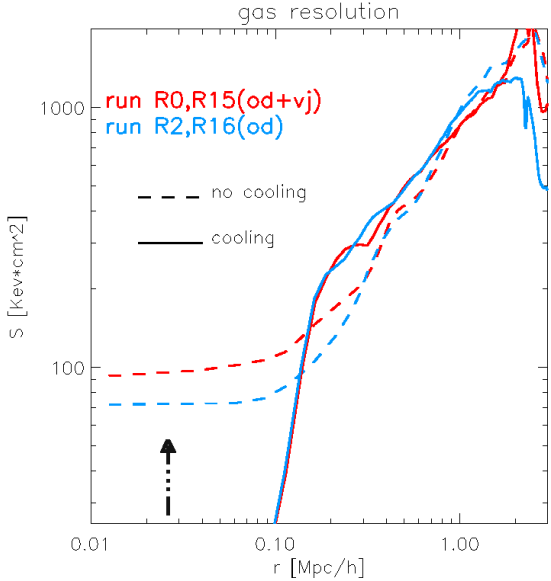}
\caption{Comparison of the entropy profiles of radiative runs, using the standard
mesh refinement (blue) and the velocity-based refinement (red). The additional
dashed lines show the profile of the corresponding non-radiative runs (R0 and R2). The vertical arrow shows the softening length adopted in run R15 and R16.}
\label{fig:prof_cool}
\end{figure}

\subsection{Where and when entropy is build in a forming cluster.}
\label{subsec:tracers}

Passive tracers are a useful tool to follow the average trajectory of 
accreted baryons in a growing galaxy clusters, and to study the 
mixing pattern driven by accretion phenomena in clusters (Vazza, Gheller \& Brunetti 2010).
Here we use tracer particles to track
the exact origin in time and space of the gas entropy deposited in the cluster core
of run R0, and to explain the emergence of a flat entropy core.
We inject passive (mass-less) tracers in the
simulation and let them be passively advected in the ICM using the information of the
3--D velocity stored for
each snapshot of the simulation, in a post-processing phase. 
Convergence tests on the interpolation technique to assign velocities to tracers, on
the time and space sampling are discussed in detail in Vazza, Gheller \& Brunetti (2010).

In this particular case, we injected $N=10^{5}$ tracers at $z=2$ in the 
fiducial run (R0), within a cubic volume of $\approx (10Mpc/h)^{3}$ 
centered in the AMR region.
At this time the total
virial mass of the forming cluster is only $10-15$ per cent of its 
final mass at $z=0$ (Fig.\ref{fig:mass_evol}) and only small proto-clusters are present
within the AMR region.

Tracers are initially placed with a random sampling of the volume, 
and then
their positions are updated with the time steps finely saved in time of the original 
simulation, using a Nearest Grid Cell interpolation scheme (e.g. Hockney \& Eastwood 1981). At all time steps, the tracers record the thermodynamic values of the closest cell in the grid distribution, and the whole thermodynamic history 
along the trajectory of every tracers can be recovered for analysis.

In Figure \ref{fig:trac1} we show the radial distribution of tracers at $z=0$, plotting on the
vertical axis the entropy of the nearest cell at each location. As a comparison, we overplot
the (density weighted and volume weighted) entropy profile of the cluster, to 
confirm that tracers sample the underlying Eulerian distribution in an accurate way.
At $z=0$ we selected 5 shells of tracers with width $25kpc/h$, located at the radii of $r<50kpc/h$, $r=150kpc/h$, $r=300kpc/h$, $r=600kpc/h$ and $r=1200kpc/h$ (shown as different colors in Fig.\ref{fig:trac1}).
This allowed us to study the mean evolutive history of the parcels of gas ending at the different level
in the cluster entropy profile at the end of the simulation.
In Fig.\ref{fig:trac2} we show the behavior of the mean gas entropy, gas entropy and velocity (or chaotic velocity, as measured in Sec.\ref{subsec:amr}) modulus for the 
5 different shells, as a function of evolving cosmic epoch.
Except for the first bin (which contains $N=20$ tracers) for all the other shells the number is of the order of $\sim 200$
and thus the mean values are very robust.

The sharp spikes in gas density (top panel) trace the main
merging episodes which involve the different ``shells'' of tracers.
The cluster entropy profile at $z=0$ is manifestly produced
by a uniform and regular in time mechanism of ``sorting in entropy'', which affects every tracers 
at the moment of its entrance in the virial region of the forming 
cluster (middle panel).
 The level of gas entropy in the cluster
core is set, on average, during intense shock heating 
at $z \sim 1$ (which also corresponds to the epoch of the most
net increase in cluster mass, as shown in Fig.\ref{fig:mass_evol}). On the other hand the gas tracers with a larger entropy ($S \sim 1000 keV cm^{2}$) at the final epoch are found to be intensely
shock heated at more recent epochs, $z \sim 0.6-0.8$. This analysis show that on average they belong to smooth, low density gas environment at $z=2$, and that they are subject to smaller bulk and chaotic velocity fields at the final epoch, since they do not belong 
to bound in-falling structures (lower panel). 

A complementary test was run by directly selecting 5 families of tracers at $z=2$ and
sorting them according to their initial gas overdensity; their evolution
is followed in Fig.\ref{fig:trac3}.
The ``sorting in entropy'' among tracers is even more evident with this 
setup: the average values of gas density and gas entropy for the different
families never overlap for $z<1$, meaning that the main variable 
which sets
the final entropy of a gas parcel ending up into a massive galaxy 
cluster is its initial overdensity. Say it differently, the fact 
that a gas particle is in a overdense (clumpy) environment around the forming cluster, determines on average the time at which it gains 
the bulk of the final entropy (in its first
impact on the volume of the cluster under virialization), and its 
final distance from the cluster center, through the mechanism of
 entropy sorting in the main cluster atmosphere. 
This is in excellent agreement with the spherical analytic models of clusters forming in a hierarchical
scenario, that prescribes a raising entropy at increasing radius from the
cluster center, following the progressive deposition of shells undergoing
stronger and stronger shock heating during the hierarchical growth of a cluster,
(e.g. Tozzi \& Norman 2001; Cavaliere, Lapi \& Fusco-Femiano 2009).

This simple scenario may of course be more complicated in the case of
a major merger, which can mix the intra cluster medium in a more
efficient way (e.g. Mc Carthy et al.2007; Poole et al.2008; 
Vazza, Gheller \& Brunetti 2010; ZuHone 2010).

The above tests show where and when the entropy profile is produced during
the cluster evolution, but do
not necessary imply the emergence of a small inner region 
of size $r_{core} \sim 0.1 R_{vir}$
where the stratification is broken and the gas sets to the constant
value of $S \sim 100keV cm^{2}$ observed in the previous runs.
A viable mechanism naturally produced by the accretion of matter onto the cluster is mixing of the inner gas layers, in response to chaotic motions in the ICM. 
To better compute the degree of 
mixing between Lagrangian tracers after their injection at $z=2$, using
a formalism introduced in Vazza, Gheller \& Brunetti (2010).
We introduce the mixing parameter of an ``s'' family of tracers, ${\it M_{s}}$, respect to all existing ``i'' families, by computing: 

\begin{equation}
{\it M_{s}} = 1-\frac{|n_{s}-\sum_{i}{n_i}(i \neq s)|}{\sum_{i} n_{i}},
\label{eq:mix}
\end{equation}

where $n_{s}$ is the number density of the ``s'' tracers within a cell (at the highest resolution
level) and the sum refers to all the species of ``i'' tracers (included the ``s'' specie). 
This formula generalizes the more simple case of mixing between two species
(e.g. Ritchie \& Thomas 2002) and has a simple interpretation: for a cell where $n_{1} \approx
n_{2} \approx n_{3} ...\approx n_{s}$ the different families are well mixed 
and we have ${\it M_{s}} \rightarrow 1$, while
${\it M_{s}} \rightarrow 0$ implies no mixing within the cell.

The total mixing in each cell, {\it M}, is the volume average between all
species, ${\it M_{i}} = \sum_{s} {\it M_{s}}/N_{s}$, where $N_{s}$ is the number
of families considered. To obtain better statistics, the cell considered to compute 
${\it M_{i}}$ have a size which is double than the maximum gas resolution ($50kpc/h$).

We sorted the density of all tracers at $z=2$ and divided the sample in 5 bins, each 
with 2000 tracers; then we computed the number density of each
``s'' family of tracers within the cluster, and computed the mean mixing 
${\it M_{i}}$ at different redshifts (Fig.\ref{fig:mix_prof}), only for those
cells containing at least 1 tracer. 

The profile of the mean mixing presents a regular trend in time, with 
a inner core of mixing material which is build as the cluster accretes matter, until an almost
flat profile (with mean mixing ${\it M_{i}} \approx 0.3$) is found for 
 $r<100kpc/h$ at $z=0$. Outside of this radius, ${\it M_{i}}$ has a sharp decline towards
the virial radius of the cluster, where the mixing is very poor ${\it M_{i}} < 0.05$. 
We verified that also different initial choices for the setup of the tracers sampling (e.g. by adopting a different number of tracers, or a different number of species) do not affect the above trend in any significant way.

Fig.\ref{fig:mix_prof} shows the evolution of the gas entropy profile (solid lines) and of the mean mixing profile (dashed lines) for three redsfhits.
Quite clearly, the formation of the central entropy floor happens together
with the formation of the prominent mixing pattern in the center of the
cluster (the similarity of the two profiles is better indicated by 
overplotting the radial profile of  $|-{\it M_{i}}|$ at $z=0$, long
dashed line), and the  $S(r) \propto r^{1.1-1.2}$ scaling expected
from SPH simulations (Voit et al.2005, Fig.\ref{fig:prof_ref}) is broken starting
from the same radius for which ${\it M_{i}}$ sharply increases to its maximum
in the center. In Vazza, Gheller \& Brunetti (2010) we provided evidence
that in general the building of cluster entropy and of prominent mixing
patterns are correlated features also in major merger systems (where, however,
large and transient plumes of efficient mixing can be found also at
large cluster radii); also recent FLASH AMR simulations (Zu Hone 2010) 
has lead to
similar conclusions, based on a parametric study of binary cluster mergers.
Usually the innermost region of cosmological cluster runs is characterized 
by small scale subsonic motions (on scales $<500kpc/h$ and $\sigma_{v} \sim 0.3-0.5 c_{s}$, where $c_{s}$ is the gas sound speed) continuously excited by the crossing
of gas/DM material accreted within the cluster (e.g. Norman \& Bryan 1999; Dolag et al.2005; Iapichino \& Niemeyer 2008; Vazza et al.2009). 
In the case of ENZO AMR simulations these
motions are well characterized by a power law spectrum for nearly two
orders of magnitude in spatial scales (Vazza et al.2009; Xu et al.2009; Vazza, Gheller \& Brunetti 2010); also the tracers pair-dispersion statistics show
a well defined power-law dependence on time ($P(t) \sim t^{3/2}$, where $P(t)$ is
the distance between couple of tracers initially located at a small distance) compatible with a fairly fast transport motions in the turbulent ICM (Vazza, Gheller \& Brunetti 2010).

When combined together, all the above evidences lead to the consistent
conclusion that the emergence of a flat entropy floor
in clusters simulated with grid codes is mainly due to the integrated effect of
(mostly subsonic) turbulent motions in the evolving ICM. 
These motions are quite effective in mixing parcels 
of gas, which have been initially sorted according
to the ``entropy sorting'' mechanism described above, generally 
inside $r<0.1 R_{vir}$ and for late $z<1$ redshift, when 
that the bulk of cluster mass has been assembled.
The same mechanism also applies when the standard mesh refinement 
is adopted, provided that the turbulent energy in the cluster core
is smaller (Fig.\ref{fig:prof_vel}) and that the mixing and the transport
of gas particles is significantly smaller (e.g. Fig.17 in Vazza, Gheller \& Brunetti 2010).

It is well known that grid codes are prone to numerical mixing (i.e. different
gas phases are forced to combine into an average cell value when their 
separation is smaller then the minimum available cell size) while in SPH
gas particles do not mix entropy by construction, unless ad-hoc diffusion term
is considered in the SPH equation (e.g. Agertz et al.2007; Price 2008; Wadsley et al.2008; Merlin et al.2010). 
Numerical mixing must be considered as an additional source of mixing also
fro the ENZO AMR resimulations presented here. 
However the ubiquitous finding
of evident mixing motions on scales much larger than the cell size 
discussed above, combined with the evidence that the entropy floor presents
a very small evolution with the grid resolution, 
for peak resolution $\leq 25kpc/h$, indicate that numerical mixing cannot
be the main responsible for the production of the entropy floor. 
According to this interpretation, the entropy floor is thus a result of mixing
of the baseline steep $S(r) \sim r^{1.1}$ profile (produced by the progressive
shock heating of infalling shells of matter) within the region where
the turbulent energy is maximum within cluster, $<0.1 R_{vir}$, as a result
of the continuous excitement of a hierarchy of chaotic motions in the ICM
driven accretion after accretion.

This is true for an unviscid treatment of the ICM, while the presence 
of magnetic fields and plasma viscosity may alter this picture in a significant way  (e.g. Parrish \& Quataert 2008; Ruszkowski \& Oh 2010 and references therein).

In the following Sections, we will explore more realistic modelization of the same cluster,
where non-gravitational
mechanisms of entropy {\it decrease}, such as radiative cooling, and entropy {\it increase}, such as heating 
from energy feedback by stars of active galactic nuclei, are computed in run time.

\begin{figure}
\includegraphics[width=0.45\textwidth]{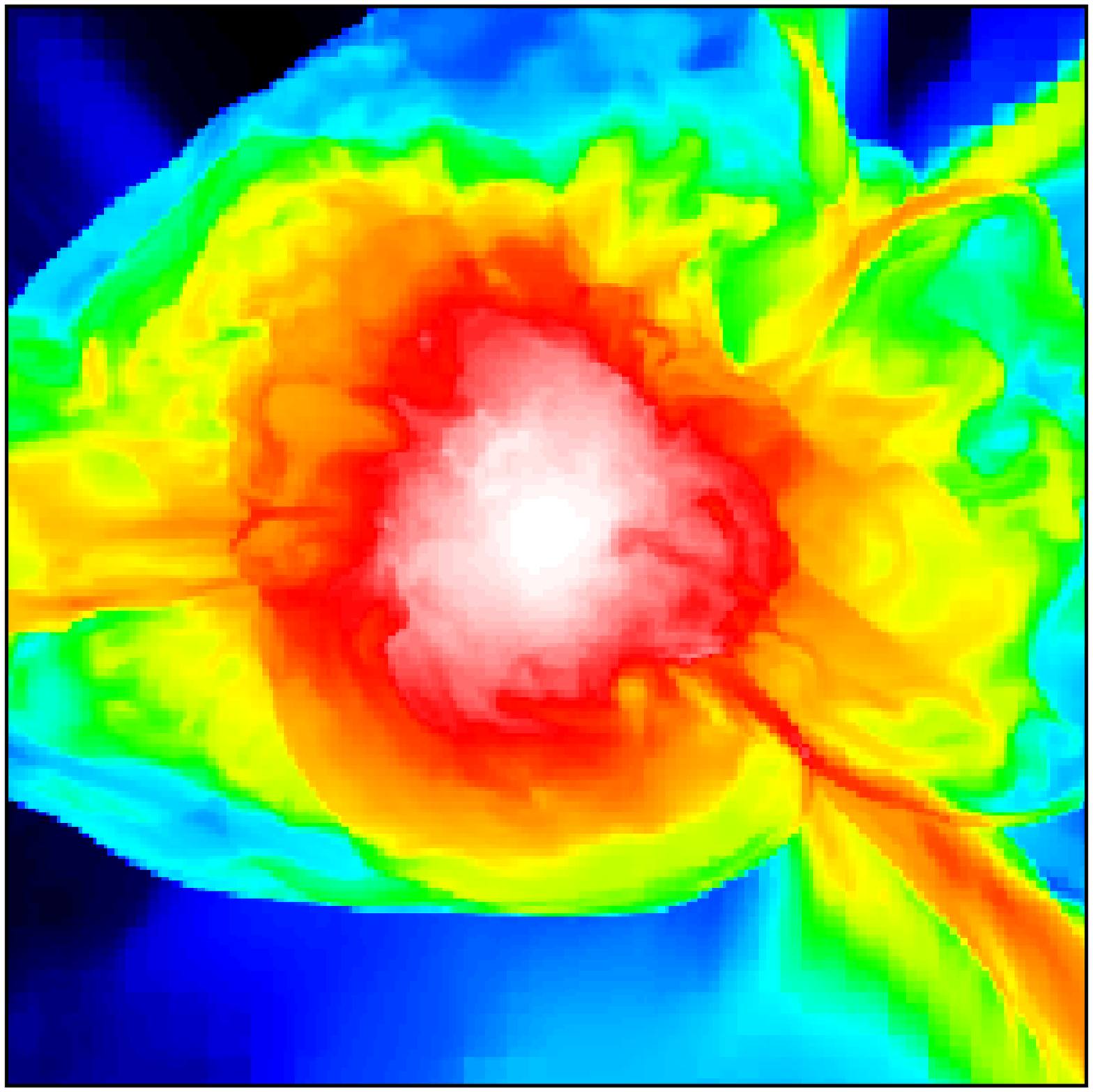}
\includegraphics[width=0.45\textwidth]{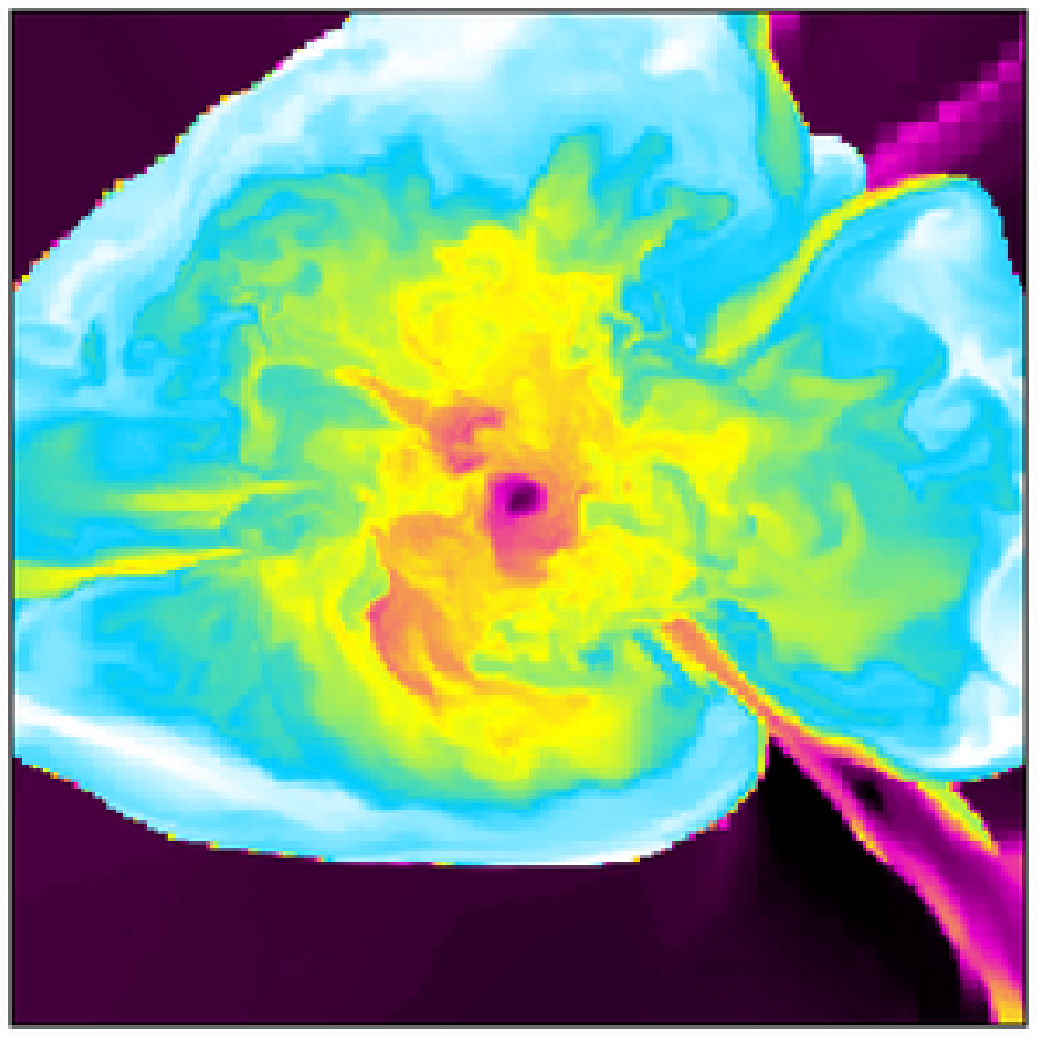}
\caption{Maps of gas density (top panel) and gas entropy (bottom panel) for a slice
taken in run B2 (cooling, pre-heating and 
refinement on velocity jumps). The side of the image and the meaning of colors are as in Fig.\ref{fig:map_r0}.)}
\label{fig:ph_map}
\end{figure}

\subsection{The role of radiative cooling and pre-heating mechanisms.}
\label{subsec:cool}

Radiative cooling in galaxy clusters
have a dramatic impact in the thermodynamics of the ICM, if
no heating mechanism other than shock heating is available to balance the cooling catastrophe, for those cluster regions in which the gas cooling time is $<<$ of the cluster age (e.g. Katz \& White 1993).
Figure \ref{fig:prof_cool} illustrates the radial profiles of gas density, gas temperature
and gas entropy for the two re-simulations with radiative cooling (assuming
a fully ionized H-He plasma with a constant metallicity of $Z=0.3 Z_{\odot}$) and
with standard mesh refinement (R15) or velocity-based refinement
(R16). In both cases, a steeply decreasing entropy profile develops
towards the center of the cluster for $ r < 100kpc/h$, with very
low entropy values, $S << 1 keV cm^{2}$ in the center, a massive gas condensation
peaking at $10^{-24}gr/cm^{3}$ and a temperature dip with $T<<0.1keV$, 
similar to the 
classic theoretical cooling flow scenario (e.g. Fabian 1994).

\begin{figure}
\includegraphics[width=0.45\textwidth]{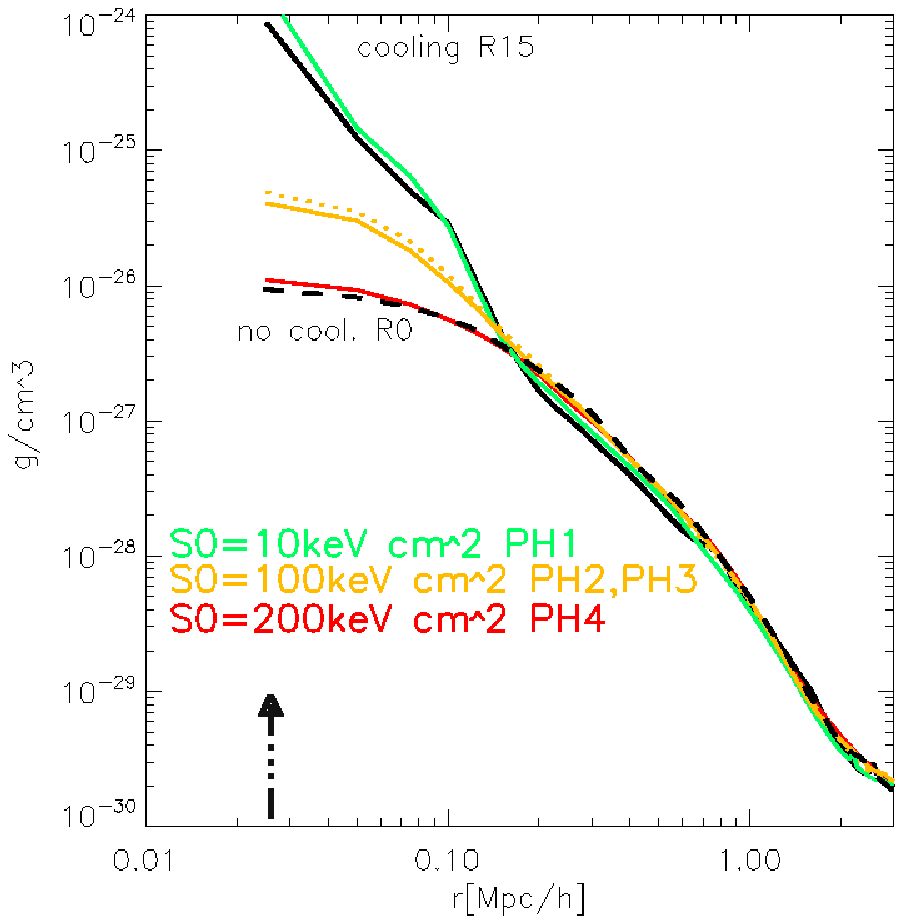}
\includegraphics[width=0.45\textwidth]{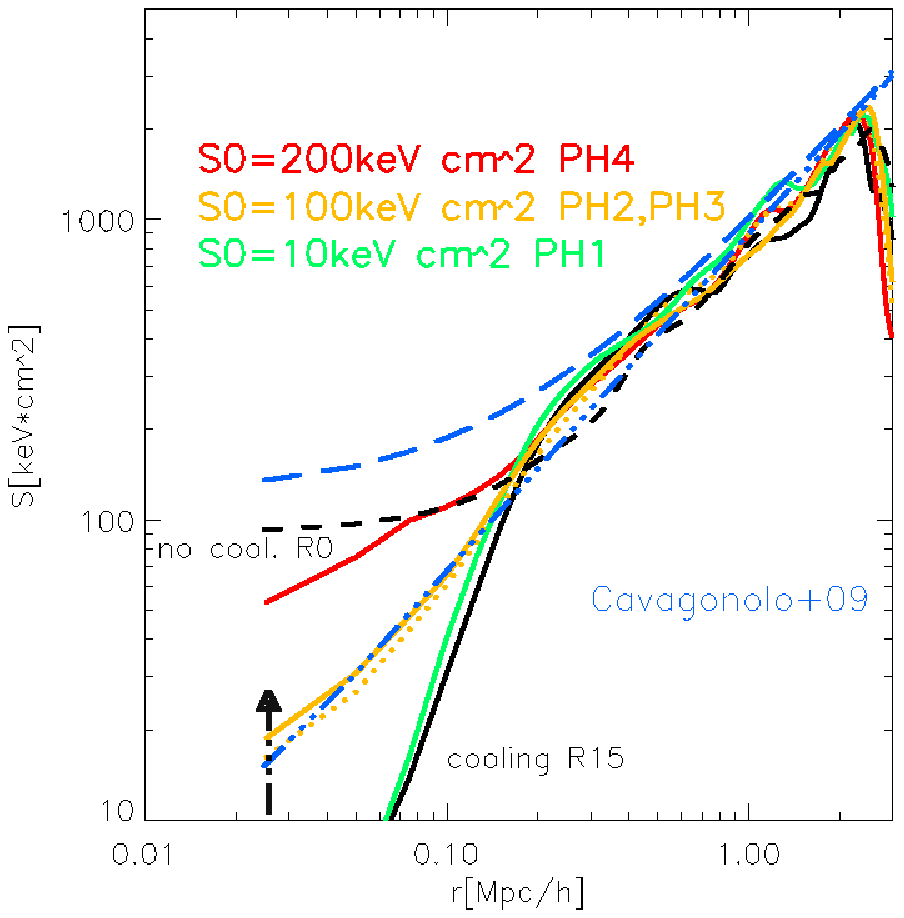}
\caption{Gas density and entropy radial profiles for the run adopting pre-heating
(run PH1 $S_{0}=10keV cm^{2}$, run PH2 $S_{0}=100keV cm^{2}$ and run PH4  $S_{0}=200keV cm^{2}$).
The dotted lines show the result for the run PH3, adopting  $S_{0}=100keV cm^{2}$ and 
standard mesh refinement.
The additional black lines shows the profile for the non-radiative run (run R0, long dashed)
and for the cooling run (run R15, solid). We also overplot in blue the best fit
profiles of Cavagnolo et al.(2009), with a core entropy of $S_{0}=15keV cm^{2}$ (dot-dashed) and $S_{0}=150 keV cm^{2}$ (long-dashed). The vertical arrow in both panel show the softening lenght adopted in the runs with cooling and pre-heating.}
\label{fig:prof_ph}
\end{figure}

We note that these simulations do not consider any prescription for star formation from the dense and cold phase of the ICM, and therefore the central 
condensation produced
by run R15, R16 (see the solid black line in the top panel of Fig.\ref{fig:prof_ph}) would be significantly reduced by modeling star formation in a self-consistent way (e.g. Pearce et al.2000; Yoshikawa, Jing \& Suto 2000; Valdarnini 2002). 
However this would make the pure cooling runs and the other explored in the
next Sections more computationally expensive. We thus preferred to 
defer to the future the study of this issue, and to use this
simplified cooling model without star formation as the framework to 
study the effects of thermal energy feedbacks in the ICM, and the modifiation
they cause to the gas entropy distribution.
 
In general major and moderate mergers (e.g. Burns et al.2008; Poole et al.2008) or gas sloshing triggered by the passage of DM/gas sub-clumps
(e.g. Ascasibar \& Markevitch 2006; ZuHone et al.2009)  may significantly reduce the cooling
catastrophe in radiative simulations, by exciting internal merger shocks or turbulent mixing in the clusters core. 
However, in the particular cluster simulated here the amount of chaotic motions excited for $z<1$ is never 
powerful enough to slow down the cooling flow in any significant way, even when 
the additional mesh refinement criterion is turned on.
We report that also in the case of the merger system studied in the Appendix, the 
major merger at $z \sim 0.85$ does not reduce the cooling catastrophe at the end
of the simulation.

As widely known, cluster configurations according the "pure cooling" scenario are 
not observed in the real Universe, and the gas temperature in real 
clusters is never observed below $\sim 0.1 keV$ (e.g. Rossetti \& Molendi 2010 and references therein).
For this reason, additional sources of gas heating were considered in order to reconcile simulations 
with observations.  In what follows, we will apply some of the most 
promising models of extra-heating developed in the literature to our cluster
simulations, and study their
impact on the gas entropy profile at $z=0$.

\subsubsection{Early pre-heating.}
\label{subsubsec:ph}

Motivated by the early evidences of significant departures from self-similar
scalings expectations
in observed galaxy clusters (e.g. White 1991; Evrard \& Henry 1991; Ponman , Cannon \& Navarro 1999; Lloyd-Davies, Ponman \& Cannon 2000),
several authors proposed a "pre-heating"
scenario, in which an energy input of non-gravitational origin is injected
in the Intra Galactic Medium at early redshift ($z \sim 3-10$). This smoothens 
the gas of in-falling primordial halos 
of low mass and supplying the innermost region of 
massive cluster of an amount of high entropy gas, to 
reconcile with observations.
(e.g. White 1991; David, Forman \& Jones 1991; Kaiser 1991; Evrard \& Henry 1991; 
Cavaliere et al.1997; Voit et al.2005).
This early heating may be provided by a number of sources, such as star formation and SNe explosion, radiative and mechanical heating from AGNs, radiative heating from hard X-ray background, etc (e.g. Tozzi \& Norman 2001). 
The estimated needed amount of extra entropy at high redshifts falls in 
the range $100 \leq \Delta S \leq 300 keV cm^{2}$, and viable sources of it in the
early Universe can be supernovae explosions, star formation and galactic winds.
From the numerical viewpoint, several
group attempted to model this process either by imposing
an impulsive entropy injection at a given epoch (e.g. Bialek, Evrard \& Mohr 2001;
Borgani et al.2005; Kay et al.2007; Romeo et al.2006; Younger \& Bryan 2007) 
or in a redshift-modulated way (e.g. Borgani 
et al.2002; Sijacki et a.2007; Mc Carthy et al.2009). 

An important drawback of early pre-heating models could be that they tend
to remove to much 
low entropy gas from lower mass halos, without obtaining a realistic 
galaxy population (e.g. Donahue et al.2006). In addition, recent results based on XMM-Newton
analysis presented by Rossetti \& Molendi (2010) proved that
most of non-cool core clusters host regions with low entropy and
high metallicity, suggesting the possibility of a recent transition between cool 
core and non-cool core systems, contrary to the pre-heating scenario (e.g. 
Guo \& Mathews 2010).

As a first step to investigate the role played by non-gravitational heating on the entropy level
of our cluster, we tested early pre-heating models following the prescription
of Younger \& Bryan (2007). 

The thermal energy within each gas cell in our cluster run is selectively
increased at $z=10$ by:

\begin{equation}
k_{B} \Delta T= S_{0} [\frac{\rho}{\mu m_{p}}]^{\gamma-1}.
\end{equation}

In detail we re-simulated run R15 (velocity based refinement and cooling) by imposing the entropy 
floor of $S_{0}=10keV cm^{2}$ (run PH1), $S_{0}=100keV cm^{2}$ (run PH2)  and $S_{0}=200keV cm^{2}$ (run PH4). 
Smoothing the gas density distribution of halos at high redshift is expected to affect the shock heating
process in forming structures, through "entropy amplification" at strong acrretion shocks (Voit et al.2005).
To highlight the dependence of this mechanism  on the accuracy with which accretion shocks are 
modeled in our runs, we also tested the intermediate pre-heating
 scenario of run PH2 ($S_{0}=100 keV cm^2$) in the standard mesh refinement strategy alone (run PH3). 

Figure \ref{fig:ph_map} shows the effect of pre-heating in extreme case of run PH4 at $z=0$. Compared to 
the non pre-heated scenario (run R0) the global cluster morphology is smoother, and the accretion
pattern are more regular in shape, since most of the accreted gas clumps were smoothed by the 
early heating episode. 
In Fig.\ref{fig:prof_ph} we compare the gas density and gas entropy profiles for the runs employing
pre-heating,  against the profiles of the fiducial run (R0) and of the simple cooling run (run R12).

Consistently with the literature, we find that energy inputs corresponding to $S_{0}=100-200keV cm^{2}$  
are capable of keeping the in-falling gas on an higher adiabat, preventing the core gas to cool
below $\sim 0.5 keV$. In particular, the pre-heating prescription of run PH4 is suitable to
recover the gas density and the gas entropy of the non-radiative
run (R0), within a $\sim 10-20$ per cent at all radii.
A similar result is also found for the major merger system studied in the Appendix.
On the other hand, the uniform pre-heating model with $S_{0}=10keV cm^{2}$  is 
found to be insufficient to prevent the cooling catastrophe and the resulting cluster profile is almost identical to that of run R15.

The trend found is qualitatively in agreement with the results of Borgani et al.(2005) and Younger \& Bryan (2007),
even if our study is based on a single object and no conclusion about the most suitable values of $S_{0}$ needed to
reconcile with observations can be derived in a statistical sense.
We also note that the final entropy configurations of run PH2, PH3 and PH4 
are compatible with the 
bimodal distributions of entropy profiles of obtained with CHANDRA (Cavagnolo et al.2009,
additional blue lines in Fig.\ref{fig:prof_ph}). This work recently suggested
the existence of two broad population of clusters, characterized by an inner entropy value
of $S \sim 15 keV cm^{2}$ or $S \sim 100-150 keV cm^{2}$, and a large radial behavior
scaling as $S(r) \propto r^{1.1}$.
Our cluster is a relaxed one at $z=0$, and the fact that it is more similar to the "low entropy core" class of
CHANDRA clusters, even when $S_{0}=200keV cm^{2}$ is applied, is fully compatible with
the idea that "high entropy core" class is produced only by those clusters with
a sufficiently violent merger in their past (e.g. Rossetti \& Molendi 2010).

\begin{figure*}
\includegraphics[width=0.95\textwidth]{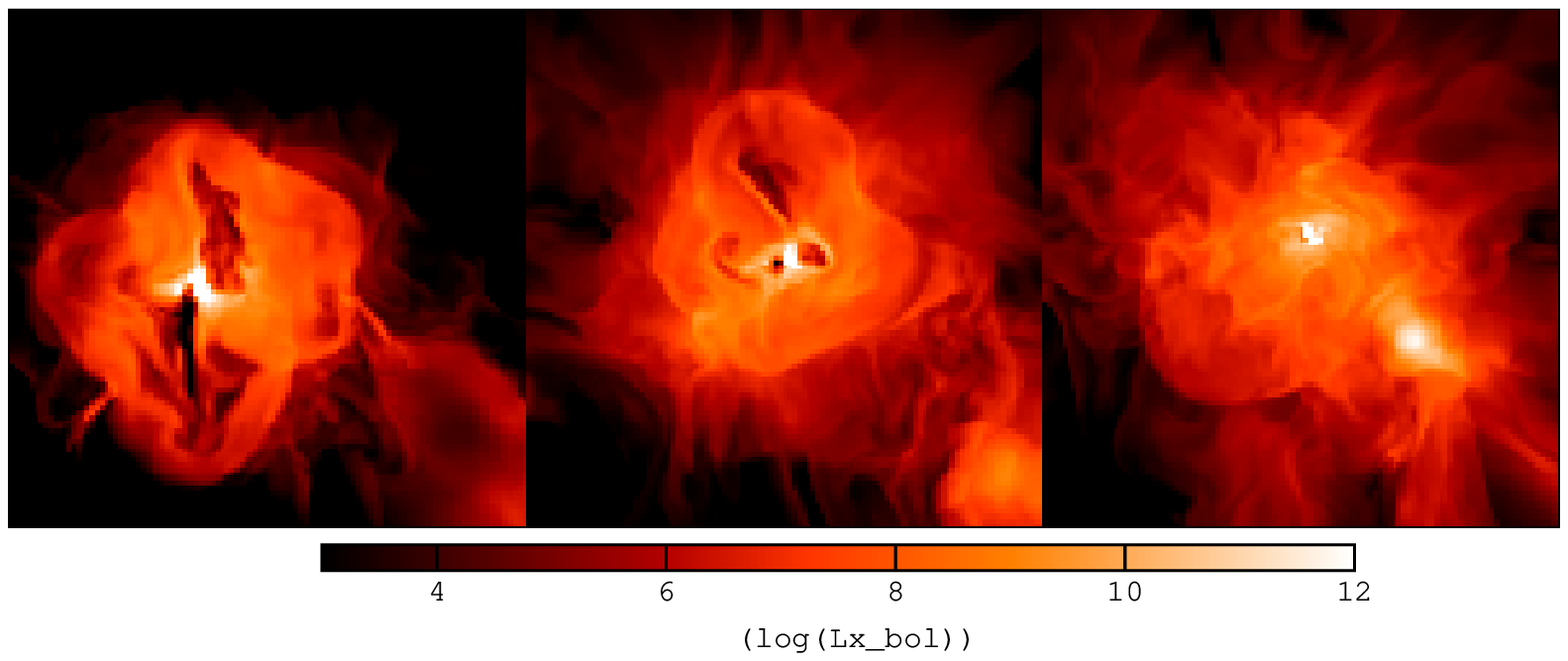}
\includegraphics[width=0.95\textwidth]{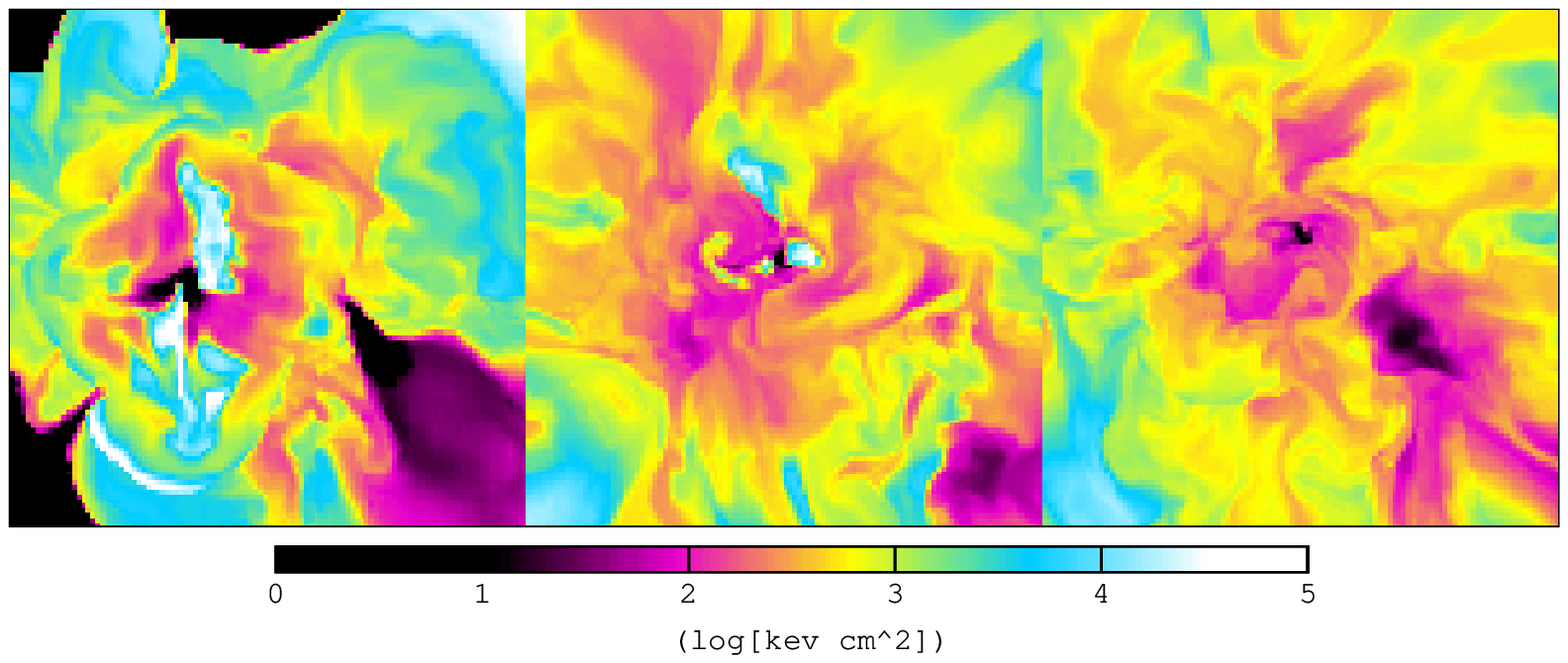}

\caption{Cuts through the center of the simulated cluster, showing
the evolution of the bolometric x-ray luminosity (top panels, arbitrary units) and gas entropy (bottom panels) for
run B2. From the left to right, the epoch shown are $z=0.96$, $z=0.85$ and $z=0.75$. The side of the images is $2.7Mpc/h$.}
\label{fig:bub_mov}
\end{figure*}

\begin{figure}
\begin{center}
\includegraphics[height=0.45\textheight]{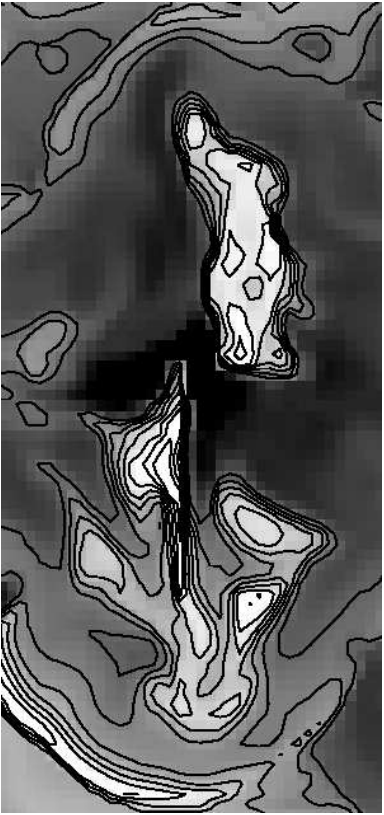}
\caption{Zoomed view of the internal ICM region after the first AGN outflows in run B2 at $z=0.96$ (logarithmic view
of gas entropy). The isocontours are drawn only for gas with $S>500 keV cm^2$ to highlight the contribution
from the AGN burst. The scale of the images is $\sim 1.5 \times 0.5 Mpc/h$.}
\end{center}
\label{fig:jet_zoom}
\end{figure}

\begin{figure}
\includegraphics[width=0.45\textwidth]{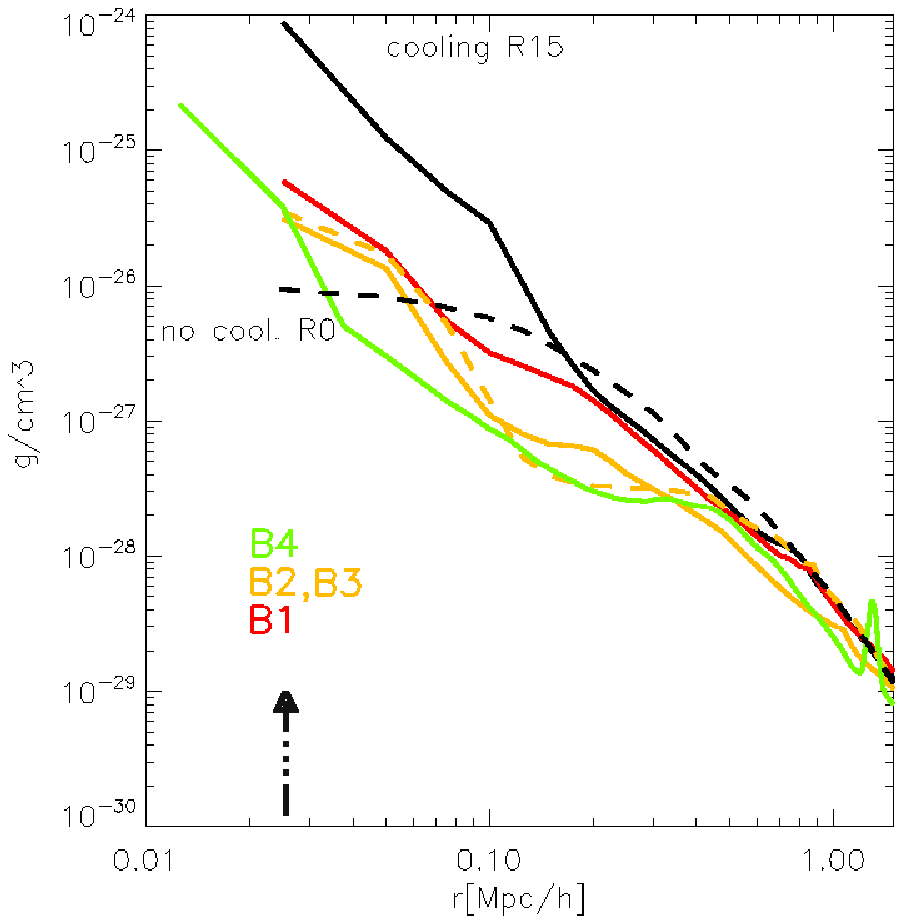}
\includegraphics[width=0.45\textwidth]{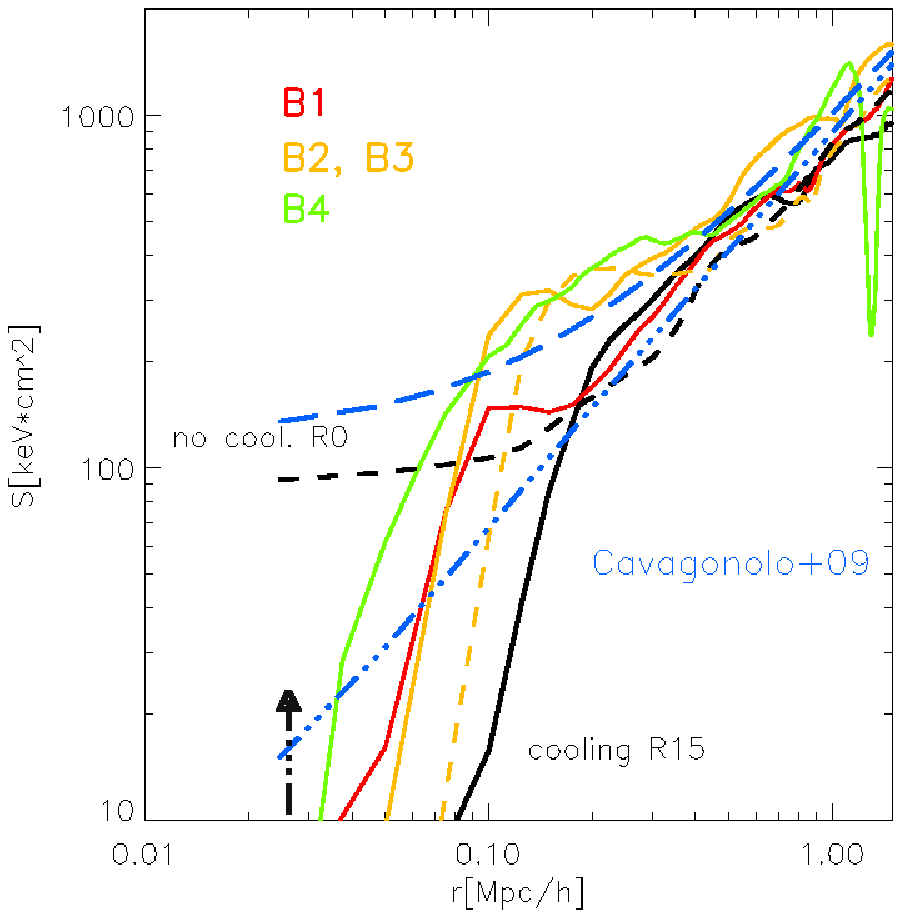}
\caption{Gas density and entropy radial profiles for the run adopting AGN jets feedback
(run B1,B4 $\epsilon_{jet}=10^{58}ergs$, runs B2 $\epsilon_{jet}=10^{59}ergs$).
The dashed lines show the result for the run B3, adopting  $\epsilon_{jet}=10^{59}ergs$ and
standard mesh refinement.
The additional black lines shows the profile for the non-radiative run (run R0, long dashed)
and for the cooling run (run R15, solid). We also overplot in blue the best fit
profiles of Cavagnolo et al.(2009), with a core entropy of $S_{0}=15keV cm^{2}$ (dot-dashed) and $S_{0}=150 keV cm^{2}$ (long-dashed). The additional arrow in both panel shows the softening length adopted in the runs with cooling and AGN feedback.}
\label{fig:prof_jet}
\end{figure}

\subsubsection{Heating from AGN jets.}
\label{subsubsec:agn}

The most successful models to achieve a balance with radiative
cooling during the simulated evolution of clusters rely on 
heating by outflows from an AGN hosted by the central massive galaxy (e.g.
Churazov et al.2000; Bruggen \& Kaiser 2002; Brighenti \& Mathews 2003; Dalla 
Vecchia et al.2004; Zanni et al.2005; Heinz et al.2006).

The self-consistent modeling of AGN heating in evolving galaxy
clusters, in connection with the matter accretion history 
of the central massive cD galaxy has become only
quite recently within the capability of full cosmological simulations (e.g. Sijacki \& Springel 2006; Dubois 
et al.2010). 
In general, this issue is made complex by the (still unclear) role played by other physical
mechanisms relevant to the thermodynamic evolution of the cluster plasma:
plasma viscosity, magnetic fields, Cosmic Rays, etc (e.g. Bruggen \& Kaiser 2001; 
De Young 2003; Brighenti \& Mathews 2003; Ruskowski et al.2007; Sijacki et al.2008; Xu et al.2008; O'Neill \& Jones 2010; 
Scannapieco \& Bruggen 2010).

In this Section, we study the energy budget necessary to quench the cooling
catastrophe developed in our radiative run (R15), using the {\it thermal} feedback from an assumed central AGN (e.g. Booth \& Schaye 2009; Teyssier et al.2010)
This is done by injecting
localized inputs of extra-thermal energy, starting at $z=1$ in run R15, in the region of the cluster density
peak, where the AGN is assumed.

This is motivated to mimic the thermal feedback response from a central AGN which releases as  feedback a fraction of the rest mass energy of the accreted cold gas, with an efficiency which
is generally assumed to be in the range of $\sim 0.01-0.1$ of the total energy radiated by the 
super massive black hole (e.g. Booth \& Schaye 2009; Giodini et al.2010 and references therein).
In our setup, we assume that the feedback manifests itself at the scale of the maximum gas resolution 
in the cluster center ($25kpc/h$), as the injection of two point-like over-pressurized ``bubbles'', 
produced by the interaction between the launched
jets and the surrounding cold ICM.
The thermal  energy at the location of the ``bubbles'' is updated as:

\begin{equation}
 E_{thermal}= \frac{3k_{B} \rho T \Delta x^{3}}{2 \mu m_{p}} + \frac{\epsilon_{jet}^{'}}{2}
\label{eq:thermal}
\end{equation}

where $\mu$ is the mean molecular mass, $m_{p}$ is the proton mass,
where $\Delta x^{3}$ is the volume of the cell, and 
$\epsilon_{jet}^{'}$ is the fraction of $\epsilon_{jet}$ released 
at every injection episode. The 3--D 
velocity field at the injected bubbles is left
unchanged. Therefore in our treatment $\epsilon_{jet}^{'}$ represents the thermal energy released in the 
ICM by the the two jets at a given time step, after the thermalisation of a 
part of their mechanical energy, 
which is assumed to happen on a sub-grid scale.
The ``bubbles'' are initially located at the distance of $d_{jet}=50kpc/h$, 
at two opposite sides of the 
gas density peak of the cluster, starting from 
$z=1$ in run R15 (or run R16 for the standard refinement strategy). 
The injection is performed before the hydro step of the PPM scheme, 
by updating the gas internal energy following Eq.\ref{eq:thermal} of two cells 
(at the maximum 
resolution level) according to \ref{eq:thermal}; then the 
 Riemann solver in ENZO is evolved in the usual way.

Preliminary tests showed that the impulsive injection of the whole $\epsilon_{jet}$ in a single time
step of the simulation produces unrealistically strong (e.g. $M>10$) shock waves in the cold and dense cooling flow cluster region.
Since only mild shocks are observed in  jets/bubbles interaction with the ICM of
real clusters (e.g.
 Simionescu et al.2009; Werner et al.2010), we preferred to adopt a more gradual release of energy from the 
central gas peak, by distributing $\epsilon_{jet}$ in $\sim 20$  injection episodes,
across a total time of $\sim 3 Gyr$, preserving the same orientation for the "jets" axis.
In principle, idealized but more self-consistent recipes to link the feedback energy with the matter accretion
rate within the cooling region can be applyed to cosmological simulations (e.g. Sijacki \& Springel 2007; 
Booth \& Schaye 2009); however here we want to investigate how different re-simulations of the same
object react to a {\it constant} model of extra thermal energy release from a central AGN.
We defer to the future any study of more self-consistent setup of the feedback energy, and 
of the way it is released within the ICM (e.g. by varying the jets orientations in time, by assuming an "quasar mode" and "radio mode" feedback, etc).

With our setup, only  $M<5$ shocks are produced, even in the most extreme scenario investigated, and only in the 
starting phase of the jet, when the surrounding ICM is in its coldest phase. 
Our trials adopted $\epsilon_{jet}=10^{58} erg$ (run B1) and 
$\epsilon_{jet}=10^{59} erg$ (run B2);
this makes the typical power of our jets in injection phase of about
 $W_{jet} \sim \epsilon_{jet}^{'}/t_{step} \sim 10^{42}-10^{43} erg/s$ 
($t_{step}$ is $\sim 4 \cdot 10^{7} yr$ at that epoch).
We note that the power for the energy release of our jets in the surrounding medium, and the assumed duty cycle and duration are within most of the estimated energy budget provided 
by the observations of AGN activity reported by many authors (e.g. Birzan et al.2004; Dunn \& Fabian 2006; Wise et al.2007; Giacintucci et al.2008; Bird, Martini \& Kaiser 2008; Worrall et al.2009; Liuzzo et al.2009;  Sanders \& Fabian 2009; Gu, Cao \& Jiang 2009; Gitti et al.2010; Giodini et al.2010).

Scannapieco \& Bruggen (2008) have recently shown that a proper treatment of turbulence
on $<10kpc$ scales is mandatory to model the full interaction 
between jet-inflated bubbles and the ICM, because 
this may change the rate of energy transfer to the 
surrounding cold phase of the ICM. 
Therefore, it is unlikely that our simulations are fully converged, and further tests at higher resolution
will be needed in the cosmological framework.
In any case, we assess the role played by numerical resolution here by  running two additional re-simulations, using only standard mesh refinement, 
(run B3) and using an additional level of refinement in the velocity based strategy, up to a maximum resolution
of $\Delta x = 12.5kpc/h$ (run B4). 

\bigskip
 
The evolution of bolometric X-ray luminosity and gas entropy in a slice crossing the cluster
 center for run B2 is shown in Fig.\ref{fig:bub_mov}.
 Soon after the first injection, two vertical outflows has developed
for $\sim 400-500kpc$ along the axis of "bubbles" injection and has pushed the dense and 
low entropy material out to 
larger radii. The first feedback episode drives a mild shock in the cold ICM, 
with $M \sim 3.5$, while along the outflows Kelvin-Helmoltz instabilities develop 
and favors the mixing between the cold uplifted gas from the core and the
surrounding hotter ICM (Fig.\ref{fig:jet_zoom}).
The outflows that follow inflate more stable "bubbles" (central panel in 
Fig.\ref{fig:bub_mov}), which are initially less overpressurized
compared to the surrounding ICM, heated by the previous feedback episode. 
These bubbles only drive weak $M \sim 2$ shocks around the central gas condensation. 
In addition, the sloshing motions in the cluster center are powerful enough to 
bend the initial orientation for the bubble launching, and to partially provide azimuthal mixing of the
injected entropy.

Compared to the fiducial run (R0) or to the pre-heated runs (PH2,PH4), we found
that the large scale accretion patterns are modified by the
outgoing propagating shocks that follow the AGN activity.

In Figure \ref{fig:prof_jet} we compare the the gas density and gas entropy profiles for all 
trials at $z=0$.
We found impossible to recover an
entropy profile similar of the fiducial run (and also with the results of Cavagnolo et al.2009, overplotted in the same Figure):
the extra energy of shocks is very efficiently 
delivered to larger cluster radii by the shocks, which develop in the
interaction between the cooling ICM and the hot bubbles phase. As a result, in these
configurations we find an excess of gas entropy for $r>75kpc/h$, compared to the
radiative and non-radiative cases and a flat entropy profile inside the cluster
core. 
However these runs produce a much smaller cooling region ($r_{cool}<50kpc/h$) 
compared to the $r_{cool} \approx 100kpc/h$ of pure cooling models (run R15, R16).
The observed steepening of the internal gas entropy after the AGN energy
released is in good agreement with the semi-analytical prediction of a quasar-driven blasts in clusters 
discussed in Lapi, Cavaliere \& Menci (2005), further suggesting
that in the trials investigated here the shock heating mechanism is the main mechanism
which interchanges energy between the AGN and the surrounding ICM.

When the mesh
refinement is triggered uniquely by gas/DM over-density (run B3), the final entropy profile
at $z=0$ present a larger cooling region (in-between the pure cooling case and the
B2 re-simulation). In this case it is difficult to disentangle the effect of the 
under-sampling issues of satellites-driven mixing  (Sec.\ref{subsec:tracers}) from that of the under-sampling
of jet-driven turbulence around the cluster core. 
The re-simulation with an additional level of mesh refinement (run B4) shows that 
as resolution is increased the final size of the cooling region is reduced, 
and the entropy at larger radii is increased, due to a better modeling of shocks induced
by the outflows and of the driven turbulent motions.
Further studies will be needed in the future to fix the best resolution needed 
for full numerical convergence in these features.

Our general conclusion is that, even if the action of feedback from 
outflows in our simplified implementation 
efficiently reduces the size of the cooling flow region compared to a pure radiative run, 
it remains difficult to reproduce a flat inner entropy profile as in the non-radiative case.
The problems is not in the energy budget assumed in the outflow
(which is a reasonable energy budget available to observed AGNs) 
but rather in the mechanism which transfers to the surrounding medium, 
which is mainly shock
heating of the cold central ICM (see also Zanni et al.2004; Lapi, Menci \& Cavaliere 2005). 
It is likely that more gentle mechanisms of feedback from the central AGN, 
such as a 
more gradual deposition of many by inflated bubbles(e.g. Churazov et al.2001; Bruggen et al.2007), can 
be more efficient in stopping the catastrophic cooling in our run.
However in that case a significantly larger resolution than the one available here 
must be considered, which is presently difficult for AMR runs with our mesh refinement
scheme. In addition,
sub-grid modeling of turbulence may necessary (e.g. Scannapieco \& Bruggen 2010) and
also physical energy component, such 
as magnetic field and relativistic particles, should be 
important to attach this problem (e.g. Bruggen \& Kaiser 2001; 
De Young 2003; Brighenti \& Mathews 2003; Ruskowski et al.2007; Sijacki et al.2008; Xu et al.2008; O'Neill \& Jones 2010; 
Scannapieco \& Bruggen 2010; De Young 2010).

\subsubsection{Hybrid external and internal extra-heating models.}
\label{subsubsec:ph_agn}

Early ($z \sim 3-10$) pre-heating and late ($z<2$) AGN feedback models account for
a variety of energy exchanges between active
galaxies and the diffuse baryon gas finally forming a galaxy cluster. The main physical
difference between the two regimes is that pre-heating acts as an {\it external} heating
mechanism, modifying the entropy of baryons in a pre-collapse phase,
while AGN feedback is an {\it internal} heating mechanism acting within the already
formed DM potential well of a massive halo. The two scenarios imply a very different
energetic budget, since for the same given entropy level a larger energy per particle
is required at higher cosmic density (e.g. Tozzi \& Norman 2001; Mc Carthy et al.2008).
Only quite recently cosmological numerical simulations have achived sufficient resolution
and complexity to follow the interplays between the ICM and the populations of galaxies in a self-consistent
way, 
along the whole
cosmic evolution (e.g. Sijacki et al.2008; Teyssier et al.2010; Dubois et al.2010; 
Mc Carthy et al.2009).

In order to match the two approaches in the same cluster run, we investigated a 
re-simulation adopting the intermediate ($S_{0}=100keV cm^{2}$, run PH2) scenario
for early pre-heating, and less powerful jets at $z=1$, with
$\epsilon_{jet} \approx 2 \cdot 10^{57} ergs$ (run B5).

\begin{figure}
\includegraphics[width=0.45\textwidth]{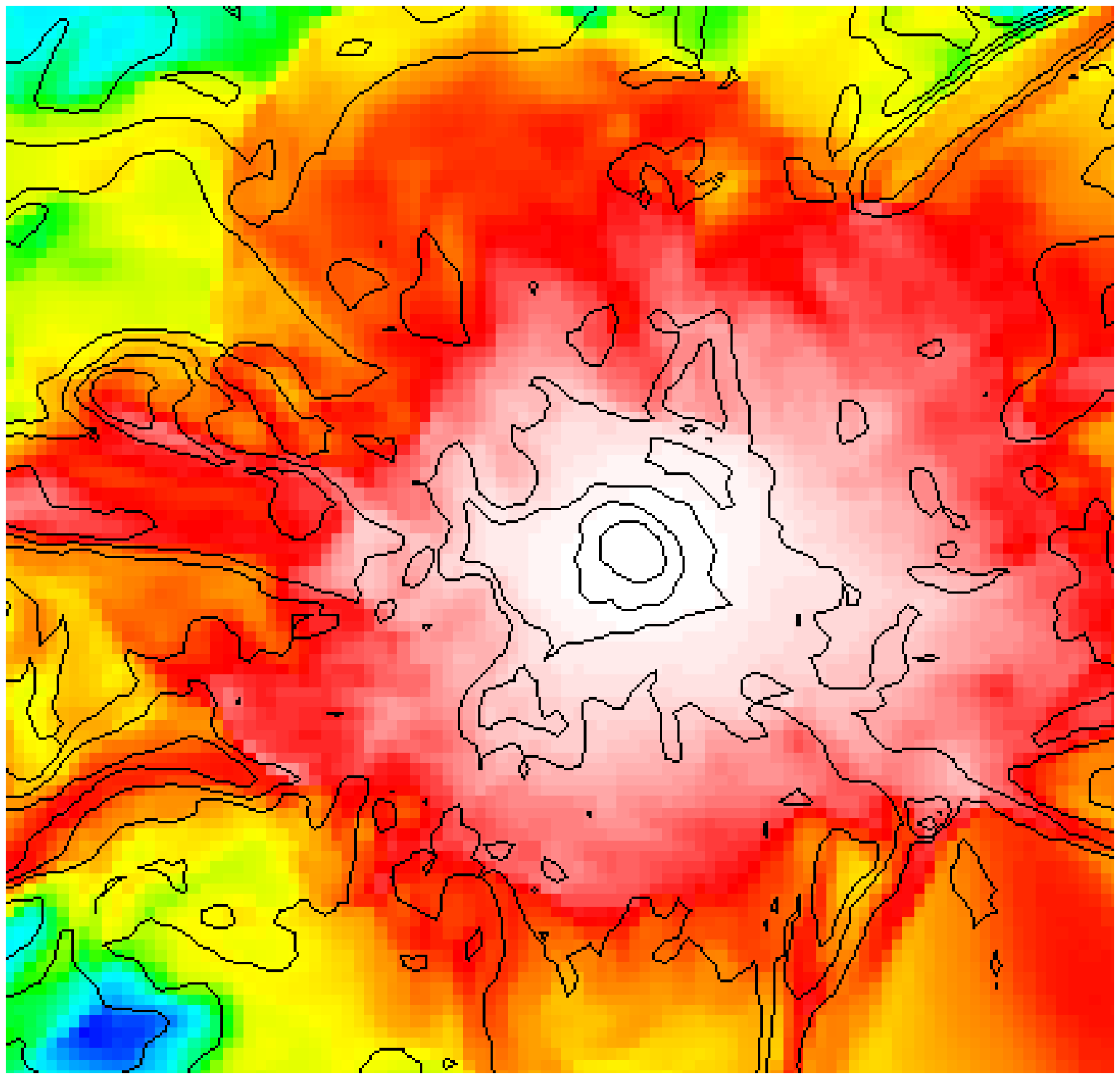}
\includegraphics[width=0.45\textwidth]{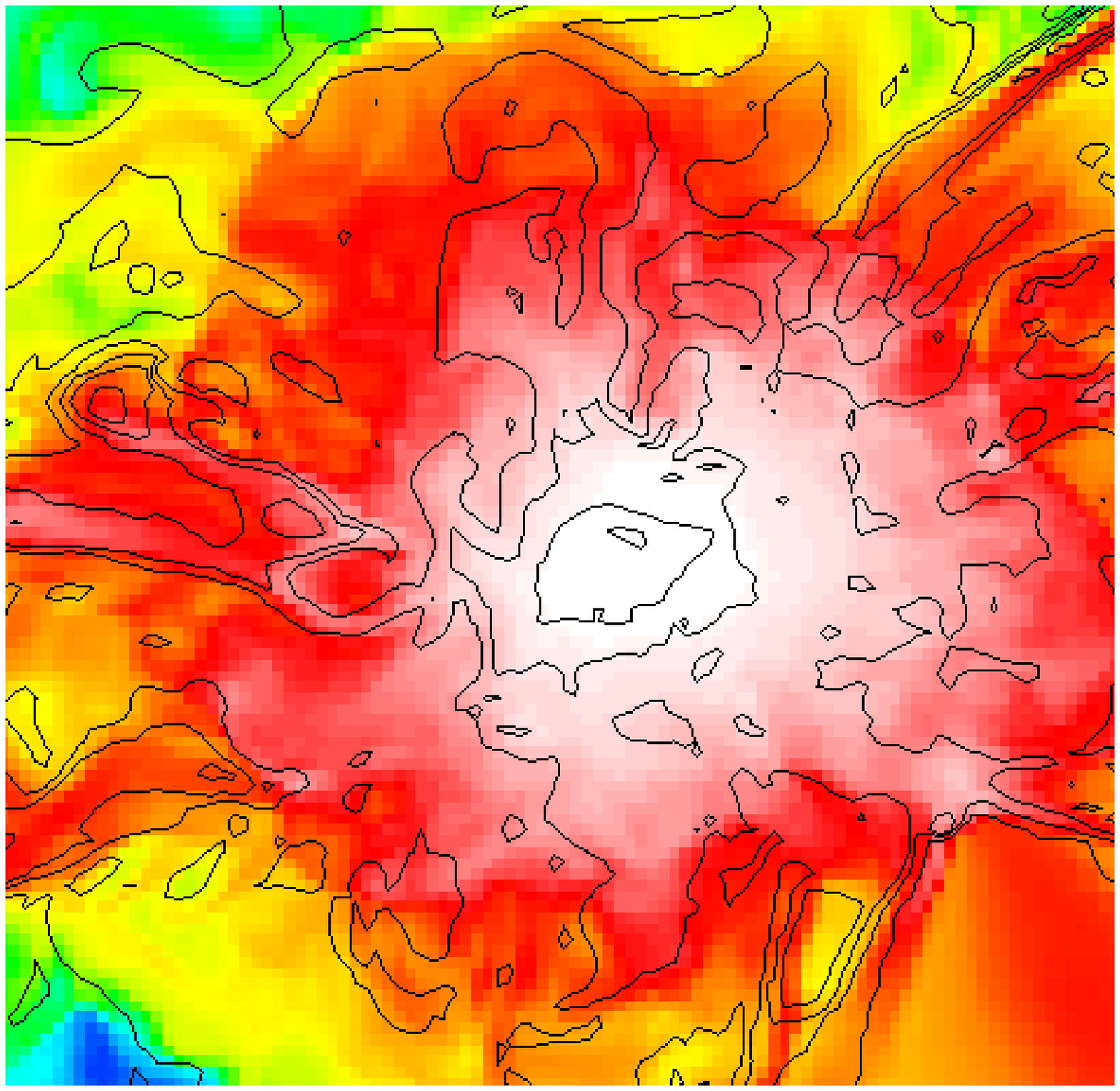}
\caption{Gas density (colors) and gas entropy (contours spaced in $\Delta log(S) = 0.2$)
for central region of run PH2 (top panel) and run B5 (bottom panel). The side
of the images is $2.5Mpc/h$; the colors
are as in Fig.\ref{fig:map_r0}.}
\label{fig:map_ph_jets}
\end{figure}

In Fig.\ref{fig:map_ph_jets} we show how the inner gas density and gas entropy 
of run PH2 (top panel) are modified by the late jets activity (bottom panel):
a vertical structure of gas with entropy $S \sim 100 keV cm^{2}$ is found 
at the opposite sides of the cluster center, resulting in significant entrainment of the
cold and dense gas of the cluster core, which is uplifted to larger radii at $z=0$. In consequence of this, the inner density core is significantly depleted compared to
run PH2 at the same redshift.

\begin{figure}
\includegraphics[width=0.45\textwidth]{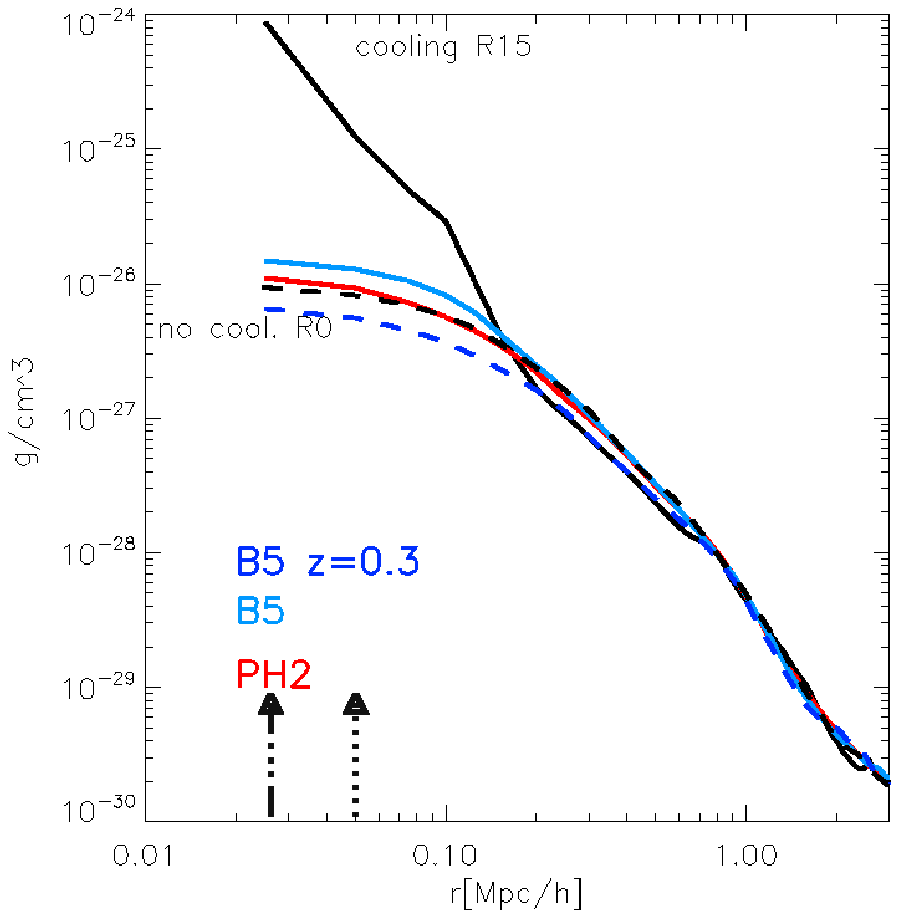}
\includegraphics[width=0.45\textwidth]{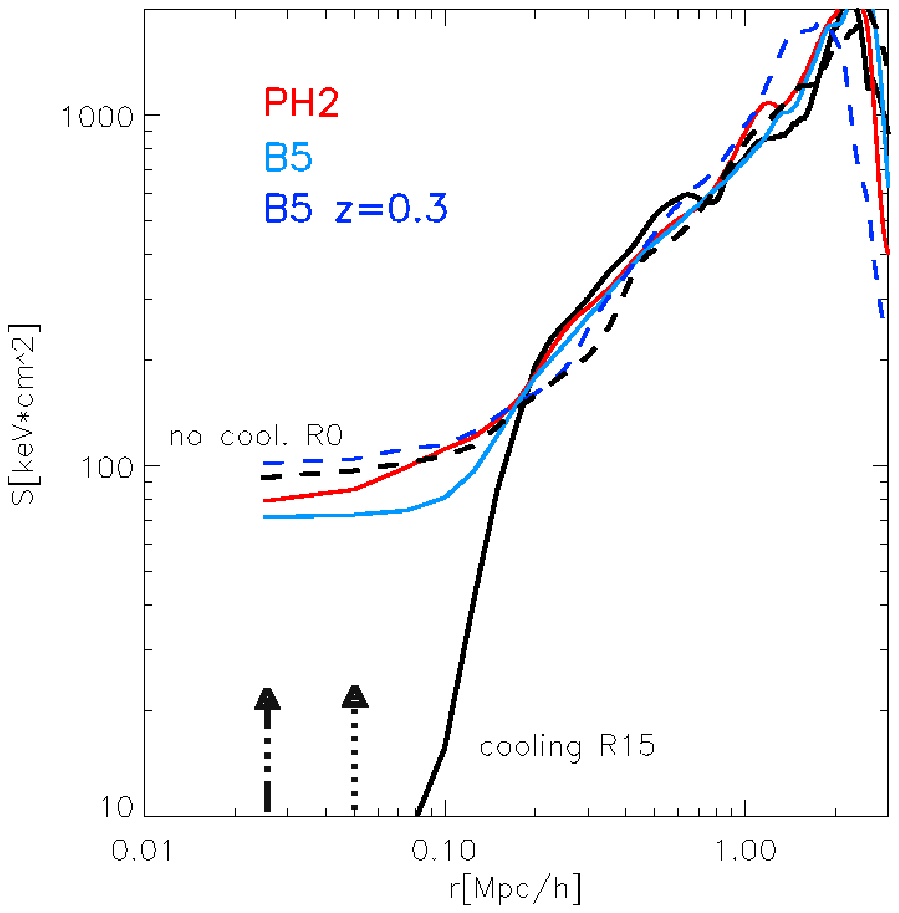}
\caption{Gas density and entropy radial profiles for the run PH2 (red), run B5 at $z=0$ (solid blue)
and $z=0.3$ (dashed blue).  
The additional black lines shows the profile for the non-radiative run (run R0, long dashed)
and for the cooling run (run R15, solid). 
The vertical dotted arrow shows the softening length adopted in run R0, while the dot-dashed one shows the softening of runs PH4 and B5.}
\label{fig:prof_ph_jets}
\end{figure}

In  Fig.\ref{fig:prof_ph_jets} we compare the profiles of gas density and gas
temperature for the non-radiative run (R0, dashed black line), of the radiative 
run with strong pre-heating (run PH4, in red) and of run B5 (blue).
Since the moderate amount of early pre-heating already prevented over-cooling of the
gas in the cluster core for $z>1$, the late injection of jets do
not drive of shocks stronger than $M>2$, and the extra entropy input 
is more uniformly released within the cluster core through mixing, rather
than through violent shock heating.
Indeed, the gas entropy profile and gas density profile are very similar to the non-radiative case  (run R0, solid blue line), and a well defined entropy floor is recovered; the similarity is even more evident at $z=0.3$ (dashed blue line), $\sim 2Gyr$ after the end of the jet injection. 

We conclude that, when the thermal properties of the cluster are concerned, 
the same 
configuration produced in a non-radiative run can be approximately achieved 
also in a radiative simulation,
thanks to the combined effect of  a uniform {\it external} pre-heating 
of $S_{0} \sim 100 keV cm^{2}$ at high ($z \sim 10$) redshift
and of a later phase of {\it internal} heating from jets injected 
by AGN at $z \sim 1$, characterized by an average power of $W_{jet} \sim 2 \cdot 10^{41} erg/s$. 

This is in line with results obtained with semi-analytical 1--D calculation
presented in Mc Carthy et al.(2008), and illustrates one of the 
(likely many) possible
combinations of external and internal heating mechanisms and radiative cooling
in realistic galaxy cluster simulations.
Of course this result is derived only from a single cluster object only, 
and additional re-simulations considering a wide range of masses
and dynamical state are necessary to investigate the above issues even in a statistical sense. 

\begin{figure}
\includegraphics[width=0.49\textwidth]{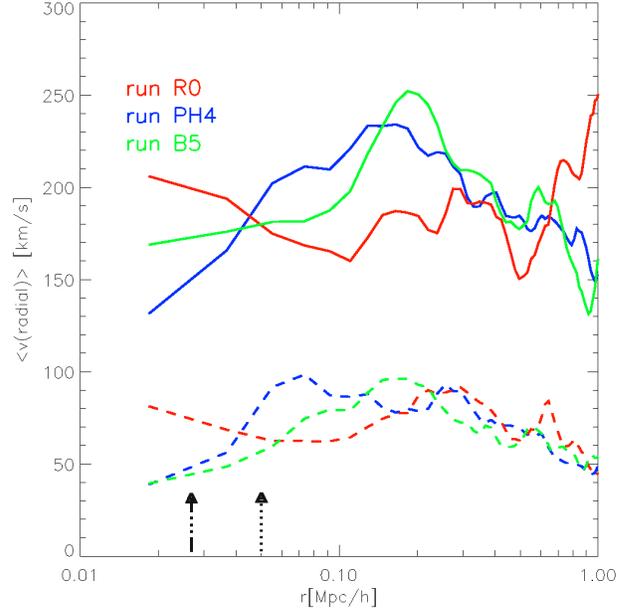}
\caption{Profiles of the total radial (solid lines) and chaotic radial (dashed lines) module of velocity for run R0 (red), for run PH4 (blue) and for run B5 (green). The dotted arrow shows the softening length adopted in run R0, while the dot-dashed one shows the softening of runs PH4 and B5.}
\label{fig:prof_vel_ph}
\end{figure}

We note however that the kinematic structures of the ICM velocity field in the various
run are very different (Fig.\ref{fig:prof_vel_ph}): the non-radiative run R0 is characterized
by a peaked velocity profile, with an infall velocity of $\sim -200km/s$ within the cluster core, 
due to the presence of a crossing satellite. On the other hand the other 
runs with additional heating and cooling do not present this feature; this is
due to the fact that the same gas clump have
been destroyed in the past by the action of early-preheating and enhanced shock
heating while crossing the main cluster virial radius. In addition, the run
with moderate jet feedback (run B5) shows a sharp velocity
structure in the profile of radial velocity with $+250km/s$ at $\sim 200-400kpc/h$, 
and a quite flat inner velocity profile, after the forcing of AGN outflows in the 
past.
We note that similar features in the radial velocity field, in response to 
AGN feedback, have been recently reported by Dubois et al.(2010) for cosmological simulations with the RAMSES code.


This kinematic differences should lead to different large scale patter of mixing/metallicity, and could be 
used to discriminate among degenerate
thermodynamical structures of galaxy clusters, to be compared with real
observations provided by future high resolution spectroscopic observations
of the ICM.

\section{Discussion and conclusion}  
\label{sec:conclusion}
In this paper 
we presented a detailed numerical study on the numerical and physical 
reasons for the presence of a flat entropy core in the innermost region
of galaxy clusters, simulated in cosmological grid simulations
adopting AMR (with customized version of the code ENZO).
To this end, we performed 30 cosmological re-simulations of the same non-major merger cluster
of final mass $M \approx 3.1 \cdot 10^{14}M_{\odot}/h$. We 
accurately studied the many parameters likely affecting
the production of physical or numerical entropy within cosmological
cluster simulations.

Evidences were presented for the {\it ubiquitous} presence of 
a well defined entropy floor inside the cluster core radius of non-radiative
cosmological runs, 
mostly independently on the numerical details of the simulations (Sec.\ref{sec:numerics}). 
This plateau-like
entropy distribution ranges from $\sim 70 keV \cdot cm^{2}$ to 
$\sim 150 keV \cdot cm^{2}$, 
and has a size of $r_{core} \sim 100-200 kpc/h$ ($\sim 0.1 R_{vir}$) at $z=0$. 
The most relevant factor affecting the level of entropy in the cluster core, among
those investigated, is the mesh refinement strategy adopted
in the simulation (Sec\ref{subsec:amr}): when mesh refinement triggered by velocity jumps is added to the 
standard mesh refinement triggered by gas/DM over-density, the level of entropy
in the core is increased by a factor $\sim 1.5 - 2$. 
This is due to the enhanced
presence of mixing motions and shocks within the cluster, which are otherwise
more dampened by under-sampling effects in a standard refinement strategy.
The other effects (N-body gravitational noise, cold unresolved flows, 
softening length in the calculation of DM dynamics, spatial 
resolution for gas dynamics) were found to introduce differences
 at the order of a few tens of percent on the final entropy profile, 
without changing the inner slope of the radial distribution in a relevant
way (Sec.\ref{subsec:dual}-Sec.\ref{subsec:gas_res}).

In Sec.\ref{subsec:tracers} we explored in detail the physical mechanism which produces
the flat entropy distribution in non radiative cosmological simulations, using Lagrangian tracers advected in the simulation. The regular accretion of shells of matter onto the forming cluster is responsible
for the strong entropy stratification found for $r>100kpc/h$ at the final epoch; this
stratification sets up already in the first phases of the collapse ($z \sim 1-2$),
and mirrors the different thermodynamic history of clumpy and smooth accretions onto the
main cluster.
Inside $r<100kpc/h$, mixing motions driven by matter accretion gradually mix lower
and higher entropy gas, producing an almost constant entropy profile in the region
where the gas+DM gravitational potential is shallow.
Our results here confirm and extend the seminal work of Mitchell et al.(2008) to 
a fully cosmological framework, and to non-major merger galaxy clusters.

We also explored more complex physical modeling of the ICM, following the effect
of radiative cooling (Sec.\ref{subsec:cool}) and of non-gravitational 
heating mechanisms, such as early uniform pre-heating (Sec.\ref{subsubsec:ph}) or
late jets-like injection by AGN (Sec.\ref{subsubsec:agn}). 
We report that, while it is feasible to recover a very similar profile of non-radiative
runs and of observed CHANDRA clusters (Cavagnolo et al.2009) with a suitable
choice of uniform extra-entropy input at $z \sim 10$ (in the range of 
$S_{0} \sim 100-200keV cm^{2}$), it results impossible to achieve similar
results with the set of trials of only jet-like injections, where we simulated
the thermal feedback from a central AGN.
In the explored configuration, the main problem is that the bulk of the energy
release from AGN outflows triggers too intense shock heating in the cooling ICM 
at $z \sim 1$,
and {\it steepens} the inner gas entropy profile as shown in semi-analytical
model presented in Lapi, Cavaliere \& Menci (2005). 
However, the adoption of a hybrid model with moderate early pre-heating and
late and modest jets feedback (Sec.\ref{subsubsec:ph_agn}) is found suitable
to recover a thermodynamic structure which extremely similar to the non-radiative
fiducial run (R0), and within observations (Cavagonolo et al.2009).
This suggests that the cluster configurations generally produced in 
cosmological non-radiative runs may be considered, at first approximation, 
a viable representation of real galaxy clusters with cooling and feedback mechanisms at work. 
However, similar thermal distributions at $z=0$ may be characterized by
quite different kinematic structure, depending on the different
feedback mechanisms at work, 
leading to potentially detectable spectroscopic X-ray features.
The study of other important observables not considered in this paper 
(such as the distribution of stars, metals, and the cluster baryon fraction) 
is expected to provide additional ways to
discriminate between similar thermal models (e.g. Booth \& Schaye 2009; Teyssier et al.2010).

Our conclusions is presently limited by the fact that we adopted so far
``ad-hoc''  models of pre-heating and AGN feedback, which are coupled to
the simulation "by hand", and that we focused on only two clusters (see also the Appendix).
It  would be interesting to investigate the same issues using a number of
clusters with different masses/dynamical history.

As a final remark, we note that the mechanism which produces the entropy floor
in cosmological grid simulations of non-radiative clusters can also explain
the long-debated difference reported for SPH and grid run (e.g. Frenk 
et al.1999; O'Shea et al.2005; Tasker et al.2008; Wadsley et al.2008; Mitchell et al.2008;
Springel 2010).
In major merger clusters, the difference in the core entropy is set by different 
efficiency in the mixing of gas at the moment of the closest encounter between
the clusters, as convincingly shown by Mitchell et al.(2008). In relaxed clusters, a similar mechanism
works on longer time scales, due to the continuous action of subsonic
chaotic motions triggered by the accretion of satellites. 
Any different in the modeling of mixing in the two numerical methods can explain 
the presence or absence of a well defined entropy core structure.
It has been shown that the presence of an artificial viscosity
term in standard SPH greatly reduces the small scale mixing in a number of realistic
cases, compared to standard Eulerian simulations (e.g. Agertz et al.2007; Wadsley
et al.2008; Springel 2010), and that the adoption of less viscous simulations
produces entropy distributions in clusters more similar to Eulerian runs (e.g. Dolag et al.2005; Mitchell et al.2009).
As shown in this work, shock heating is the leading source of entropy {\it production} 
in cluster (well beyond the role of any possible numerical artifact), while physical
mixing is the reason for the {\it spreading} of entropy in the innermost
cluster regions.
The fact that mixing in SPH is usually reduced by numerical effects, fully 
explains while the two methods are in disagreement in the center of clusters, when
mixing is maximum in grid codes, while they are
found in much better agreement at larger radii (e.g. Frenk et al.1999).

Since the buoyancy in the stratified ICM is strongly dependent on the underlying entropy distribution, the above findings emphasize
the need of having a suitable numerical representation of  
cluster cores, since this may affect also the estimated energy budget needed from
non-gravitational heating mechanism, and in their efficiency in mixing/heating
the surrounding Intra Cluster Medium.

 
\section{acknowledgments}
I am strongly indebted with my friends and advisors, G.Brunetti and C.Gheller,
for the fruitful collaboration of these years, which made this work possible.
I acknowledge R.Brunino, M.Nanni, F. Tinarelli and A. Tugnoli of helpful computational support at CINECA and at Radio Astronomy Institute in Bologna.
I thank M.Bruggen, E.Liuzzo, K.Dolag and G.Bryan for very useful technical discussions. I thank the referee of the paper, Tom Theuns, for his constructive suggestions and comments, which helped improving the quality of this work.
I acknowledge partial 
support through grant ASI-INAF I/088/06/0 and PRIN INAF 2007/2008, and the 
usage of computational resources under the CINECA-INAF 2008-2010 agreement
and the 2009 Key Project ``Turbulence, shocks and cosmic rays electrons 
in massive galaxy clusters at high resolution''.

\section{Appendix}

Cluster mergers may boost shock heating and mixing motions in the ICM for several Gyrs (e.g. Ricker \& Sarazin 2001), significanly changing the physical entropy
generation in a cluster with a dynamical evolution different than the relaxed case
explored in the main part of the paper.

We present here 
some complementary tests on the entropy distribution of a major merger cluster of final
mass $M \approx 2.1 \cdot 10^{14}M_{\odot}/h$.
The bulk of the total mass of this cluster is assembled in a major merger at $z=0.85$,
with an approximate mass ratio of $M_{1}/M_{2} \sim 3$ between the colliding halos. 
Only a subsample of the re-simulations presented in the main part of the paper were
repeated with this cluster;
the parameters of the tests run in this case are listed in Table \ref{tab:tab2}
(all the cosmological parameter are as in Sec.\ref{subsec:fiducial}).
The non radiative simulation of this cluster produces a flat entropy
core with $S \sim 80-90 keV cm^{2}$ for $r<100kpc/h$, similar to the non-radiative fiducial run (run R0, Sec.\ref{subsec:amr}).

The adoption of the additional mesh refinement based on velocity jumps (run A2) causes
a net increase of the internal entropy compared with the more standard
mesh refinement strategy based on gas/DM over-density (A1), as shown in Sec.\ref{subsec:amr}.
The role of the gravitational softening is found to be more important in this major
merger cluster, compared to the relaxed cluster studied in the main part of the paper:
the adoption of a larger softening ($50kpc/h$, run A3) produces an entropy core larger
by $\sim 20$ per cent compared to a smaller softening ($25kpc/h$, run A2). This stresses
the higher importance of having a good resolution for the computation of gravitational
forces in the case of violent oscillations of the gravitational potential driven in a
merger event, which may generate an amount of extra-entropy
production of numerical origin.

The adoption of radiative cooling (in the case of standard refinement, C1, or with the
velocity-based refinement, C2) causes a very similar trend as in the case of the relaxed
cluster explored in the paper, with the onset of catastrophic cooling for $r<200kpc/h$. This shows
that, at least for this {\it early} major merger ($z \sim 0.85$), the
action of intense heating from merger shocks is not effective in destroying
the forming cooling region, somewhat at variance with other works with 
Eulerian simulations (e.g. Burns et al.2008).

To spare computational time, most of the runs with non-gravitational heating
were performed only with the standard refinement scheme.
A uniform pre-heating  of $S_{0}=100keV cm^{2}$ at $z=10$ (run PH2) is ineffective to stop the catastrophic cooling for $r<50kpc/h$; however when applied 
to this major merger system it results in a significantly higher inner entropy
value compared to the relaxed system studied in the main part of the paper (Sec.\ref{subsubsec:ph}). A pre-heating of $S_{0}=200keV cm^{2}$ (PH3) on the
other hand almost perfectly recover the entropy distribution of the non-radiative run (PH3) for the merger cluster.

Runs with pre-heating of $S_{0}=100keV cm^{2}$ and AGN feedback were produced
for the jets energy of $\epsilon_{jet}=2\cdot 10^{57} ergs$ (run J1) and  $\epsilon_{jet}=10^{58} ergs$. The profile of the J1 run is very similar to what obtained for the relaxed cluster of the paper; the profile with a higher AGN energy
results in a flat entropy profile at $\sim 130 keV cm^{2}$ for $r<200kpc/h$.
Finally, we re-simulated run J2 adopting the velocity based refinement (run J3), finding
still a very flat profile inside $r<200kpc/h$, and a $\sim 50$ per cent larger
entropy in the center.
According to the assumed jet energy, our re-simulations with cooling, pre-heating and AGN feedback can thus provide an acceptable
match with one of the two classes of the bimodal entropy distribution reported for the CHANDRA observations of Cavagnolo et al.(2009). 

In Fig.\ref{fig:prof_appendix2} we show the time evolution for three snapshots
of run C1, J2 and J3, showing the X-ray bolometric luminosity for a region
of $\sim 1.8 \times 2.2Mpc/h$ per side and thickness $25kpc/h$, around the
epoch of the major merger.
Comparing C1 to J2, we show how the action of AGN feedback removes the central
gas condensation within the cooling region in a few $\sim 10 Myr$.
Run J3 emphasizes the role played by the mesh refinement strategy on the expanding shocks driven by the (almost contemporary) AGN burst and the major merger. We also note how at least one inflated ``bubble'' can survive for a few time steps
after the injection, due to the reduced numerical mixing in the implemented mesh refinement strategy.

The above tests suggest that the most important findings reported in the main body
of the article are general, since they apply to clusters with a similar mass but two completely different dynamical histories. 
However the efficiency of the extra-heating models applied to radiative runs may depend on the dynamical history of the host cluster, and further studies are needed to estimate the global efficiency of the proposed scenarios in a statistical sense.

\begin{table}
\label{tab:tab2}
\caption{Main characteristics of the performed runs (as in Tab.\ref{tab:tab1}).In the last row, the assumed pre-heating background is $S_{0}=100keV cm^{2}$ and the thermal energy of the jets is $2 \cdot 10^{57}ergs$}
\centering \tabcolsep 5pt 
\begin{tabular}{c|c|c|c|c}
 ID & Max Res. [kpc/h] &  soft. [kpc/h] & AMR & note \\  \hline
    A1 & 25 & 25 & D & non-radiative\\
    A2 & 25 & 25 & DV & non-radiative\\
    A3 & 25 & 50 & DV & non-radiative\\
    C1 & 25 & 25 & D & cooling\\
    C2 & 25 & 25 & DV & cooling\\
    PH2 & 25 & 25 & D & cool.+PH($100keV cm^{2}$) \\
    PH3 & 25 & 25 & D & cool.+PH($200keV cm^{2}$) \\
    J1 & 25 & 25 & D & cool.+PH+J($2\cdot 10^{57}ergs$) \\
    J2 & 25 & 25 & D & cool.+PH+J($10^{58}ergs$)\\ 
    J3 & 25 & 25 & DV & cool.+PH+J($10^{58}ergs$) \\
    \end{tabular}
    \end{table}

\begin{figure}
\includegraphics[width=0.47\textwidth]{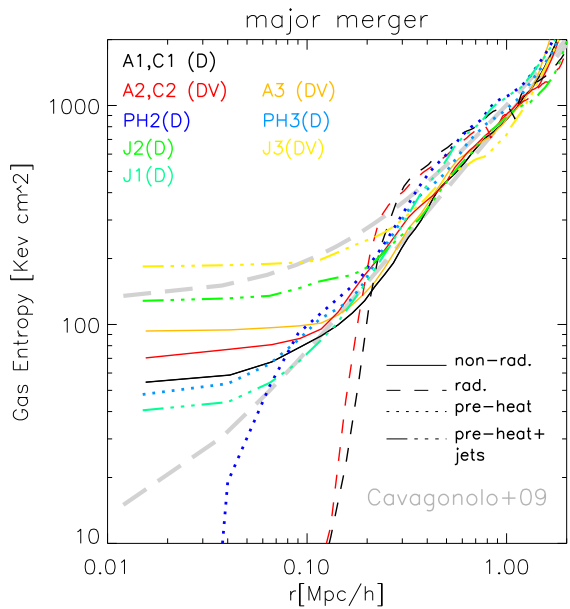}
\caption{Profiles of the gas entropy in the major merger run. 
The additional long dashed line show the bimodal gas entropy distribution
from CHANDRA observations (Cavagnolo et al.2009).}
\label{fig:prof_appendix}
\end{figure}

\begin{figure*}
\begin{center}
\includegraphics[height=0.4\textheight,width=0.96\textwidth]{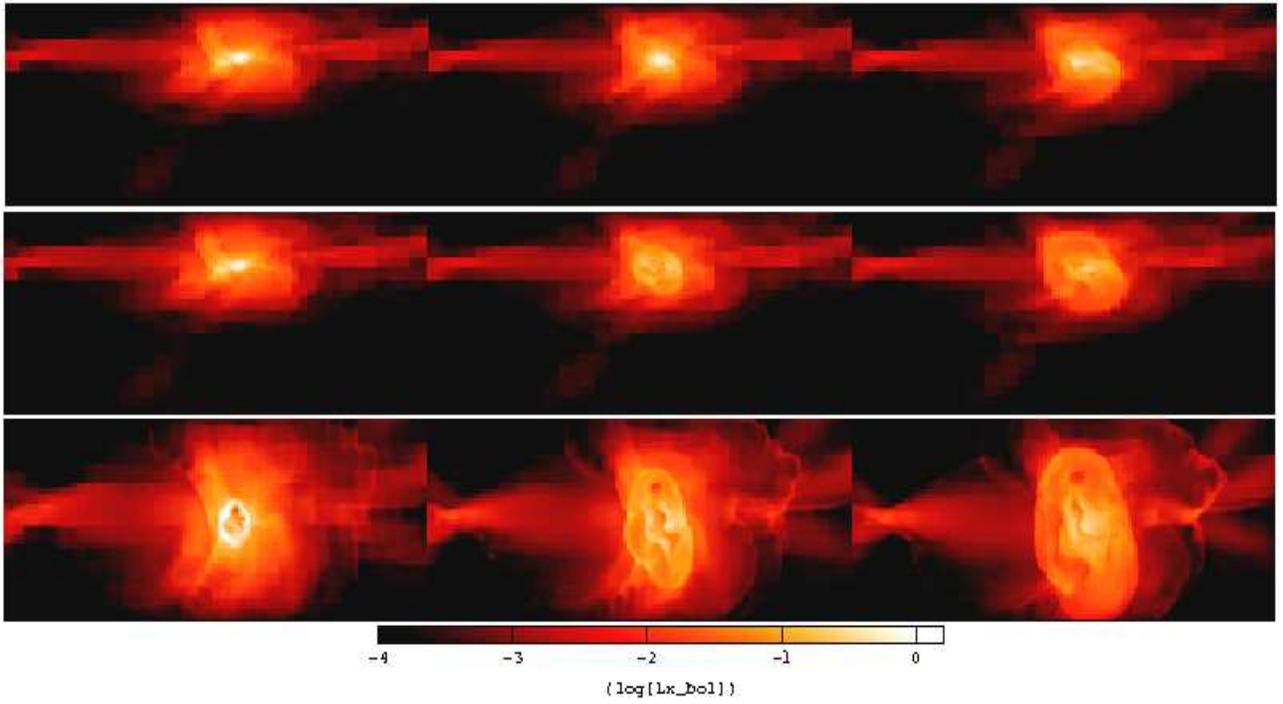}
\caption{Maps for the evolution of the X-ray bolometric luminosity for slices
of $1.8 \times 2.2Mpc/h$ and depth $25kpc/h$ for the major merger cluster studied in the Appendix, at $z=0.9$, $z=0.85$ and $z=0.81$. The top panels are for the pure cooling run (C1), the middle panels are run J1 with cooling, pre-heating ($S=100keV cm^{2}$ at $z=10$) and AGN feedback ($\epsilon_{jet}=10^{58}erg/s$), while
the bottom panel are for a re-simulation with the same setup, but mesh refinement triggered by velocity jump (run J3).}
\end{center}
\label{fig:prof_appendix2}
\end{figure*}


\begin{thebibliography} {199}
\bibliographystyle{mn2e}

\bibitem[Abel(2010)]{2010arXiv1003.0937A} Abel, T.\ 2010, arXiv:1003.0937 


\bibitem[\protect\citeauthoryear{Ascasibar 
\& Markevitch}{2006}]{2006ApJ...650..102A} Ascasibar Y., Markevitch M., 2006, ApJ, 650, 102 

\bibitem[Bialek et al.(2001)]{2001ApJ...555..597B} Bialek, J.~J., Evrard, A.~E., \& Mohr, J.~J. 2001, ApJ, 555, 597 

\bibitem[Bird et al.(2008)]{2008ApJ...676..147B} Bird, J., Martini, P., 
\& Kaiser, C. 2008, ApJ, 676, 147 

\bibitem[B{\^i}rzan et al.(2004)]{2004ApJ...607..800B} B{\^i}rzan, L., 
Rafferty, D.~A., McNamara, B.~R., Wise, M.~W., 
\& Nulsen, P.~E.~J. 2004, ApJ, 607, 800 

\bibitem[B{\^i}rzan et al.(2008)]{2008ApJ...686..859B} B{\^i}rzan, L., 
McNamara, B.~R., Nulsen, P.~E.~J., Carilli, C.~L., 
\& Wise, M.~W. 2008, ApJ, 686, 859 


\bibitem[Borgani et al.(2002)]{2002MNRAS.336..409B} Borgani, S., Governato, 
F., Wadsley, J., Menci, N., Tozzi, P., Quinn, T., Stadel, J., 
\& Lake, G. 2002, MNRAS, 336, 409 


\bibitem[Borgani et al.(2005)]{2005MNRAS.361..233B} Borgani, S., 
Finoguenov, A., Kay, S.~T., Ponman, T.~J., Springel, V., Tozzi, P., 
\& Voit, G.~M.\ 2005, MNRAS, 361, 233 

\bibitem[Booth 
\& Schaye(2009)]{2009MNRAS.398...53B} Booth, C.~M., \& Schaye, J. 2009, MNRAS, 398, 53 


\bibitem[Brighenti 
\& Mathews(2003)]{2003ApJ...587..580B} Brighenti, F., \& Mathews, W.~G. 2003, ApJ, 587, 580 



\bibitem[\protect\citeauthoryear{Burns et al.}{2008}]{2008ApJ...675.1125B} 
Burns J.~O., Hallman E.~J., Gantner B., Motl P.~M., Norman M.~L., 2008, 
ApJ, 675, 1125 

\bibitem[Br{\"u}ggen 
\& Kaiser(2002)]{2002Natur.418..301B} Br{\"u}ggen, M., \& Kaiser, C.~R.\ 2002, Nat., 418, 301 

\bibitem[Br{\"u}ggen et al.(2007)]{2007MNRAS.380L..67B} Br{\"u}ggen, M., 
Heinz, S., Roediger, E., Ruszkowski, M., 
\& Simionescu, A. 2007, MNRAS, 380, L67 



\bibitem[\protect\citeauthoryear{Bryan et al.}{1995}]{1995CoPhC..89..149B} 
Bryan G.~L., Norman M.~L., Stone J.~M., Cen R., Ostriker J.~P., 1995, 
CoPhC, 89, 149 

\bibitem[Cavaliere et al.(1997)]{1997ApJ...484L..21C} Cavaliere, A., Menci, 
N., \& Tozzi, P. 1997, ApJl, 484, L21 

\bibitem[Cavagnolo et al.(2009)]{2009ApJS..182...12C} Cavagnolo, K.~W., 
Donahue, M., Voit, G.~M., \& Sun, M. 2009, ApJs, 182, 12 

\bibitem[Churazov et 
al.(2000)]{2000A&A...356..788C} Churazov, E., Forman, W., Jones, C., Boehringer, H. 2000, A\&A, 356, 788 

\bibitem[Dalla Vecchia et al.(2004)]{2004MNRAS.355..995D} Dalla Vecchia, 
C., Bower, R.~G., Theuns, T., Balogh, M.~L., Mazzotta, P., 
\& Frenk, C.~S. 2004, MNRAS, 355, 995 


\bibitem[\protect\citeauthoryear{David, Forman, 
\& Jones}{1991}]{1991ApJ...380...39D} David L.~P., Forman W., Jones C., 1991, ApJ, 380, 39 

\bibitem[David et al.(2009)]{2009ApJ...705..624D} David, L.~P., Jones, C., 
Forman, W., Nulsen, P., Vrtilek, J., O'Sullivan, E., Giacintucci, S., 
\& Raychaudhury, S. 2009, ApJ, 705, 624 


\bibitem[De Young(2003)]{2003MNRAS.343..719D} De Young, D.~S. 2003, 
MNRAS, 343, 719 

\bibitem[De Young(2010)]{2010ApJ...710..743D} De Young, D.~S. 2010, ApJ, 
710, 743 

\bibitem[Dolag et al.(2005)]{2005MNRAS.364..753D} Dolag, K., Vazza, F., 
Brunetti, G., \& Tormen, G. 2005, MNRAS, 364, 753 

\bibitem[Donahue et al.(2006)]{2006ApJ...643..730D} Donahue, M., Horner, 
D.~J., Cavagnolo, K.~W., \& Voit, G.~M. 2006, ApJ, 643, 730 


\bibitem[Dubois et al.(2010)]{2010arXiv1004.1851D} Dubois, Y., Devriendt, 
J., Slyz, A., \& Teyssier, R.\ 2010, arXiv:1004.1851 

\bibitem[Dunn 
\& Fabian(2006)]{2006MNRAS.373..959D} Dunn, R.~J.~H., \& Fabian, A.~C. 2006, MNRAS, 373, 959 

\bibitem[\protect\citeauthoryear{Evrard 
\& Henry}{1991}]{1991ApJ...383...95E} Evrard A.~E., Henry J.~P., 1991, ApJ, 383, 95 


\bibitem[\protect\citeauthoryear{Evrard 
\& Henry}{1991}]{1991tpsu.rept..127E} Evrard A.~E., Henry J.~P., 1991, tpsu.rept, 127 

\bibitem[\protect\citeauthoryear{Fabian}{1994}]{1994ARA&A..32..277F} Fabian A.~C., 1994, ARA\&A, 32, 277 


\bibitem[Frenk et al.(1999)]{1999ApJ...525..554F} Frenk, C.~S., et al. 
1999, ApJ , 525, 554 

\bibitem[Fryxell et al.(2000)]{2000ApJS..131..273F} Fryxell, B., et al.
2000, ApJs, 131, 273 

\bibitem[Giacintucci et al.(2008)]{2008ApJ...682..186G} Giacintucci, S., et 
al. 2008, ApJ, 682, 186 


\bibitem[Giodini et al.(2010)]{2010ApJ...714..218G} Giodini, S., et al. 
2010, ApJ, 714, 218

\bibitem[Gitti et al.(2010)]{2010ApJ...714..758G} Gitti, M., O'Sullivan, 
E., Giacintucci, S., David, L.~P., Vrtilek, J., Raychaudhury, S., 
\& Nulsen, P.~E.~J. 2010, ApJ, 714, 758 

\bibitem[Gu et al.(2009)]{2009MNRAS.396..984G} Gu, M., Cao, X., 
\& Jiang, D.~R. 2009, MNRAS, 396, 984 

\bibitem[Guo 
\& Mathews(2010)]{2010arXiv1004.2258G} Guo, F., \& Mathews, W.~G. 2010, arXiv:1004.2258 

\bibitem[\protect\citeauthoryear{Haardt \& 
Madau}{1996}]{1996ApJ...461...20H} Haardt F., Madau P., 1996, ApJ, 461, 20 

\bibitem[Heinz et al.(2006)]{2006MNRAS.373L..65H} Heinz, S., Br{\"u}ggen, 
M., Young, A., \& Levesque, E. 2006, MNRAS, 373, L65 

\bibitem[Hockney 
\& Eastwood(1981)]{1981csup.book.....H} Hockney, R.~W., \& Eastwood, J.~W. 1981, Computer Simulation Using Particles, New York: McGraw-Hill, 1981, 

\bibitem[\protect\citeauthoryear{Iapichino 
\& Niemeyer}{2008}]{2008MNRAS.388.1089I} Iapichino L., Niemeyer J.~C., 2008, MNRAS, 388, 1089 

\bibitem[\protect\citeauthoryear{Kaiser}{1991}]{1991ApJ...383..104K} Kaiser 
N., 1991, ApJ, 383, 104 

\bibitem[\protect\citeauthoryear{Katz 
\& White}{1993}]{1993ApJ...412..455K} Katz N., White S.~D.~M., 1993, ApJ, 412, 455 

\bibitem[Kay et al.(2007)]{2007MNRAS.377..317K} Kay, S.~T., da Silva, 
A.~C., Aghanim, N., Blanchard, A., Liddle, A.~R., Puget, J.-L., Sadat, R., 
\& Thomas, P.~A. 2007, MNRAS, 377, 317 



\bibitem[\protect\citeauthoryear{Lapi, Cavaliere, 
\& Menci}{2005}]{2005ApJ...619...60L} Lapi A., Cavaliere A., Menci N., 2005, ApJ, 619, 60 


\bibitem[Lin et al.(2006)]{2006ApJ...651..636L} Lin, W.~P., Jing, Y.~P., 
Mao, S., Gao, L., \& McCarthy, I.~G. 2006, ApJ, 651, 636 

\bibitem[Liuzzo et 
al.(2009)]{2009A&A...501..933L} Liuzzo, E., Taylor, G.~B., Giovannini, G., \& Giroletti, M. 2009, A\&A, 501, 933 



\bibitem[\protect\citeauthoryear{Lloyd-Davies, Ponman, 
\& Cannon}{2000}]{2000MNRAS.315..689L} Lloyd-Davies E.~J., Ponman T.~J., Cannon D.~B., 2000, MNRAS, 315, 689 


\bibitem[\protect\citeauthoryear{McCarthy et 
al.}{2007}]{2007MNRAS.376..497M} McCarthy I.~G., et al., 2007, MNRAS, 376, 
497 

\bibitem[McCarthy et al.(2009)]{2009arXiv0911.2641M} McCarthy, I.~G., et 
al.\ 2009, arXiv:0911.2641 

\bibitem[Merlin et 
al.(2010)]{2010A&A...513A..36M} Merlin, E., Buonomo, U., Grassi, T., Piovan, L., \& Chiosi, C. 2010, A\&A, 513, A36 

\bibitem[Mitchell et al.(2009)]{2009MNRAS.395..180M} Mitchell, N.~L., 
McCarthy, I.~G., Bower, R.~G., Theuns, T., 
\& Crain, R.~A. 2009, MNRAS, 395, 180 


\bibitem[Norman 
\& Bryan(1999)]{1999LNP...530..106N} Norman, M.~L., \& Bryan, G.~L. 1999, The Radio Galaxy Messier 87, 530, 106 

\bibitem[\protect\citeauthoryear{Norman et al.}{2007}]{2007arXiv0705.1556N} 
Norman M.~L., Bryan G.~L., Harkness R., Bordner J., Reynolds D., O'Shea B., 
Wagner R., 2007, arXiv, 705, arXiv:0705.1556 


\bibitem[O'Neill 
\& Jones(2010)]{2010ApJ...710..180O} O'Neill, S.~M., \& Jones, T.~W. 2010, ApJ, 710, 180 



\bibitem[O'Shea et al.(2005)]{2005ApJS..160....1O} O'Shea, B.~W., Nagamine, 
K., Springel, V., Hernquist, L., \& Norman, M.~L. 2005, ApJ, 160, 1 

\bibitem[Parrish 
\& Quataert(2008)]{2008ApJ...677L...9P} Parrish, I.~J., \& Quataert, E. 2008, ApjL, 677, L9 

\bibitem[Pearce et al.(2000)]{2000MNRAS.317.1029P} Pearce, F.~R., Thomas, 
P.~A., Couchman, H.~M.~P., \& Edge, A.~C. 2000, MNRAS, 317, 1029 



\bibitem[\protect\citeauthoryear{Ponman, Cannon, 
\& Navarro}{1999}]{1999Natur.397..135P} Ponman T.~J., Cannon D.~B., Navarro J.~F., 1999, Natur, 397, 135 

\bibitem[\protect\citeauthoryear{Poole et al.}{2008}]{2008MNRAS.391.1163P} 
Poole G.~B., Babul A., McCarthy I.~G., Sanderson A.~J.~R., Fardal M.~A., 
2008, MNRAS, 391, 1163 

\bibitem[Price(2008)]{2008JCoPh.22710040P} Price, D.~J. 2008, Journal of 
Computational Physics, 227, 10040 


\bibitem[Ricker 
\& Sarazin(2001)]{2001ApJ...561..621R} Ricker, P.~M., \& Sarazin, C.~L. 2001, ApJ, 561, 621 


\bibitem[\protect\citeauthoryear{Ritchie 
\& Thomas}{2002}]{2002MNRAS.329..675R} Ritchie B.~W., Thomas P.~A., 2002, MNRAS, 329, 675 


\bibitem[Robertson et al.(2010)]{2010MNRAS.401.2463R} Robertson, B.~E., 
Kravtsov, A.~V., Gnedin, N.~Y., Abel, T., 
\& Rudd, D.~H. 2010, MNRAS, 401, 2463 



\bibitem[Romeo et al.(2006)]{2006MNRAS.371..548R} Romeo, A.~D., 
Sommer-Larsen, J., Portinari, L., 
\& Antonuccio-Delogu, V. 2006, MNRAS, 371, 548 


\bibitem[\protect\citeauthoryear{Rossetti 
\& Molendi}{2010}]{2010A&A...510A..83R} Rossetti M., Molendi S., 2010, A\&A, 510, A83 

\bibitem[Ruszkowski et al.(2007)]{2007MNRAS.378..662R} Ruszkowski, M., 
En{\ss}lin, T.~A., Br{\"u}ggen, M., Heinz, S., 
\& Pfrommer, C 2007, MNRAS, 378, 662 

\bibitem[Ruszkowski 
\& Oh(2010)]{2010ApJ...713.1332R} Ruszkowski, M., \& Oh, S.~P. 2010, ApJ , 713, 1332 


\bibitem[Ryu et al.(1993)]{1993ApJ...414....1R} Ryu, D., Ostriker, J.~P., 
Kang, H., \& Cen, R. 1993, ApJ, 414, 1 



\bibitem[\protect\citeauthoryear{Ryu et al.}{2003}]{2003ApJ...593..599R} 
Ryu D., Kang H., Hallman E., Jones T.~W., 2003, ApJ, 593, 599 


\bibitem[Sanders et al.(2009)]{2009MNRAS.396.1449S} Sanders, J.~S., Fabian, 
A.~C., \& Taylor, G.~B. 2009, MNRAS, 396, 1449 


\bibitem[Scannapieco 
\& Br{\"u}ggen(2008)]{2008ApJ...686..927S} Scannapieco, E., \& Br{\"u}ggen, M.\ 2008, ApJ, 686, 927 


\bibitem[Scannapieco 
\& Br{\"u}ggen(2010)]{2010MNRAS.tmp..583S} Scannapieco, E., \& Br{\"u}ggen, M.\ 2010, MNRAS, 583 

\bibitem[Schwarzschild 1959] {} Schwarzschild, M. 1959, Ap J., 130,345.


\bibitem[Sijacki 
\& Springel(2006)]{2006MNRAS.366..397S} Sijacki, D., \& Springel, V. 2006, MNRAS, 366, 397 


\bibitem[Sijacki et al.(2007)]{2007MNRAS.380..877S} Sijacki, D., Springel, 
V., Di Matteo, T., \& Hernquist, L. 2007, MNRAS, 380, 877 

\bibitem[Sijacki et al.(2008)]{2008MNRAS.387.1403S} Sijacki, D., Pfrommer, 
C., Springel, V., \& En{\ss}lin, T.~A. 2008, MNRAS, 387, 1403 


\bibitem[\protect\citeauthoryear{Springel}{2010}]{2010MNRAS.401..791S} 
Springel V., 2010, MNRAS, 401, 791 

\bibitem[\protect\citeauthoryear{Springel}{2010}]{2010MNRAS.401..791S} 
Springel V., 2010, MNRAS, 401, 791 

\bibitem[Tasker et al.(2008)]{2008MNRAS.390.1267T} Tasker, E.~J., Brunino, 
R., Mitchell, N.~L., Michielsen, D., Hopton, S., Pearce, F.~R., Bryan, 
G.~L., \& Theuns, T.\ 2008, MNRAS, 390, 1267 

\bibitem[Teyssier et al.(2010)]{2010arXiv1003.4744T} Teyssier, R., Moore, 
B., Martizzi, D., Dubois, Y., \& Mayer, L. 2010, arXiv:1003.4744 



\bibitem[\protect\citeauthoryear{Tozzi 
\& Norman}{2001}]{2001ApJ...546...63T} Tozzi P., Norman C., 2001, ApJ, 546, 63 

\bibitem[Valdarnini(2002)]{2002ApJ...567..741V} Valdarnini, R.\ 2002, ApJ, 
567, 741 


\bibitem[Vazza et 
al.(2009)]{2009A&A...504...33V} Vazza, F., Brunetti, G., Kritsuk, A., Wagner, R., Gheller, C., \& Norman, M. 2009, A\&A, 504, 33 

\bibitem[Vazza et 
al.(2010)]{2010A&A...513A..32V} Vazza, F., Gheller, C., \& Brunetti, G. 2010, A\&A, 513, A32 

\bibitem[\protect\citeauthoryear{Vazza et al.}{2010}]{2010arXiv1003.5658V} 
Vazza F., Brunetti G., Gheller C., Brunino R., 2010, arXiv, arXiv:1003.5658 



\bibitem[Voit et al.(2005)]{2005MNRAS.364..909V} Voit, G.~M., Kay, S.~T., 
\& Bryan, G.~L. 2005, MNRAS, 364, 909 


\bibitem[Wadsley et al.(2008)]{2008MNRAS.387..427W} Wadsley, J.~W., 
Veeravalli, G., \& Couchman, H.~M.~P. 2008, MNRAS, 387, 427 

\bibitem[\protect\citeauthoryear{White}{1991}]{1991ApJ...367...69W} White 
R.~E., III, 1991, ApJ, 367, 69 

\bibitem[Wise et al.(2007)]{2007ApJ...659.1153W} Wise, M.~W., McNamara, 
B.~R., Nulsen, P.~E.~J., Houck, J.~C., 
\& David, L.~P. 2007, ApJ, 659, 1153 


\bibitem[\protect\citeauthoryear{Woodward \& 
Colella}{1984}]{1984JCoPh..54..115W} Woodward P., Colella P., 1984, JCoPh, 
54, 115 

\bibitem[Worrall(2009)]{2009A&ARv..17....1W} Worrall, D.~M. 2009, A\&Ar, 17, 1 

\bibitem[Xu et al.(2009)]{2009ApJ...698L..14X} Xu, H., Li, H., Collins, 
D.~C., Li, S., \& Norman, M.~L. 2009, ApJL, 698, L14 


\bibitem[Younger 
\& Bryan(2007)]{2007ApJ...666..647Y} Younger, J.~D., \& Bryan, G.~L. 2007, ApJ, 666, 647 

\bibitem[Yoshikawa et al.(2000)]{2000ApJ...535..593Y} Yoshikawa, K., Jing, 
Y.~P., \& Suto, Y.\ 2000, ApJ , 535, 593 


\bibitem[Zanni et 
al.(2005)]{2005A&A...429..399Z} Zanni, C., Murante, G., Bodo, G., Massaglia, S., Rossi, P., \& Ferrari, A. 2005, A\&A, 429, 399 




\bibitem[ZuHone et al.(2009)]{2009arXiv0912.0237Z} ZuHone, J.~A., 
Markevitch, M., \& Johnson, R.~E. 2009, arXiv:0912.0237 

\bibitem[\protect\citeauthoryear{ZuHone}{2010}]{2010arXiv1004.3820Z} ZuHone 
J., 2010, arXiv, arXiv:1004.3820 


\end{thebibliography}
\end{document}